\documentclass[twocolumn]{aastex63}

\usepackage{amsmath}

\newcommand{\Msun}{\,\mathrm{M}_\odot}

\newcommand{\Zsun}{\,\mathrm{Z}_\odot}
\newcommand{\au}{\,\textsc{au}}
\newcommand{\kms}{\,\mathrm{km\,s^{-1}}}
\newcommand{\Hz}{\,\mathrm{Hz}}

\newcommand{\hr}{\,\mathrm{hr}}
\newcommand{\yr}{\,\mathrm{yr}}
\newcommand{\Myr}{\,\mathrm{Myr}}
\newcommand{\Gyr}{\,\mathrm{Gyr}}

\newcommand{\Gpc}{\,\mathrm{Gpc}}
\newcommand{\perpc}{\,\mathrm{pc^{-3}}}
\newcommand{\Gpcyr}{\,\mathrm{Gpc^{-3}\,yr^{-1}}}
\newcommand{\MsunGpcyr}{\,\mathrm{M}_\odot\,\mathrm{Gpc^{-3}\,yr^{-1}}}

\received{XXX}
\revised{YYY}
\accepted{ZZZ}

\submitjournal{AAS Journals}

\shorttitle{GW progenitors in quadruple-star systems}
\shortauthors{Vynatheya \& Hamers}

\begin{document}

\title{How important is secular evolution for black hole and neutron star mergers in 2+2 and 3+1 quadruple-star systems?}

\author[0000-0003-3736-2059]{Pavan Vynatheya}
\affiliation{Max Planck Institut für Astrophysik \\
Karl-Schwarzschild-Straße 1  \\
85748 Garching bei München, Germany}

\author[0000-0003-1004-5635]{Adrian S. Hamers}
\affiliation{Max Planck Institut für Astrophysik \\
Karl-Schwarzschild-Straße 1  \\
85748 Garching bei München, Germany}

\begin{abstract}

Mergers of black holes (BHs) and neutron stars (NSs) result in the emission of gravitational waves that can be detected by LIGO. In this paper, we look at 2+2 and 3+1 quadruple-star systems, which are common among massive stars, the progenitors of BHs and NSs. We carry out a detailed population synthesis of quadruple systems using the MSE code, which seamlessly takes into consideration stellar evolution, binary and tertiary interactions, $N$-body dynamics, and secular evolution. We find that, although secular evolution plays a role in compact object (BH and NS) mergers, (70--85) \% (depending on the model assumptions) of the mergers are solely due to common envelope (CE) evolution. Significant eccentricities in the LIGO band (higher than 0.01) are only obtained with zero supernova (SNe) kicks and are directly linked to the role of secular evolution. A similar outlier effect is seen in the $\chi_{\mathrm{eff}}$ distribution, with negative values obtained only with zero SNe kicks. When kicks are taken into account, there are no systems that evolve into a quadruple consisting of four compact objects. For our fiducial model, we estimate the merger rates (in units of $\Gpcyr$) in 2+2 quadruples (3+1 quadruples) to be 10.8 $\pm$ 0.9 (2.9 $\pm$ 0.5), 5.7 $\pm$ 0.6 (1.4 $\pm$ 0.4) and 0.6 $\pm$ 0.2 (0.7 $\pm$ 0.3) for BH-BH, BH-NS and NS-NS mergers respectively. The BH-BH merger rates represent a significant fraction of the current LIGO rates, whereas the other merger rates fall short of LIGO estimates.

\end{abstract}



\section{Introduction} \label{sec:intro}

In the past few years, there have been extensive studies of gravitational wave (GW) sources and their progenitors. This has been motivated by recent detections of GWs by LIGO/VIRGO, starting in 2015. \cite{2021PhRvX..11b1053A} introduced the second and latest version of the Gravitational Wave Transient catalog (GWTC-2), which also includes the GW detections from the previous catalog (GWTC-1) of \cite{2019PhRvX...9c1040A}.

GWs are emitted during the merger of neutron stars (NSs) and black holes (BHs). These compact objects are the final stages in the evolution of massive stars ($\gtrsim 8 \Msun$, assuming solar metallicity and single star evolution). Hence, for individual BHs and NSs to merge, the progenitor massive stars must avoid merging before evolving into a compact object binary. Various stages in a star's life (radius expansion in giant phases, mass-loss due to stellar winds, external encounters, supernova kicks) tend to destroy binary systems before compact object formation. Therefore, systems with binary BHs and NSs are expected to be very rare. Mergers within a Hubble time ($\sim 14 \Gyr$) are even rarer.

Any proposed channel for the merger of compact objects (henceforth used to refer only to BHs and NSs, and not white dwarfs or WDs) must, hence, explain the presence of such systems and their merger within a Hubble time. There have been a number of merger channels proposed in the recent past, which can be divided into: (1) isolated binary evolution \citep[e.g.,][]{2002ApJ...572..407B,2012ApJ...759...52D,2016Natur.534..512B,2016MNRAS.460.3545D,2018MNRAS.474.2937C,2018MNRAS.474.2959G,2018MNRAS.480.2011G,2019MNRAS.485..889S}; (2) dynamical interactions in star clusters \citep[e.g.,][]{2000ApJ...528L..17P,2010MNRAS.402..371B,2016PhRvD..93h4029R,2017ApJ...836L..26C,2017MNRAS.467..524B,2019MNRAS.487.5630H}, galactic nuclei \citep[e.g.,][]{2009MNRAS.395.2127O,2012ApJ...757...27A,2016ApJ...831..187A,2017ApJ...846..146P,2018ApJ...856..140H,2018MNRAS.477.4423A,2018ApJ...865....2H,2019MNRAS.488...47F}, and triple and quadruple systems \citep[e.g.,][]{2017ApJ...836...39S,2017ApJ...841...77A,2018ApJ...863...68L,2019MNRAS.483.4060L,2019MNRAS.486.4443F,2019MNRAS.486.4781F,2020ApJ...895L..15F,2021A&A...650A.189A,2021MNRAS.506.5345H}; (3) in AGN disks \citep[e.g.,][]{2017MNRAS.464..946S,2017ApJ...835..165B,2018ApJ...866...66M,2019ApJ...878...85S,2020ApJ...898...25T}; or (4) primordial BH mergers \citep[e.g.,][]{2016PhRvL.116t1301B,2016PhRvL.117f1101S,2017JCAP...09..037R,2017PhRvD..96l3523A}.

Compact object mergers in triples are interesting for several reasons. Firstly, studies \citep{2010ApJS..190....1R,2017ApJS..230...15M} have found that massive stars, which are the progenitors of BHs and NSs, are most likely found in high-multiplicity star systems. \cite{2017ApJS..230...15M} showed that for stellar systems in the field with primary components more massive than $10 \Msun$, the triple and quadruple fractions each exceed $20\%$. Secondly, the presence of companion stars can significantly affect the dynamics of triple and quadruple-star systems. Unlike isolated binaries, triple star systems can undergo eccentricity enhancements (if mutual inclinations are large) in the inner binary orbits due to the presence of tertiary companions. These perturbations, to the lowest order, are known as Lidov-Kozai (LK) oscillations\footnote{Also referred to as von Zeipel-Lidov-Kozai oscillations after \cite{2019MEEP....7....1I} noted the contribution of \cite{1910AN....183..345V}} (\citealp{1962P&SS....9..719L,1962AJ.....67..591K}; see for reviews \citealp{2017ASSL..441.....S,2016ARA&A..54..441N}). LK oscillations can accelerate compact object mergers since enhanced eccentricity shortens the coalescence time due to GW-driven orbital inspiral \citep{2002ApJ...578..775B,2011ApJ...741...82T}. Thus, the study of GW progenitors is incomplete without accounting for multiple-star systems.

Population synthesis is a useful tool to study the statistical properties of such systems. \cite{2017ApJ...841...77A} used a population synthesis code \textsc{TRES} \citep{2016ComAC...3....6T} to combine the effects of orbital dynamics and stellar evolution, in order to estimate merger rates in triples. They derived BH-BH merger rates of $\sim$ (0.3--1.3)$\Gpcyr$, and showed that mergers from the triple channel have much higher eccentricities in the LIGO band ($10 \Hz$) compared to the isolated binary channel.

\begin{figure}
\plotone{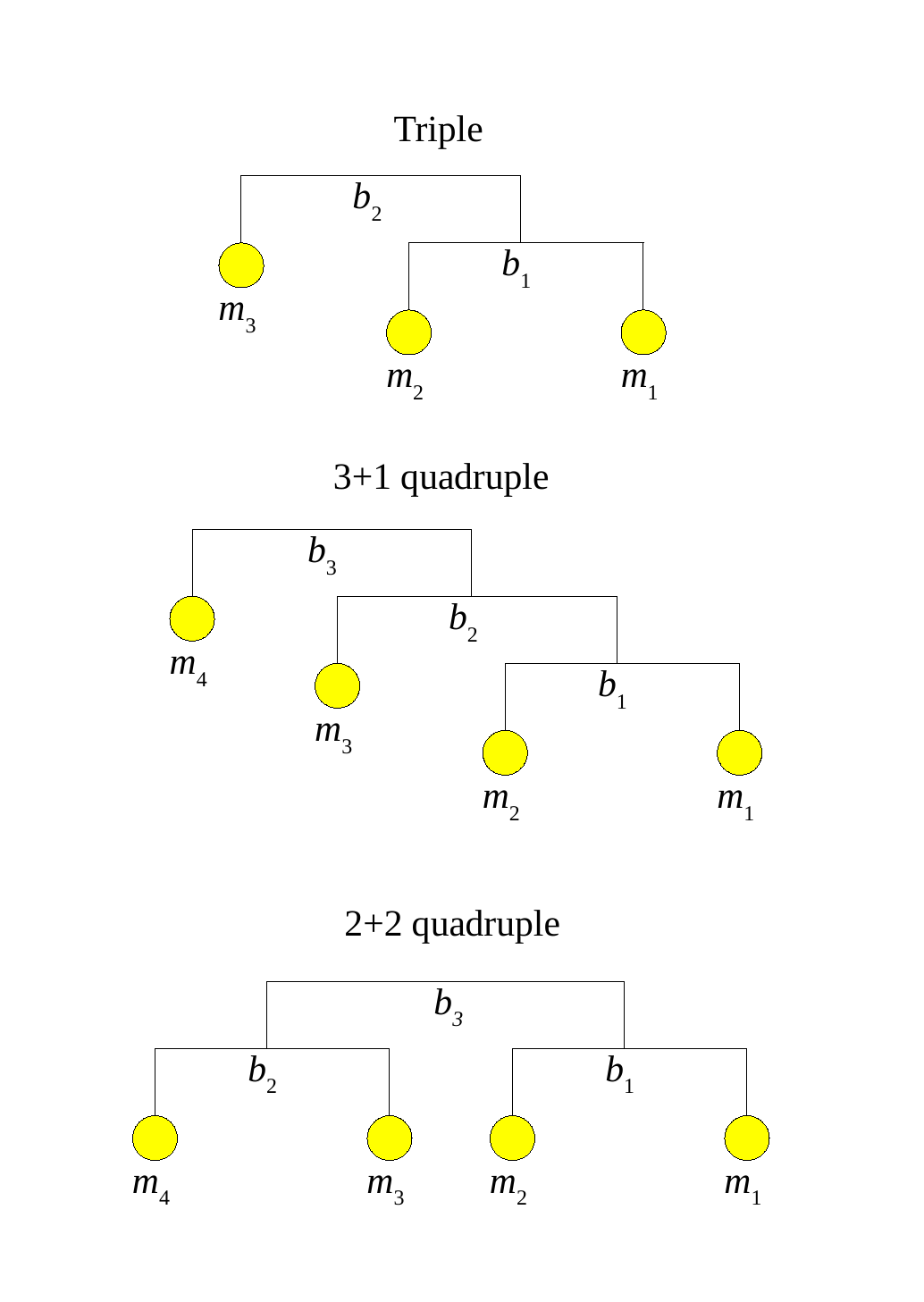}
\caption{Mobile diagrams of the two types of quadruples, with a triple for comparison. Here, the $m_i$s are masses and $b_i$s are nested binaries. \label{fig:hierarchy}}
\end{figure}

In this paper, we concentrate on quadruple-star systems. Quadruples allow for a larger parameter space than triples for eccentricity excitation due to secular (long-term) evolution \citep{2013MNRAS.435..943P,2015MNRAS.449.4221H,2017MNRAS.466.4107H,2017MNRAS.470.1657H,2018MNRAS.474.3547G,2018MNRAS.478..620H,2019MNRAS.482.2262H,2019MNRAS.483.4060L}. Smaller mutual inclinations can lead to chaotic behavior and extreme eccentricity enhancements if various secular timescales are commensurate. This, coupled with the fact that massive stars are likely to reside in triple and quadruple systems, justifies a detailed investigation into quadruples. Based on their hierarchical configuration, quadruples can be classified into two types -- (1) the 2+2, where two binaries orbit each other; and (2) the 3+1, where a triple system is orbited by a distant fourth body (see Figure \ref{fig:hierarchy}). 

There have been a few studies on 2+2 quadruples. \cite{2019MNRAS.486.4781F} carried out a population synthesis study assuming four BHs in a 2+2 configuration. In reality, the survival of bound quadruple systems consisting of four BHs is rare since gravitational dynamics and stellar evolution tend to destabilize orbits. Supernova (SNe) natal kicks are a major cause of the destruction of potential BH quadruple systems. Therefore, a thorough study self-consistently should combine both these effects to predict merger rates. Nevertheless, the authors found that merger fractions in quadruples can be $\sim$ (3--4) times higher than in triples. Thus, the quadruple channel cannot be ignored. \cite{2021MNRAS.506.5345H} performed a simplified evolution of 2+2 quadruples, where two binaries are evolved independently, and secular evolution is considered only after compact object formation. They inferred a compact object merger rate of $\sim$ (10--100)$\Gpcyr$, which they mentioned is most likely an overestimation.

In this paper, we go a step further and use the recently-developed population synthesis code \textit{Multiple Stellar Evolution} (MSE), which combines stellar evolution, binary interaction, gravitational dynamics, fly-bys in the field, and other processes seamlessly \citep{2021MNRAS.502.4479H}. Additionally, we study both 2+2 and 3+1 quadruples and compare their merger rates with other channels.

The structure of this paper is as follows. Section \ref{sec:method} describes the methods used, Section \ref{sec:example} has a few examples of compact object mergers in quadruple systems selected from a large set of population synthesis calculations, Section \ref{sec:popsyn} discusses the initial conditions and assumptions made in the population synthesis in detail, Section \ref{sec:result} presents the results, Section \ref{sec:discuss} is the discussion, and Section \ref{sec:conclude} concludes.

\section{Methods} \label{sec:method}

For our population synthesis simulations, we use the MSE code (Version 0.84), described thoroughly in \cite{2021MNRAS.502.4479H}. A brief overview of MSE is given in the following sub-sections.

\subsection{Gravitational dynamics} \label{subsec:dynamics}

MSE uses two methods to model the dynamics of a multi-body system: secular and direct $N$-body integration. The code dynamically switches between these two modes depending on the stability of the configuration at a given time step. 

The secular (orbit-averaged) approximation is used when the orbit is dynamically stable. It is faster than direct $N$-body integration since an orbit-averaged and expanded Hamiltonian is used and orbital phases are not resolved. MSE uses the code \textsc{SecularMultiple} \citep{2016MNRAS.459.2827H,2018MNRAS.476.4139H,2020MNRAS.494.5492H} for this purpose. Tides are included following the equilibrium tide model (\citealp{1981A&A....99..126H,1998ApJ...499..853E}, with the efficiency of tidal dissipation determined by the prescription of \citealp{2002MNRAS.329..897H}). Post-Newtonian (PN) terms up to orders 1PN and 2.5PN (ignoring orbit-orbit interaction terms) are considered in the secular integration mode.

In certain situations, however, the secular approximation can break down. This can be due to changes in the orbital parameters due to wind mass-loss from evolving stars, SNe natal kicks, fly-bys, or secular evolution in multiple-star systems (the latter applies particularly to 3+1 quadruple systems). The code then switches to the direct $N$-body integration mode, where Newton's equations of motions are solved using the code MSTAR \citep{2020MNRAS.492.4131R}. When the switch occurs, the positions and velocities of all bodies are computed under the assumption that the mean anomalies of all orbits evolve linearly with time. The MSTAR code uses algorithmic chain regularization for highly accurate integration for a wide range of mass ratios. It includes PN terms, although tidal effects are currently not included.

MSE switches modes from secular to direct $N$-body in the following cases.
\begin{enumerate}
    \item The system becomes dynamically unstable according to the stability criterion of \cite{2001MNRAS.321..398M}.
    \item The system enters the `semisecular regime' \citep{2012ApJ...757...27A,2016MNRAS.458.3060L,2018MNRAS.474.3547G,2018MNRAS.481.4602L,2020MNRAS.494.5492H} i.e. the timescale of angular momentum change due to secular evolution is shorter than the orbital timescale.
    \item Any orbits become unbound due to SNe kicks or mass-loss.
    \item After common envelope (CE) evolution and directly following collisions.
\end{enumerate}

The code switches back to the secular mode if it is deemed stable (see \citealp{2021MNRAS.502.4479H} for details).

\subsection{Stellar evolution} \label{subsec:evolution}

The evolution of isolated stars follows the stellar tracks from the code Single Star Evolution (SSE) \citep{2000MNRAS.315..543H}. The evolution track of a star with given mass and metallicity is fit analytically from a grid of pre-computed tracks of standard masses and metallicities. MSE uses SSE at each time step, and the orbital response to stellar mass-loss is calculated, assuming adiabatic wind mass-loss. When a star evolves into an NS or a BH, MSE accounts for the mass-loss (assumed to be instantaneous, and with no feedback of the mass lost on the rest of the system) and any natal kicks from the SNe explosion.

In multiple-star systems, however, interactions between stars can become important and binary evolution can play an important role. Many of the assumptions for mass transfer and CE evolution in MSE are based on the code Binary Star Evolution (BSE) \citep{2002MNRAS.329..897H}. An exception to this is the way that mass transfer is treated in eccentric orbits. Instead of enforcing circular orbits at the onset of mass transfer, we assume the following model for the secular orbital changes due to mass transfer in an orbit $k$:
\begin{subequations}
\begin{align}
    \frac{\dot{a}_k}{a_k} &= -2\frac{\dot{m}_\mathrm{d}}{m_\mathrm{d}} \left (1-\beta \frac{m_\mathrm{d}}{m_\mathrm{a}} \right ) \sqrt{1-e_k^2} \\
    \dot{e}_k &= -2\frac{\dot{m}_\mathrm{d}}{m_\mathrm{d}} \left (1-\beta \frac{m_\mathrm{d}}{m_\mathrm{a}} \right ) \sqrt{1-e_k^2} (1-e_k) e_k
\end{align}
\label{eq:mtransfer}
\end{subequations}
where $a_k$ and $e_k$ are the orbital semi-major axis and eccentricity, respectively, $m_\mathrm{d}$ and $m_\mathrm{a}$ are the donor and accretor mass, respectively, and $\beta$ is the mass transfer efficiency. This model is adopted from \citet{2007ApJ...667.1170S,2016ApJ...825...71D}, ignoring finite-size effects, and with an additional factor of $e_k$ in the equation for $\dot{e}_k$ to resolve the problem that the equations of motion would otherwise break down as the orbit circularizes due to mass transfer. Our model, although simplified, accommodates the onset and self-consistent treatment of eccentric mass transfer in multiple-star systems.

In addition, MSE includes prescriptions for triple mass transfer and triple CE evolution in the case when an outer star fills its Roche lobe around an inner binary, motivated by more detailed simulations. It also takes into consideration the effect of fly-bys in the field, under the impulsive approximation (see \citealp{2021MNRAS.502.4479H} for details).

\section{Examples} \label{sec:example}

In this section, we provide examples of quadruple systems which undergo compact object mergers, using the MSE code. These examples are taken from the population synthesis simulations (Section \ref{sec:popsyn}). It is important to note that the same systems can evolve differently if different random numbers are generated by the code (for example, the magnitudes and directions of the SNe natal kicks). We describe three qualitatively different scenarios of mergers, in 2+2 quadruples, with examples 1--3. For completeness, we also provide an example of a 3+1 quadruple system undergoing a merger (Scenario 3). Other examples of mergers are briefly mentioned in Section \ref{sec:discuss}.

\begin{figure*}
\centering
\includegraphics[width=\textwidth,trim={5.6cm 11.0cm 4.5cm 3.1cm},clip]{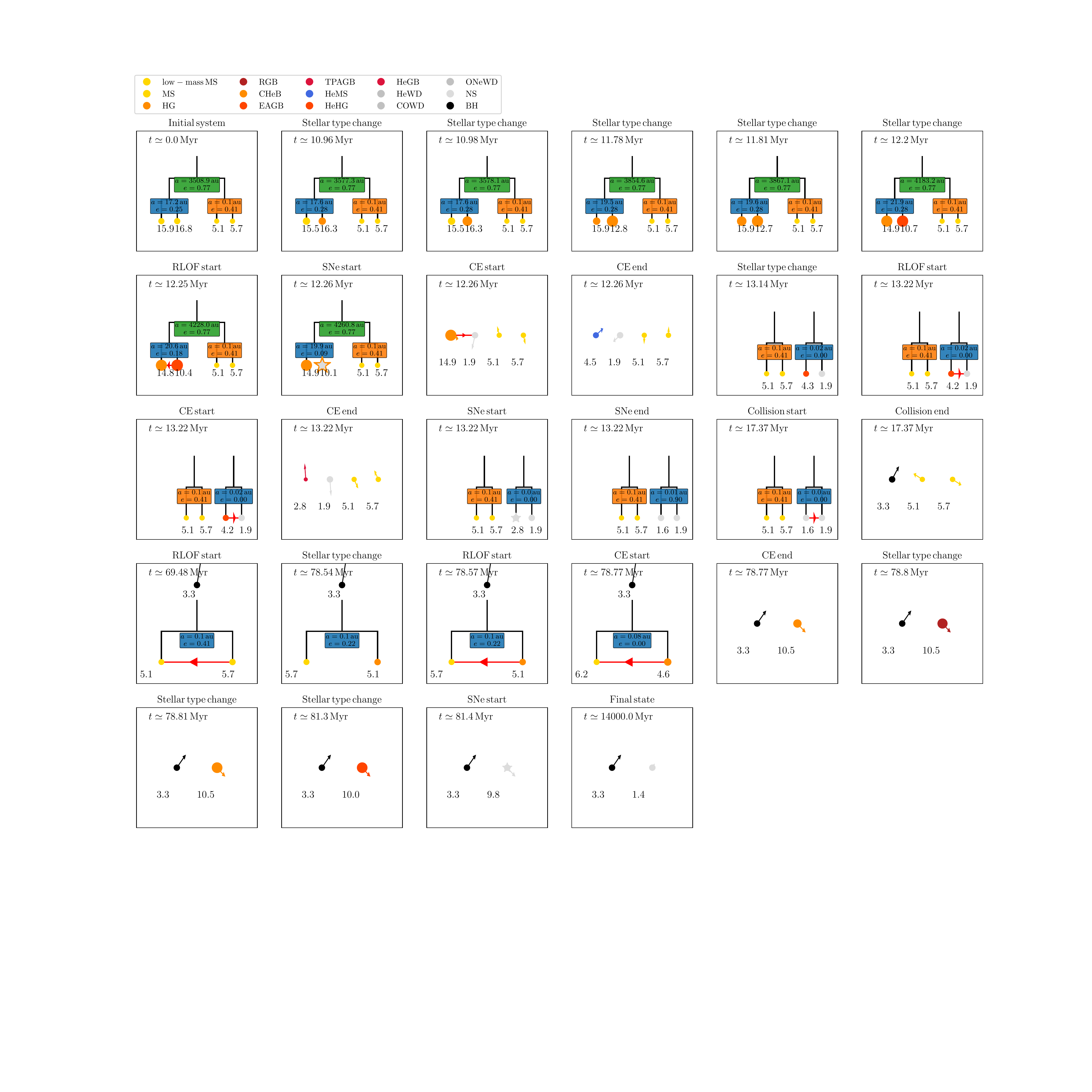}
\caption{\textit{Example 1}: (Model 0) A 2+2 quadruple system undergoing essentially isolated binary evolution with CE formation and eventually leading to a NS-NS merger. \label{fig:ex1}}
\end{figure*}

\begin{figure*}
\centering
\includegraphics[width=\textwidth,trim={6.5cm 11.8cm 5.3cm 4.0cm},clip]{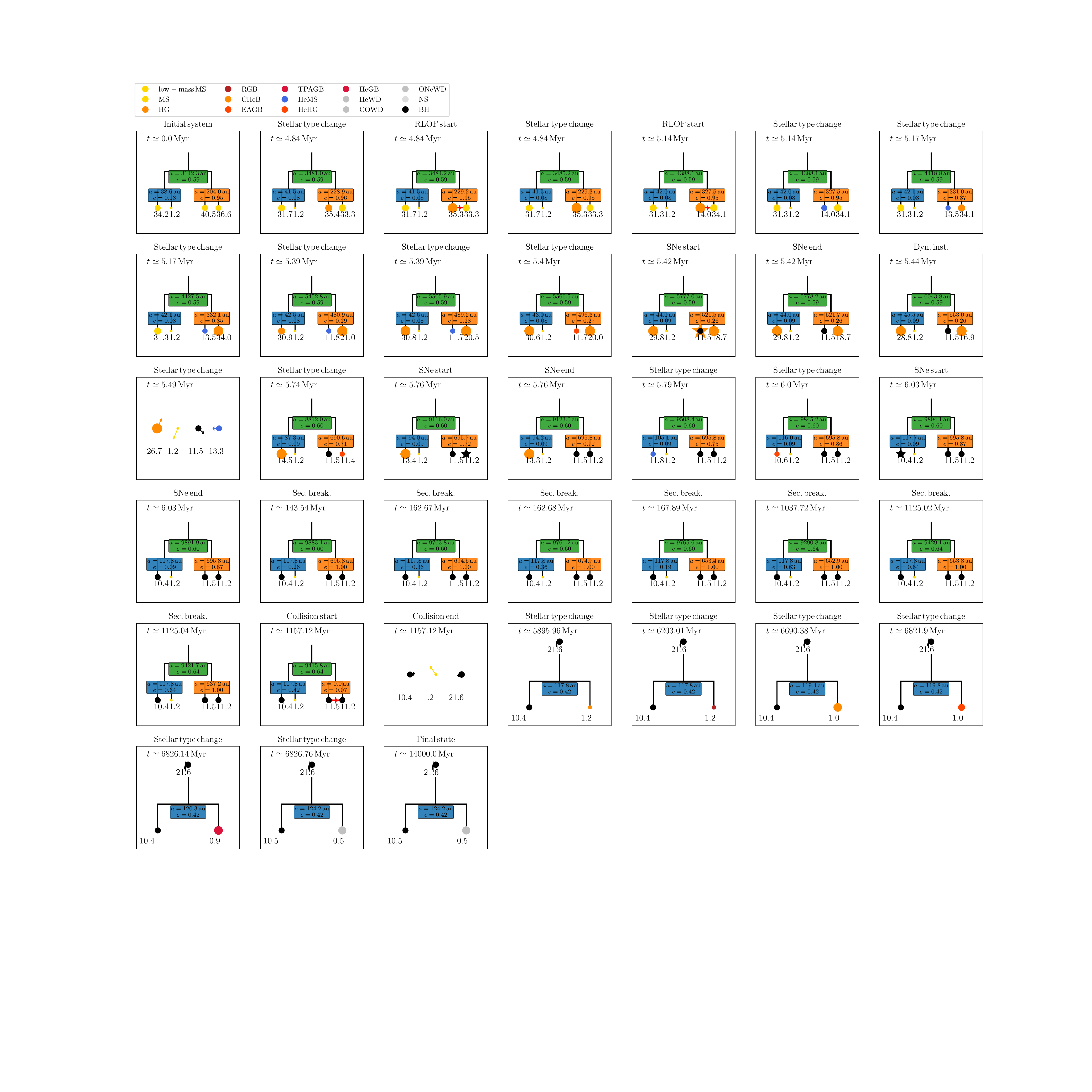}
\caption{\textit{Example 2}: (Model 1) A 2+2 quadruple system where a CE does not form, and the eventual BH-BH merger is solely due to secular evolution. \label{fig:ex2}}
\end{figure*}

\begin{figure*}
\centering
\includegraphics[width=\textwidth,trim={5.6cm 5.0cm 4.5cm 3.1cm},clip]{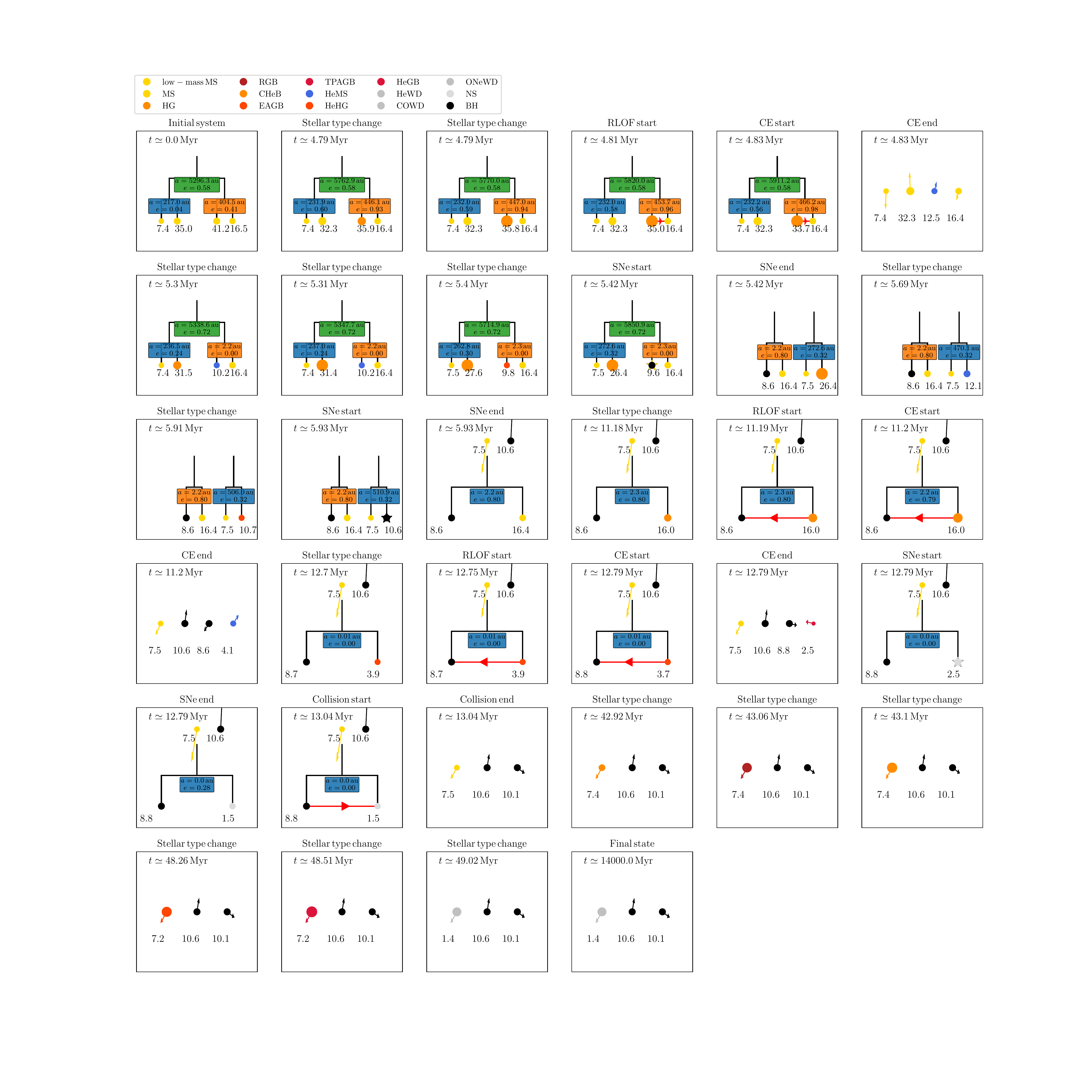}
\caption{\textit{Example 3}: (Model 0) A 2+2 quadruple system in which both secular evolution and CE evolution play a key role in the eventual BH-NS merger. \label{fig:ex3}}
\end{figure*}

\begin{figure*}
\centering
\includegraphics[width=\textwidth,trim={5.6cm 11.0cm 4.5cm 3.1cm},clip]{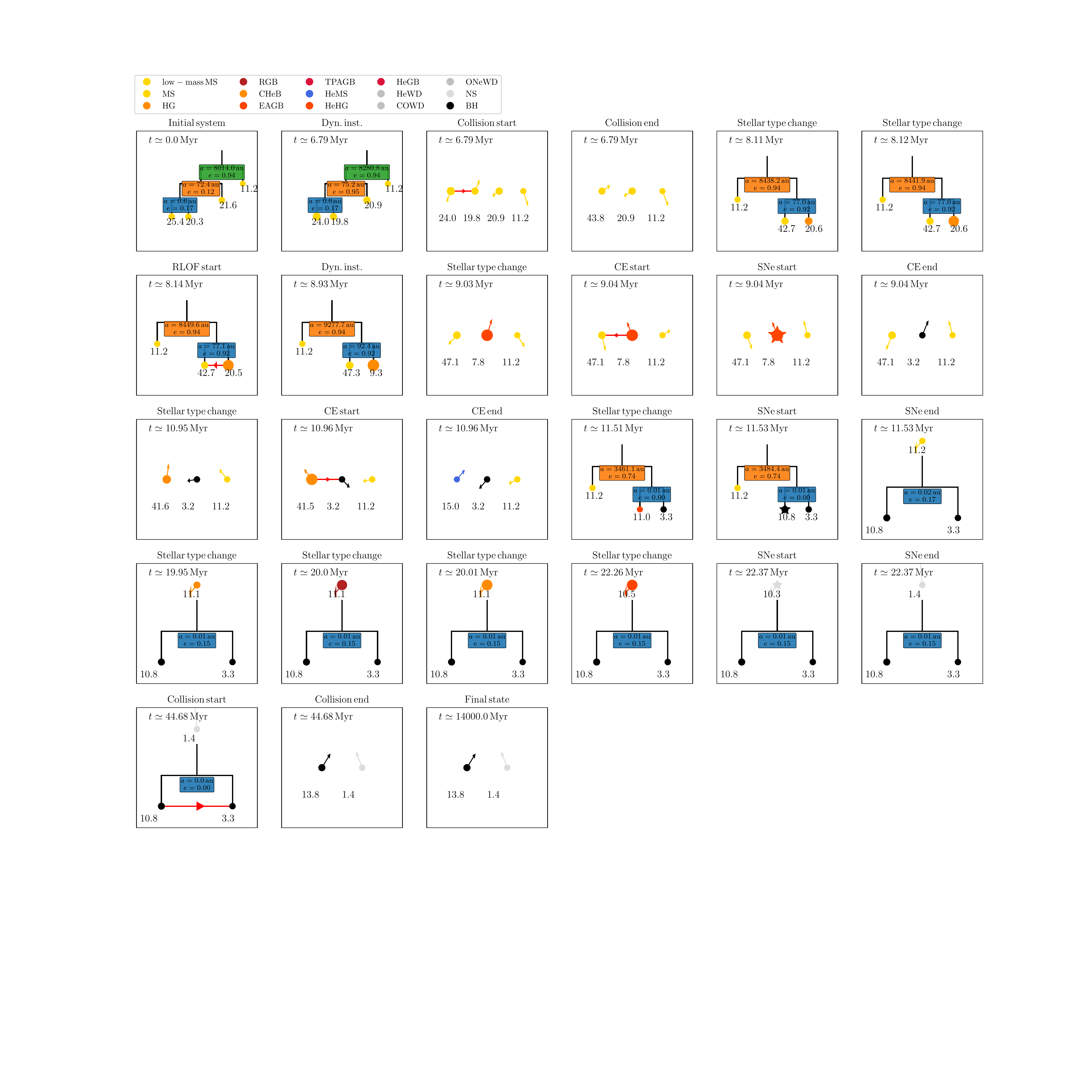}
\caption{\textit{Example 4}: (Model 0) A 3+1 quadruple system in which both secular evolution and CE evolution play a key role in the eventual BH-BH merger. \label{fig:ex4}}
\end{figure*}

\begin{enumerate}
    \item \textit{Scenario 1:} Only CE evolution \\ (most of the cases). Figure \ref{fig:ex1} (Model 0) shows a 2+2 quadruple with the inner binaries having relatively small initial semi-major axes ($a_{\mathrm{in_1}} \simeq 17 \au$ and $a_{\mathrm{in_2}} \simeq 0.1 \au$), but a much larger outer separation ($a_{\mathrm{out}} \simeq 3509 \au$, with a periapsis of $807 \au$). Owing to the very hierarchical configuration of this quadruple, the inner binaries more or less evolve independently of each other. The `interesting' inner binary, in which the merger occurs, has two massive stars of masses $16 \Msun$ and $17 \Msun$. Within $t \simeq 12.2 \Myr$, both stars reach their giant phases, and an RLOF event ensues shortly. After two SNe explosions (which unbind the two binaries from each other, but not the binaries themselves) and CE evolution, the separation between the two new neutron stars reduces significantly. Finally, at $t \simeq 17.4 \Myr$, the NS-NS binary merges due to GW emission. In the meantime, the two stars in the other binary evolve into giants, go through a CE phase of their own and merge into a single massive star. Eventually, this star evolves into an NS. The bound 2+2 quadruple therefore finally evolves into two single compact objects.
    \item \textit{Scenario 2:} Only secular evolution \\ (extremely rare, only possible with zero SNe kicks). Figure \ref{fig:ex2} (Model 1) shows a 2+2 quadruple with inner binaries with semi-major axes $a_{\mathrm{in_1}} \simeq 39 \au$ and $a_{\mathrm{in_2}} \simeq 204 \au$ (but very high eccentricity of $0.95$), and an outer orbit with semi-major axis $a_{\mathrm{out}} \simeq 3142 \au$ and a periapsis of $1288 \au$. In the evolution of this system, there are RLOF events, but no CE event to bring inner companions close to each other. The `interesting' binary has two very massive stars of $41 \Msun$ and $37 \Msun$. Since SNe kicks are disabled in this model, the quadruple system remains bound with three BHs and one low-mass main-sequence (MS) star at $t \simeq 6.0 \Myr$. By this time, the BH-BH inner binary is still wide, with a semi-major axis of $696 \au$ and a periapsis of $90 \au$. This is too wide for a merger in a Hubble time. However, secular evolution now shows its capability. The orbit eccentricity is enhanced significantly from $0.87$ to almost $1.0$ ($1-e \simeq 5\times10^{-7}$) after $t \simeq 143 \Myr$, which leads to a secular breakdown. The two BHs end up making extremely close passes to each other for many million years and ultimately collide at $t \simeq 1157 \Myr$. The LIGO band eccentricity $e_{\mathrm{LIGO}} = 5\times10^{-3}$ is higher than that of Scenario 1. The GW recoil unbinds the merger remnant from the companion binary without strongly interacting with it (i.e., the scenario proposed by \citealp{2020ApJ...895L..15F} and \citealp{2021MNRAS.506.5345H} does not occur here). The companion binary finally becomes a wide BH+WD binary without further interaction (when it was still bound to its companion, it experienced eccentricity excitation).
    \item \textit{Scenario 3:} (2+2 quadruple) Mixture of both \\ (intermediary in occurrence; see Section \ref{subsec:scenario}). Figure \ref{fig:ex3} (Model 0) shows a 2+2 quadruple with wide inner binaries (semi-major axes of $a_{\mathrm{in_1}} \simeq 217 \au$ and $a_{\mathrm{in_2}} \simeq 405 \au$) with a large outer separation ($a_{\mathrm{out}} \simeq 5296 \au$, with a periapsis of $2224 \au$). The mutual inclination between the inner and outer orbits is high enough to excite the inner eccentricities from $e_{\mathrm{in_1}} \simeq 0.04$ to $0.60$ and from $e_{\mathrm{in_2}} \simeq 0.41$ to $0.96$ respectively. Chaotic quadruple secular evolution could also play a role here since the two inner semimajor axes are not too distinct from each other (so the LK timescales are similar). This eccentricity enhancement is a key factor for the eventual compact object merger. The periapsis of the `interesting' binary (with masses $41 \Msun$ and $17 \Msun$) reduces drastically from $243 \au$ to a mere $18 \au$, which is close enough for RLOF and CE evolution to occur. This is what happens after $t \simeq 4.8 \Myr$, and the orbit shrinks even further and circularizes. Again, SNe kicks unbind the quadruple companions, but the inner binary remains intact. The final CE phase brings the two objects to a close separation of $0.01 \au$, and a BH-NS merger occurs at $t \simeq 13.0 \Myr$. Meanwhile, the other inner binary does not survive an SNe kick when its more massive star evolves into a BH. The two stars then evolve independently, with the companion becoming a high mass O-Ne WD. 
    \item \textit{Scenario 3:} (3+1 quadruple) \\ Figure \ref{fig:ex4} (Model 0) shows a 3+1 quadruple with a very close inner binary ($a_{\mathrm{in}} \simeq 0.6 \au$), a fairly distant intermediate star ($a_{\mathrm{mid}} \simeq 72 \au$) and a very distant, but very eccentric, outer star ($a_{\mathrm{out}} \simeq 8014 \au$, with a periapsis of $480 \au$). Secular evolution excites the intermediate eccentricity from a low $0.12$ to a very high $0.95$, with the intermediate periapsis decreasing from $64 \au$ to just $4 \au$. This triggers a phase of dynamical instability; due to the tight inner semi-major axis, the massive inner stars (of masses $25 \Msun$ and $20 \Msun$) collide early and become a $44 \Msun$ star. The resulting system is a triple, with a high inner (previously intermediate) eccentricity. The previously intermediate $20 \Msun$ star becomes a giant and transfers mass to its very massive companion. Note that this type of evolution (the onset of RLOF of a star onto a companion which is twice as massive) is not expected in isolated binary evolution, and is unique to higher-order multiple systems. After another dynamical instability phase, a CE event follows and the two stars end up in a near-circular orbit with a semi-major axis $0.01 \au$. The inner stars survive two SNe events (although the outer companion is kicked out), become a BH-NS binary, and merge within $t \simeq 13 \Myr$. Meanwhile, the ejected outer $8 \Msun$ star eventually becomes an NS.
\end{enumerate}



\section{Population synthesis} \label{sec:popsyn}

We perform a population synthesis of both 2+2 and 3+1 quadruples. Each system is run for $14 \Gyr$ or a Hubble time. In the case of the 2+2 quadruples, we also compare results with isolated binaries having the same initial conditions as the inner binaries of the quadruples. Hence, we have twice as many isolated binaries as 2+2 quadruples for comparison. It is important to note that these isolated binaries are not distributed like real binaries in the field, mainly due to the quadruple stability requirements. The latter implies that the inner binaries in 2+2 quadruples are, on average, more compact compared to `truly' isolated binaries in the field. We include this additional set to directly investigate the effect of secular evolution in quadruples.

For each of the three above-mentioned cases, we use 7 different models which vary certain distributions or parameters (see Section \ref{subsec:models} for details). In each model, we run $10^5$ quadruples (or $2\times10^5$ binaries in the isolated binaries cases). Hence, in total, we run $7\times10^5$ 2+2 and 3+1 quadruples each, and $14\times10^5$ isolated binaries, making a total of $2.8\times10^6$ systems.

The following sub-sections describe the different models used, the initial conditions, and the other important parameters.

\subsection{Different models} \label{subsec:models}

To better comprehend our results, we vary different parameters to see how they affect compact object merger rates. We use seven different models, numbered 0, 1, 2, 3a, 3b, 4a, 4b (see Table \ref{tab:modelparam}). Model 0 is taken to be our fiducial model, and the others are compared against it. In Model 1, SNe kicks during NS and BH formation are excluded. In Model 2, fly-bys, which are enabled by default, are excluded. In Models 3a and 3b, the stellar metallicities $Z$ are changed from solar ($\Zsun = 2\times10^{-2}$) to one-tenth and one-hundredth of solar metallicity respectively. In Models 4a and 4b, the CE mass-loss timescale $t_{\mathrm{CE}}$, which is chosen to be $10^3 \yr$ by default, is varied to one-tenth and ten times that of the default respectively. $t_{\mathrm{CE}}$ parameterizes the timescale at which mass is lost during a CE event. Specifically, if $t_{\mathrm{CE}}$ is short compared to the orbital period of the companions to the two stars undergoing CE evolution, then the mass-loss can be considered to be instantaneous for these orbits, and the companions can become unbound. In contrast, if $t_{\mathrm{CE}}$ is long compared to the orbital motion of the outer companions, the effect can be considered to be adiabatic and the outer orbits remain bound and become wider. In MSE, these regimes, and the transitional regime, are taken into account by carrying out short term $N$-body integrations in which the mass of the binary undergoing CE evolution (modeled as a point mass) gradually loses mass (see \citealp{2021MNRAS.502.4479H} for details).

Here, we stress that the parameters discussed above are among the many which could have been altered. For example, the CE evolution prescription, the $\alpha$-CE model (\citealp{1976IAUS...73...75P,1976IAUS...73...35V,1984ApJ...277..355W,1988ApJ...329..764L,1990ApJ...358..189D}; see for review \citealp{2013A&ARv..21...59I}), is uncertain. In this paper, we choose to alter $t_{\mathrm{CE}}$ since this effect has not been studied much, whereas it can determine whether or not outer companions remain bound after a CE event (e.g. \citealp{2019MNRAS.484.4711M}). However, other parameters, most notably the $\alpha_{\mathrm{CE}}$ parameter, also significantly affect binary evolution and compact object merger rates in particular \citep{2012ApJ...759...52D,2021MNRAS.tmp.2473B,2021ApJ...918L..38F}. Another example is the choice of a model for the SNe kick distribution. We choose a Maxwell-Boltzmann distribution for both NSs and BHs. While the NS kick distribution has been widely studied \citep[e.g.,][]{2005MNRAS.360..974H}, the BH kick distribution is largely unknown. Other assumptions we make include the initial distributions for stellar masses, semi-major axes, and eccentricities. They are still poorly constrained for quadruples; here, we chose to focus on model assumptions, rather than different initial conditions.

The parameters varied in the different models are given in Table \ref{tab:modelparam}. The details of important parameters are given in Section \ref{subsec:ICs}.

\begin{deluxetable}{cccccc}
\tablecaption{Parameters varied in different models. \label{tab:modelparam}}
\tablewidth{0pt}
\tablehead{
\colhead{Model} & \colhead{SNe kicks} & \colhead{fly-bys} & \colhead{$Z$}
& \colhead{$t_{\rm CE}$}
}
\startdata
0 & non-zero & included & $2\times10^{-2}$ & $10^3 \yr$ \\
1 & \textbf{zero} & included & $2\times10^{-2}$ & $10^3 \yr$ \\
2 & non-zero & \textbf{excluded} & $2\times10^{-2}$ & $10^3 \yr$ \\
3a & non-zero & included & \boldmath$2\times10^{-3}$ & $10^3 \yr$ \\
3b & non-zero & included & \boldmath$2\times10^{-4}$ & $10^3 \yr$ \\
4a & non-zero & included & $2\times10^{-2}$ & \boldmath$10^2 \yr$ \\
4b & non-zero & included & $2\times10^{-2}$ & \boldmath$10^4 \yr$ \\
\enddata
\end{deluxetable}

\subsection{Initial conditions} \label{subsec:ICs}

For each of the quadruple systems, we sample four zero-age main-sequence (ZAMS) masses $m_i$, and three each of eccentricities $e_i$, semi-major axes $a_i$, orbital inclinations $i_i$, longitudes of the ascending node $\Omega_i$ and arguments of periapsis $\omega_i$. Stellar metallicities have constant values in each model, as given in Table \ref{tab:modelparam}. 
After sampling, we check for dynamical stability of the orbital configuration using the triple stability criterion described by \cite{2001MNRAS.321..398M}. For 2+2 quadruples, the stability criterion is evaluated for both inner binaries, considering the other binary to be the `tertiary' star. In the case of 3+1 quadruples, the two `triple' systems are (1) the innermost binary, with the intermediate star as the companion; and (2) the intermediate star-inner two stars binary, with the outermost star as the companion. Furthermore, we also check that the ZAMS stars (with a mass-radius relation given by $m \sim R^{0.7}$, in Solar units) do not fill their Roche lobes at periapsis (approximate Roche lobe radii are given by \citealp{1983ApJ...268..368E}). If any of these criteria are not fulfilled, the sampling is restarted.

\begin{figure}
\plotone{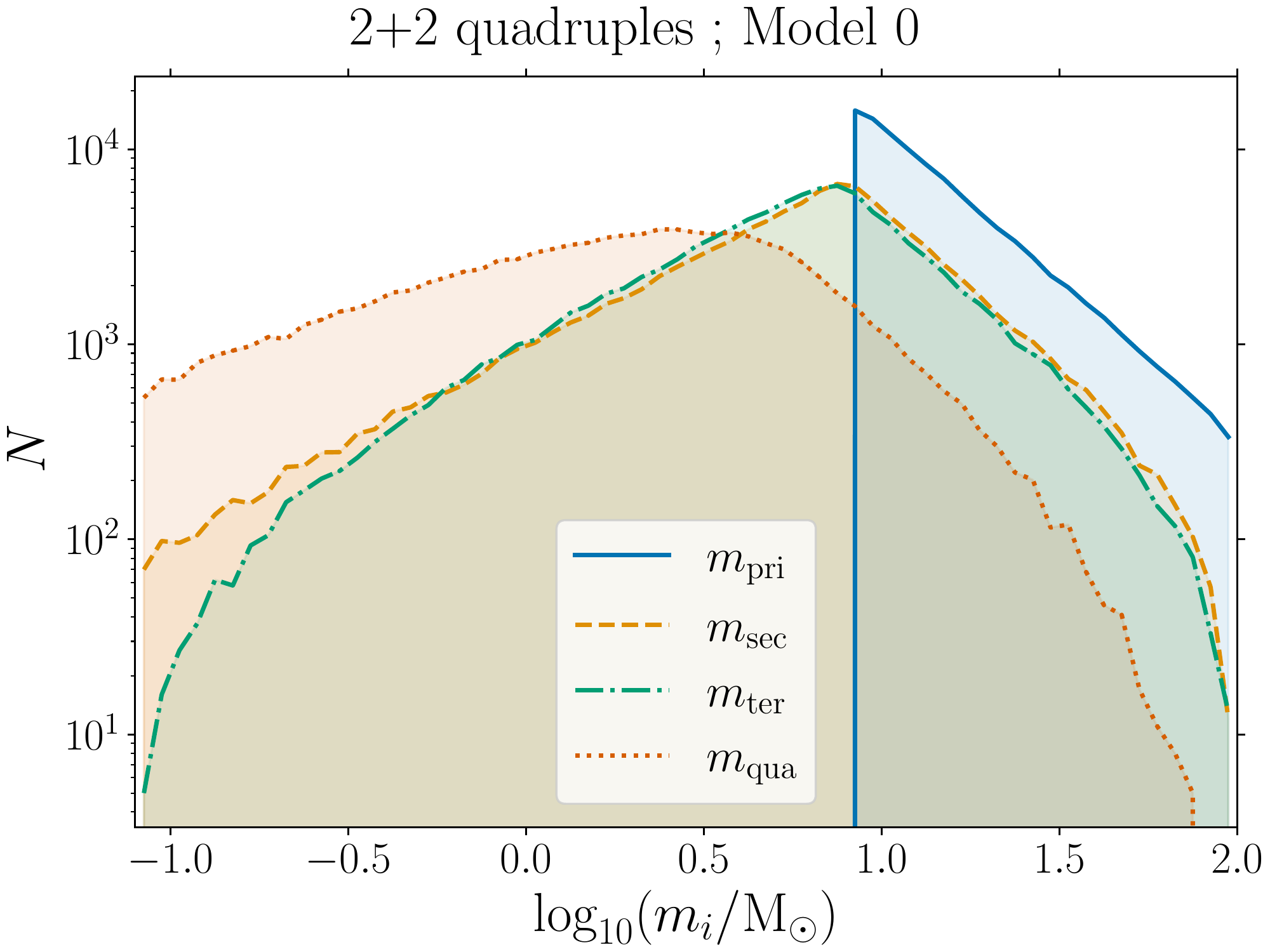}
\caption{Initial mass distribution of $10^5$ 2+2 quadruple systems for all models. $m_{\mathrm{pri}}$ is the most massive star, $m_{\mathrm{sec}}$ is its immediate companion, $m_{\mathrm{ter}}$ is the heavier star in the other binary and $m_{\mathrm{qua}}$ is its final companion. \label{fig:2p2_ICmasses}}
\end{figure}

\begin{figure}
\plotone{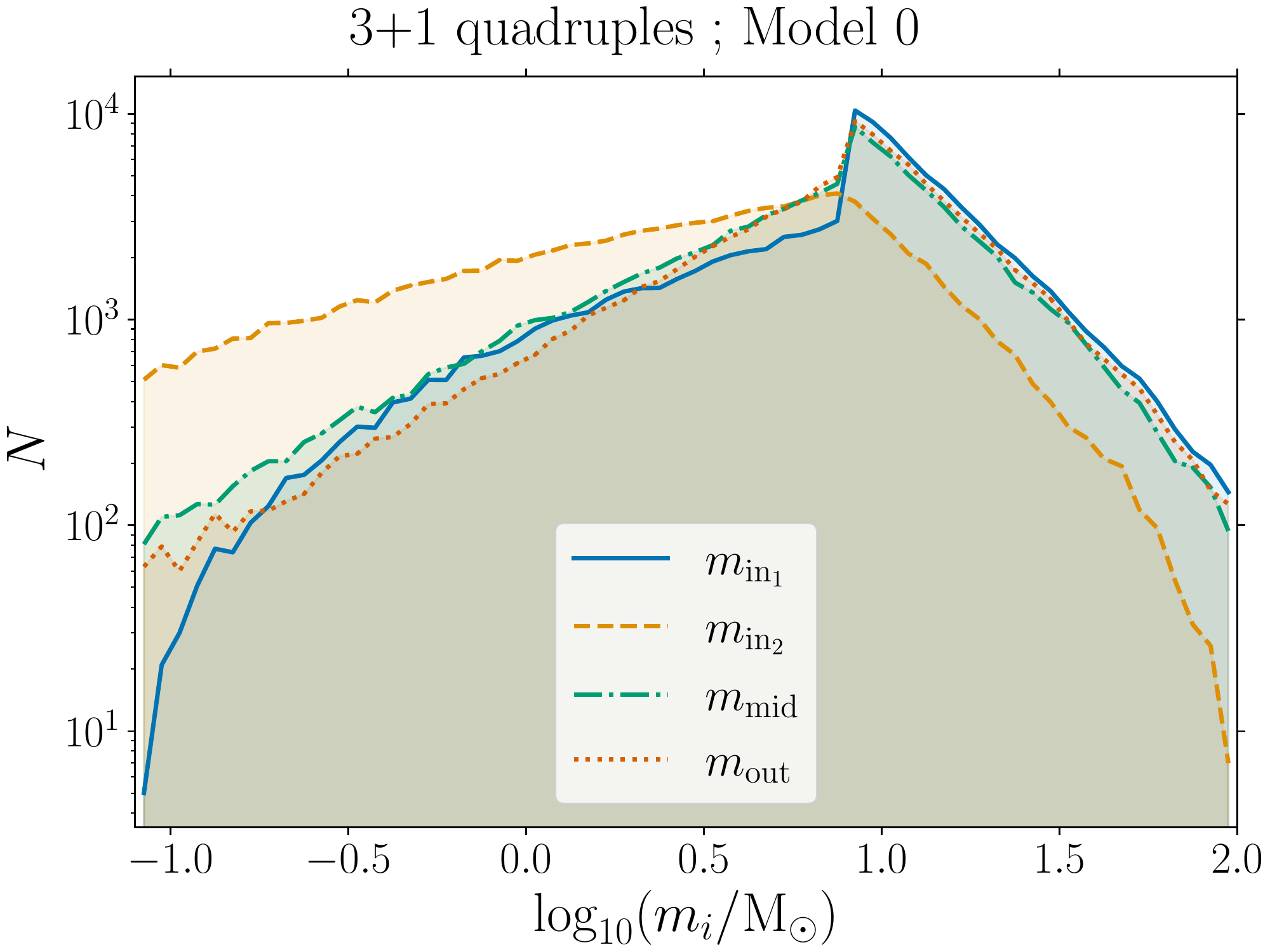}
\caption{Initial mass distribution of $10^5$ 3+1 quadruple systems for all models. $m_{\mathrm{in_1}}$ is the heavier inner star, $m_{\mathrm{in_2}}$ is its inner companion, $m_{\mathrm{mid}}$ is the intermediate star orbiting around the inner binary and $m_{\mathrm{out}}$ is the outer star orbiting around the triple. \label{fig:3p1_ICmasses}}
\end{figure}

\begin{figure}
\plotone{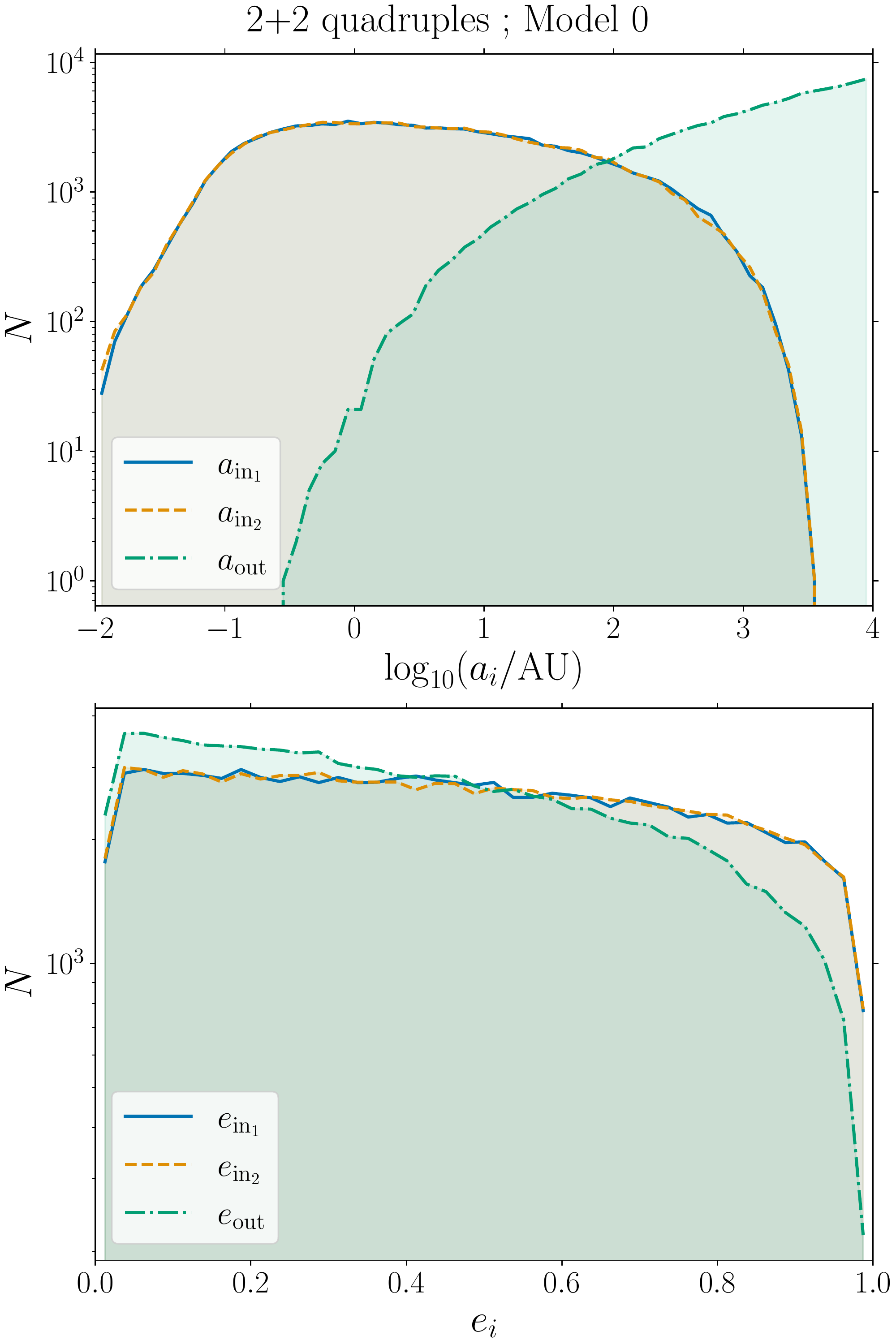}
\caption{Initial semi-major axis and eccentricity distribution of $10^5$ 2+2 quadruple systems for all models. Subscripts `${\mathrm{in_1}}$' and `${\mathrm{in_2}}$' refer to the inner binaries, and `${\mathrm{out}}$' refers to the outer binary (see Figure \ref{fig:hierarchy}). \label{fig:2p2_ICorbits}}
\end{figure}

\begin{figure}
\plotone{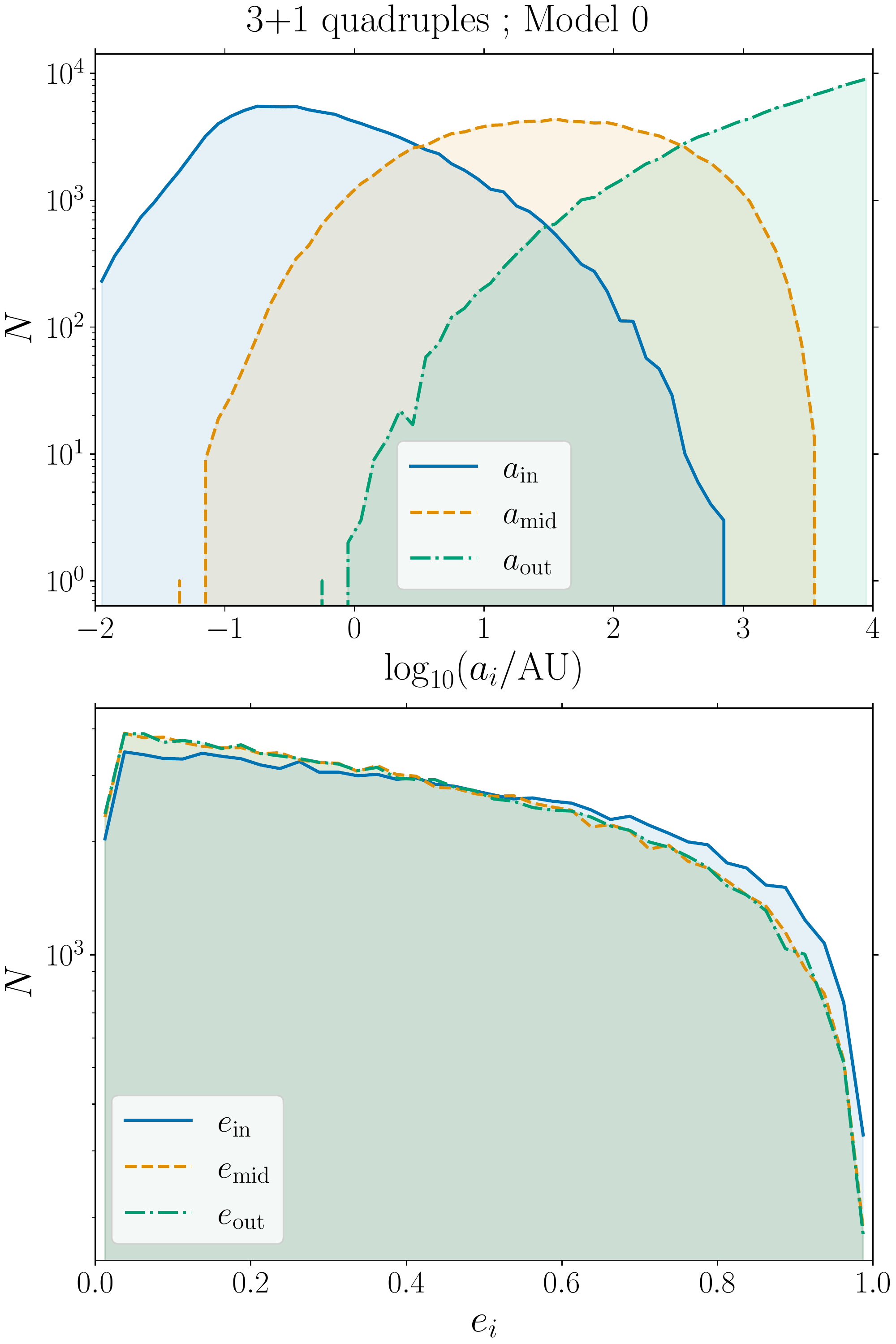}
\caption{Initial semi-major axis and eccentricity distribution of $10^5$ 3+1 quadruple systems for all models. Subscripts `${\mathrm{in}}$', `${\mathrm{mid}}$' and `${\mathrm{out}}$' refer to the inner, intermediate and outer binaries respectively (see Figure \ref{fig:hierarchy}). \label{fig:3p1_ICorbits}}
\end{figure}

\begin{itemize}
    \item \textit{Mass sampling:} The primary mass (most massive star in the quadruple system) $m_{\mathrm{pri}}$ is sampled from a Kroupa distribution ($\mathrm{d}N/\mathrm{d}m \propto m^{-2.3}$ in the high-mass tail; \citealp{2001MNRAS.322..231K}). However, we put an additional constraint that $m_{\mathrm{pri}} > 8 \Msun$, which is justified since we look specifically for BH and NS mergers. In principle, it could be possible to form BHs and NSs in multiple-star systems with primary masses $m_{\mathrm{pri}} < 8 \Msun$ if stellar mergers occur during the evolution, e.g., due to secular eccentricity excitation. However, in post-processing, we find zero compact object mergers in systems with primary masses lower than $10\,\Msun$. The maximum possible mass of each star is $100 \Msun$, a constraint set by SSE.
    
    In the case of 2+2 quadruples, the secondary mass (immediate companion star of the primary) $m_{\mathrm{sec}}$ is sampled from a flat mass ratio distribution $\sim U(0,1)$ with respect to the primary ($q = m_{\mathrm{sec}}/m_{\mathrm{pri}} < 1$). The companion binary is sampled similarly with respect to the original binary and is then split randomly to give the final two masses.
    
    In 3+1 quadruples, the primary mass can be (1) one of the inner stars (twice as likely as the other two cases), (2) the intermediate star or (3) the outer star. In the first case, we sample the immediate companion and the other two stars in a way similar to that of the 2+2 quadruples ($q = m_{\mathrm{in_2}}/m_{\mathrm{in_1}} < 1$). In the second case, the immediate companion of the primary is the inner binary, whose mass is sampled from a flat mass ratio distribution $\sim U(0,2)$ with respect to to the primary ($q = (m_{\mathrm{in_1}} + m_{\mathrm{in_2}})/m_{\mathrm{mid}} < 2$). Then, this mass is split between the two inner stars as described in the case of 2+2 quadruples. The outer mass is sampled from the total mass of the inner triple. In the third case, the companion of the primary is the inner triple, and is sampled from a flat mass ratio distribution $\sim U(0,3)$ with respect to to the primary ($q = (m_{\mathrm{in_1}} + m_{\mathrm{in_2}} + m_{\mathrm{mid}})/m_{\mathrm{out}} < 3$). The individual star masses are then split as before.
    
    In each of the samplings, we check that every star has a minimum mass of $0.08 \Msun$ and a maximum mass equal to that of the primary. If not satisfied, the sampling is done again. The obtained mass distributions are shown in Figure \ref{fig:2p2_ICmasses} (for 2+2 quadruples; all models) and \ref{fig:3p1_ICmasses} (for 3+1 quadruples; all models).
    
    \item \textit{Semi-major axis sampling:} The $a_i$s are sampled independent of each other, from a log flat distribution \citep{2012Sci...337..444S}, ranging from $10^{-2} \au$ to $10^4 \au$. However, the final stability check is biased towards more hierarchical orbits, which is seen in Figures \ref{fig:2p2_ICorbits} (for 2+2 quadruples; all models) and \ref{fig:3p1_ICorbits} (for 3+1 quadruples; all models). 
    
    \item \textit{Eccentricity sampling:} The $e_i$s are also sampled independent of each other, from a flat distribution \citep{2013ARA&A..51..269D}, ranging from $0.01$ to $0.99$. The final stability check is biased towards lower eccentricities, especially the outer ones, which is seen in Figures \ref{fig:2p2_ICorbits} (for 2+2 quadruples; all models) and \ref{fig:3p1_ICorbits} (for 3+1 quadruples; all models). The flat distribution of eccentricities is chosen over the commonly adopted thermal distribution \citep{1919MNRAS..79..408J}.
    
    \item \textit{Orbital angles sampling:} The orbital angles are sampled such that the orbits are distributed isotropically in 3D space. The $i_i$s are hence distributed uniformly in $\cos{i}$, whereas the $\Omega_i$s and the $\omega_i$s are distributed randomly from $0$ to $2 \pi$.
    
   Studies have shown that orbital alignment in multiple-star systems is not always isotropic. \cite{2017ApJ...844..103T} showed that, while low-mass triples with wider outer orbits are nearly isotropic, tighter triples are more aligned to an orbital plane. However, the study also showed that high-mass triples do not show as significant an alignment. Hence, our assumption of isotropy is justified.
\end{itemize}

\subsection{Other parameters and assumptions} \label{subsec:params}
There are various parameters, other than the initial conditions, which significantly affect MSE's results. Since MSE does not model the detailed stellar structure, many parameter values are either prescription-dependent or assumed. Some of the important parameters, and the assumptions involved, are mentioned below.

\begin{itemize}
    \item \textit{Supernova kicks:} The default SNe kick distribution in MSE is Maxwellian for both NSs and BHs.  Here, $\sigma_{\mathrm{NS}} = 265 \kms$ \citep{2005MNRAS.360..974H} and $\sigma_{\mathrm{BH}} = 50 \kms$. In our models with SNe kicks (all except Model 1), we use the default distribution.
    \item \textit{Fly-bys:} MSE samples stars passing by the multiple-star system assuming a homogeneous stellar background of solar density ($n_{\star} = 0.1 \perpc$) and a Maxwellian distribution of stellar velocities, with dispersion $\sigma_{\star} = 30 \kms$, consistent with the Solar neighborhood \citep{2008gady.book.....B,2017AJ....154..272H}. We adopt an encounter sphere radius of $R_{\mathrm{enc}} = 10^5 \au$, with the perturber masses following the Kroupa distribution. Only impulsive encounters (the orbital velocity is much lower than the velocity of the external star) are assumed to affect the orbits since the effects of secular encounters are usually unimportant in low-density systems in the field (this is different for dense stellar systems, see \citealp{1975MNRAS.173..729H,1996MNRAS.282.1064H,2019MNRAS.487.5630H}; and also Section \ref{subsec:caveats}). In our models with fly-bys (all except Model 2), we use this method.
    \item \textit{CE parameters:} MSE uses the energy argument-based $\alpha$-CE prescription. The common envelope efficiency $\alpha_{\mathrm{CE}} = 1$ by default. In this paper, we use three different values for the CE mass-loss timescale $t_{\mathrm{CE}}$, as seen in Table \ref{tab:modelparam}.
    \item \textit{Collisions:} In MSE, a `collision' between stars is assumed to have occurred when their mutual separation is lesser than the sum of their effective radii. The effective radius is the same as the stellar radius for non-compact objects, whereas it is a factor 100 more for compact objects. This is done since it is computationally expensive to integrate the equations of motion just before a compact object merger, and justified by the very short remaining merger time (see also \citealp{2021MNRAS.502.4479H}, Eq. 108).
\end{itemize}

Table \ref{tab:quantities} summarizes the initial conditions and adopted parameters.

\begin{deluxetable*}{ll}
\tablecaption{Initial conditions and parameters. \label{tab:quantities}}
\tablehead{
\colhead{Quantity} & \colhead{Distribution or value}
}
\startdata
Masses $m$ & Primary from Kroupa distribution ($> 8 \Msun$); others as mass ratios of primary \\
Metallicities $Z$ & $0.02$ \textit{(changed in models 3a and 3b)} \\
Semi-major axes $a$ & Log flat distribution ($10^{-2} \au$ -- $10^4 \au$); only stable systems \\
Eccentricities $e$ & Flat distribution ($0.01$ -- $0.99$); only stable systems \\
Inclinations $i$ & Flat in $\cos{i}$ \\
Longitudes of ascending node $\Omega$ & Flat distribution ($0$ -- $2 \pi$) \\
Arguments of periapsis $\omega$ & Flat distribution ($0$ -- $2 \pi$) \\
\hline
Supernova kicks & Maxwellian distribution; $\sigma_{\mathrm{NS}} = 265 \kms$ and $\sigma_{\mathrm{BH}} = 50 \kms$ \textit{(0 for model 1)} \\
Fly-bys & $n_{\star} = 0.1 \perpc$; $\sigma_{\star} = 30 \kms$ (Maxwellian); $R_{\mathrm{enc}} = 10^5 \au$ \textit{(0 for model 2)} \\
CE parameters & $\alpha_{\mathrm{CE}=1}$; $t_{\mathrm{CE}=10^3} \yr$ \textit{(changed in models 4a and 4b)}
\enddata
\tablecomments{See text for detailed descriptions.}
\end{deluxetable*}

\section{Results} \label{sec:result}
In Section \ref{subsec:def}, we define quantities of interest pertaining to compact object mergers. In Section \ref{subsec:res2p2}, we present the compact object merger numbers for both 2+2 quadruples and isolated binaries, for direct comparison.  In Section \ref{subsec:res3p1}, we do the same for 3+1 quadruples. In Sections \ref{subsec:scenario} and \ref{subsec:rate}, we discuss in detail merger scenarios and rates respectively. Finally, in Section \ref{subsec:toolong}, we talk about systems which are ignored in this study.

\subsection{Definitions of certain quantities} \label{subsec:def}

One of the most important features of a compact object merger is the eccentricity $e_{\mathrm{LIGO}}$ in the LIGO band. Firstly, we have the analytical relation between $a$ and $e$ due to GW emission given by \citep{1964PhRv..136.1224P}:
\begin{equation}
    a(e) = C_0 \frac{e^{12/19}}{1-e^2} \left[1 + \frac{121}{304}e^2\right]^{870/2299}
\label{eq:ae_Peters}
\end{equation}
where $C_0$ depends on the initial values $a_0$ and $e_0$. We also have a relation for the GW peak frequency for given $a$, $e$ and total mass $M = m_1 + m_2$ from \cite{2003ApJ...598..419W}:
\begin{equation}
    f_{\mathrm{GW}}(a,e,M) = \frac{\sqrt{G M}}{\pi} \frac{(1+e)^{1.1954}}{[a (1-e^2)]^{1.5}}
\label{eq:fGW_Wen}
\end{equation}
Using these equations and adopting $f_{\mathrm{LIGO}} = 10 \Hz$, we can calculate $e_{\mathrm{LIGO}}$.

Another feature of a LIGO detection is the effective spin parameter $\chi_{\mathrm{eff}}$:
\begin{equation}
    \chi_{\mathrm{eff}} = \frac{\chi_1 m_1 (\hat{\textbf{S}}_1 \cdot \hat{\textbf{L}}) + \chi_2 m_2 (\hat{\textbf{S}}_2 \cdot \hat{\textbf{L}})}{m_1 + m_2}
\label{eq:chieff}
\end{equation}
where $\hat{\textbf{S}}_1$ and $\hat{\textbf{S}}_2$ are the unit spin angular momentum vectors of the two compact objects with $\chi_i = c \| \vec{\textbf{S}}_i \| / G m_i^2$ lying between 0 and 1 ($c$ and $G$ have their usual meaning), and $\hat{\textbf{L}}$ is the unit Newtonian orbital angular momentum. We assume that the spins during compact object formation $\chi_1$ and $\chi_2$ are sampled uniformly between 0 and 1. It should be noted that assuming a different range for $\chi_i$ results only in a horizontal re-scaling, while the distribution shape remaining the same (e.g. \citealp{2021MNRAS.506.5345H} show this for the range 0 to 0.1).

We also calculate another spin parameter, the spin precession parameter $\chi_{\mathrm{p}}$ \citep{2015PhRvD..91b4043S,2020PhRvD.102d3015A}:
\begin{equation}
    \chi_{\mathrm{p}} = \mathrm{max} \left\{ \chi_1 \| \hat{\textbf{S}}_{1\perp} \|,\kappa \chi_2 \| \hat{\textbf{S}}_{2\perp} \| \right\}
\label{eq:chip}
\end{equation}
where $\hat{\textbf{S}}_{i\perp} = \hat{\textbf{S}}_i - (\hat{\textbf{S}}_i \cdot \hat{\textbf{L}}) \hat{\textbf{L}}$ (component of $\hat{\textbf{S}}_i$ perpendicular to $\hat{\textbf{L}}$) and $\displaystyle \kappa = \frac{q (4q + 3)}{4 + 3q}$.

A final important quantity is the merger mass ratio $q = m_2/m_1$, where $m_1$ and $m_2$ are the heavier and lighter compact object masses respectively ($0 < q \leq 1$).

\subsection{2+2 quadruples and isolated binaries} \label{subsec:res2p2}

Table \ref{tab:merger_num} shows the number of compact object mergers, and the Poisson errors, in $10^5$ and $2\times10^5$ sampled systems, for each model of 2+2 quadruples and isolated binaries, respectively. It also shows the merger rates for the 2+2 quadruples. It is important to note that these isolated binaries are not distributed like `real' binaries, and thus, it is irrelevant to consider their rates. Their numbers should be considered only as a direct comparison to 2+2 quadruples, and not out of this context.

\begin{deluxetable*}{cc|ccc|ccc|ccc}
\tablecaption{Number of compact object mergers in $10^5$ 2+2 and 3+1 quadruples, and $2\times10^5$ isolated binaries. \label{tab:merger_num}}
\tablehead{
\colhead{Model} & \colhead{Description} & \multicolumn3c{2+2 quadruples} & \multicolumn3c{Isolated binaries} & \multicolumn3c{3+1 quadruples} \\
\colhead{} & \colhead{} & \colhead{BH-BH} & \colhead{BH-NS} & \colhead{NS-NS} & \colhead{BH-BH} & \colhead{BH-NS} & \colhead{NS-NS} & \colhead{BH-BH} & \colhead{BH-NS} & \colhead{NS-NS}
}
\startdata
0 & Fiducial & 156 \tiny $\scriptscriptstyle \pm$ 13 & 82 \tiny $\scriptscriptstyle \pm$ 9 & 9 \tiny $\scriptscriptstyle \pm$ 3 & 168 \tiny $\scriptscriptstyle \pm$ 13 & 80 \tiny $\scriptscriptstyle \pm$ 9 & 14 \tiny $\scriptscriptstyle \pm$ 4 & 32 \tiny $\scriptscriptstyle \pm$ 6 & 15 \tiny $\scriptscriptstyle \pm$ 4 & 8 \tiny $\scriptscriptstyle \pm$ 3 \\
1 & 0 kicks & 285 \tiny $\scriptscriptstyle \pm$ 17 & 215 \tiny $\scriptscriptstyle \pm$ 15 & 351 \tiny $\scriptscriptstyle \pm$ 19 & 272 \tiny $\scriptscriptstyle \pm$ 16 & 194 \tiny $\scriptscriptstyle \pm$ 14 & 346 \tiny $\scriptscriptstyle \pm$ 19 & 164 \tiny $\scriptscriptstyle \pm$ 13 & 120 \tiny $\scriptscriptstyle \pm$ 11 & 44 \tiny $\scriptscriptstyle \pm$ 7 \\
2 & No fly-bys & 108 \tiny $\scriptscriptstyle \pm$ 10 & 79 \tiny $\scriptscriptstyle \pm$ 9 & 10 \tiny $\scriptscriptstyle \pm$ 3 & 130 \tiny $\scriptscriptstyle \pm$ 11 & 52 \tiny $\scriptscriptstyle \pm$ 7 & 8 \tiny $\scriptscriptstyle \pm$ 3 & 24 \tiny $\scriptscriptstyle \pm$ 5 & 17 \tiny $\scriptscriptstyle \pm$ 4 & 8 \tiny $\scriptscriptstyle \pm$ 3 \\
3a & 0.1 $\Zsun$ & 274 \tiny $\scriptscriptstyle \pm$ 17 & 191 \tiny $\scriptscriptstyle \pm$ 14 & 20 \tiny $\scriptscriptstyle \pm$ 4 & 309 \tiny $\scriptscriptstyle \pm$ 18 & 197 \tiny $\scriptscriptstyle \pm$ 14 & 28 \tiny $\scriptscriptstyle \pm$ 5 & 72 \tiny $\scriptscriptstyle \pm$ 8 & 64 \tiny $\scriptscriptstyle \pm$ 8 & 17 \tiny $\scriptscriptstyle \pm$ 4 \\
3b & 0.01 $\Zsun$ & 429 \tiny $\scriptscriptstyle \pm$ 21 & 525 \tiny $\scriptscriptstyle \pm$ 23 & 51 \tiny $\scriptscriptstyle \pm$ 7 & 477 \tiny $\scriptscriptstyle \pm$ 22 & 619 \tiny $\scriptscriptstyle \pm$ 25 & 45 \tiny $\scriptscriptstyle \pm$ 7 & 87 \tiny $\scriptscriptstyle \pm$ 9 & 199 \tiny $\scriptscriptstyle \pm$ 14 & 31 \tiny $\scriptscriptstyle \pm$ 6 \\
4a & 0.1 $t_{\mathrm{CE,0}}$ & 148 \tiny $\scriptscriptstyle \pm$ 12 & 72 \tiny $\scriptscriptstyle \pm$ 8 & 14 \tiny $\scriptscriptstyle \pm$ 4 & 171 \tiny $\scriptscriptstyle \pm$ 13 & 79 \tiny $\scriptscriptstyle \pm$ 9 & 14 \tiny $\scriptscriptstyle \pm$ 4 & 30 \tiny $\scriptscriptstyle \pm$ 5 & 19 \tiny $\scriptscriptstyle \pm$ 4 & 8 \tiny $\scriptscriptstyle \pm$ 3 \\
4b & 10 $t_{\mathrm{CE,0}}$ & 149 \tiny $\scriptscriptstyle \pm$ 12 & 68 \tiny $\scriptscriptstyle \pm$ 8 & 10 \tiny $\scriptscriptstyle \pm$ 3 & 171 \tiny $\scriptscriptstyle \pm$ 13 & 79 \tiny $\scriptscriptstyle \pm$ 9 & 14 \tiny $\scriptscriptstyle \pm$ 4 & 35 \tiny $\scriptscriptstyle \pm$ 6 & 17 \tiny $\scriptscriptstyle \pm$ 4 &  8 \tiny $\scriptscriptstyle \pm$ 3 \\
\enddata
\tablecomments{Refer to Table \ref{tab:modelparam} for detailed model specifications.}
\end{deluxetable*}

The table shows that the corresponding number of mergers in both cases (bound vs. unbound 2+2 quadruples) are mostly within the Poisson error margin of each other. However, we can see that isolated binaries consistently produce a higher number of mergers in all models except Model 1 (where SNe kicks are disabled). This may be due to secular evolution driving mergers in 2+2 quadruples in the pre-compact object phases. For example, an isolated inner binary of a 2+2 quadruple could lead to a BH-BH merger, whereas the same inner binary in a bound system could see a merger before the component stars evolve into BHs due to eccentricity excitation.

\begin{figure*}
\gridline{\fig{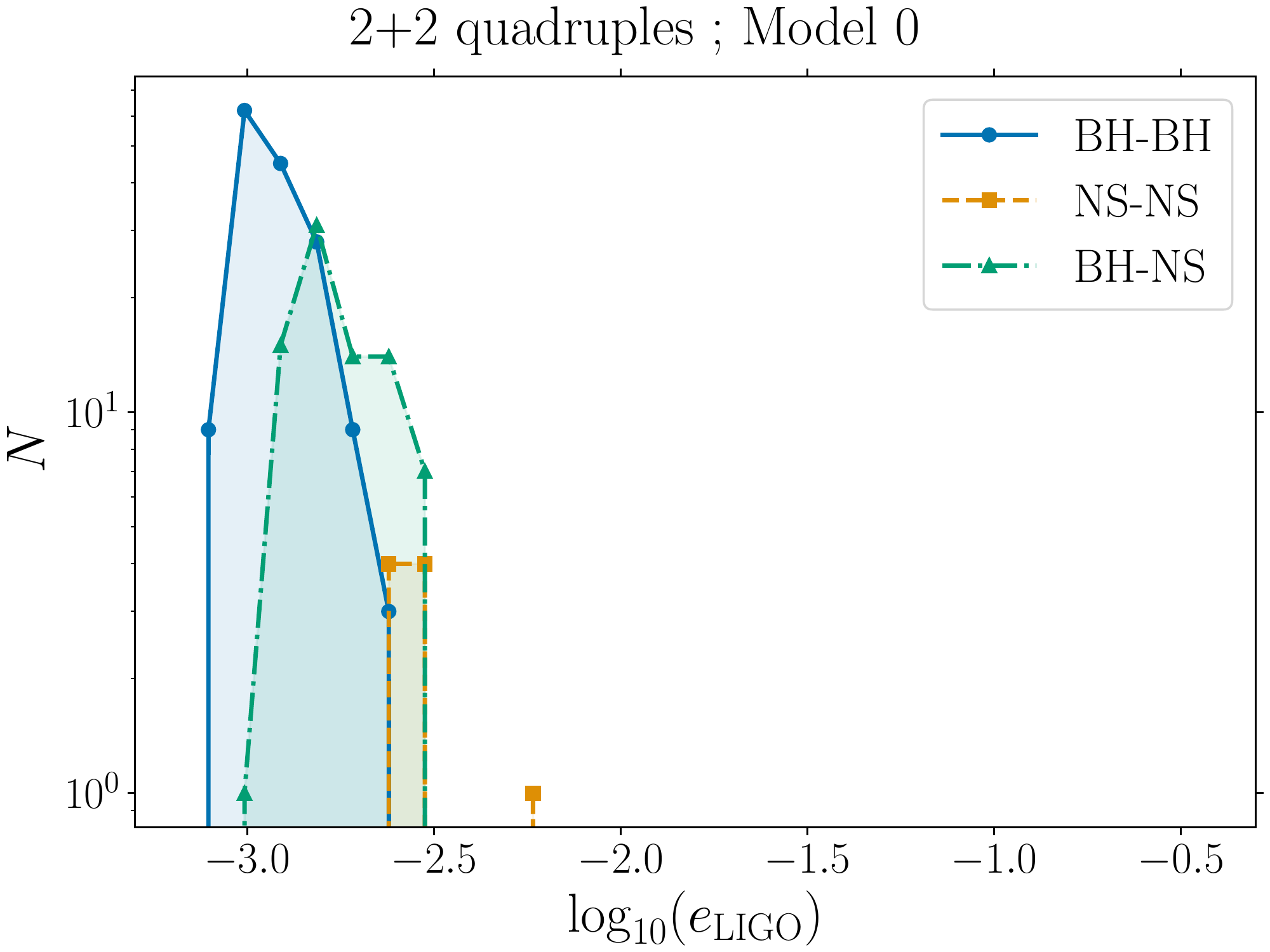}{0.45\textwidth}{(a)}
          \fig{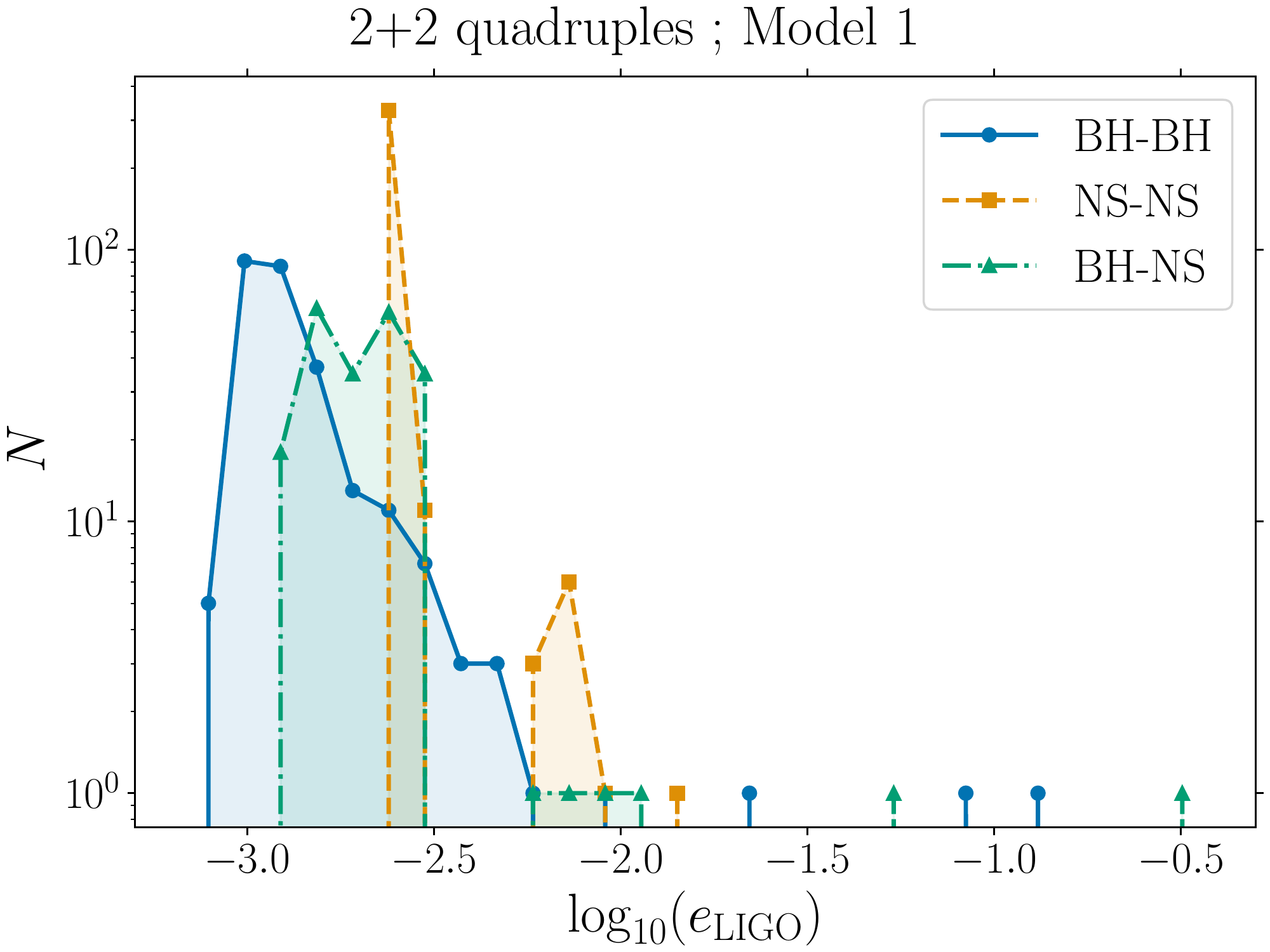}{0.45\textwidth}{(b)}
          }
\gridline{\fig{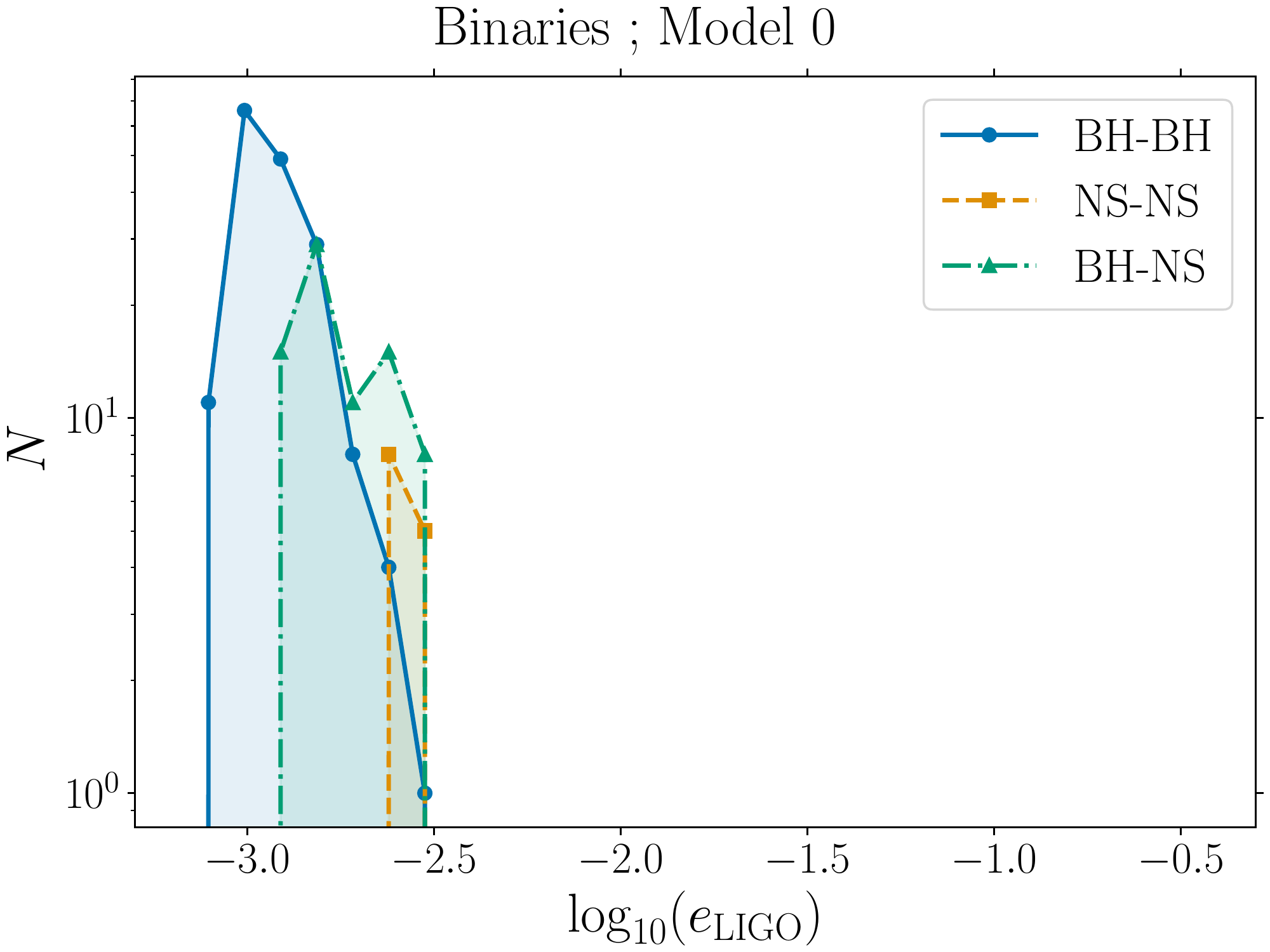}{0.45\textwidth}{(c)}
          \fig{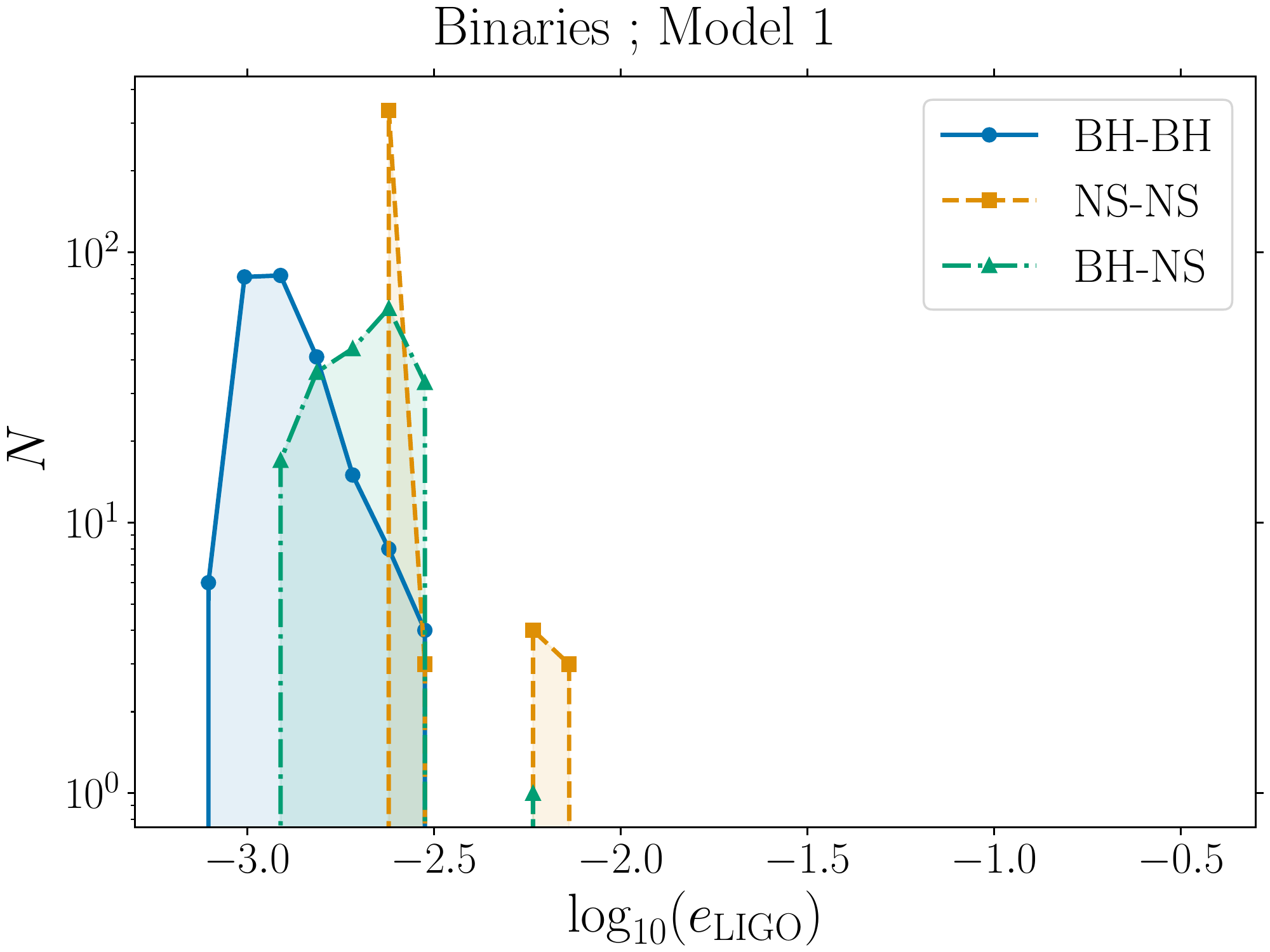}{0.45\textwidth}{(d)}
          }
\gridline{\fig{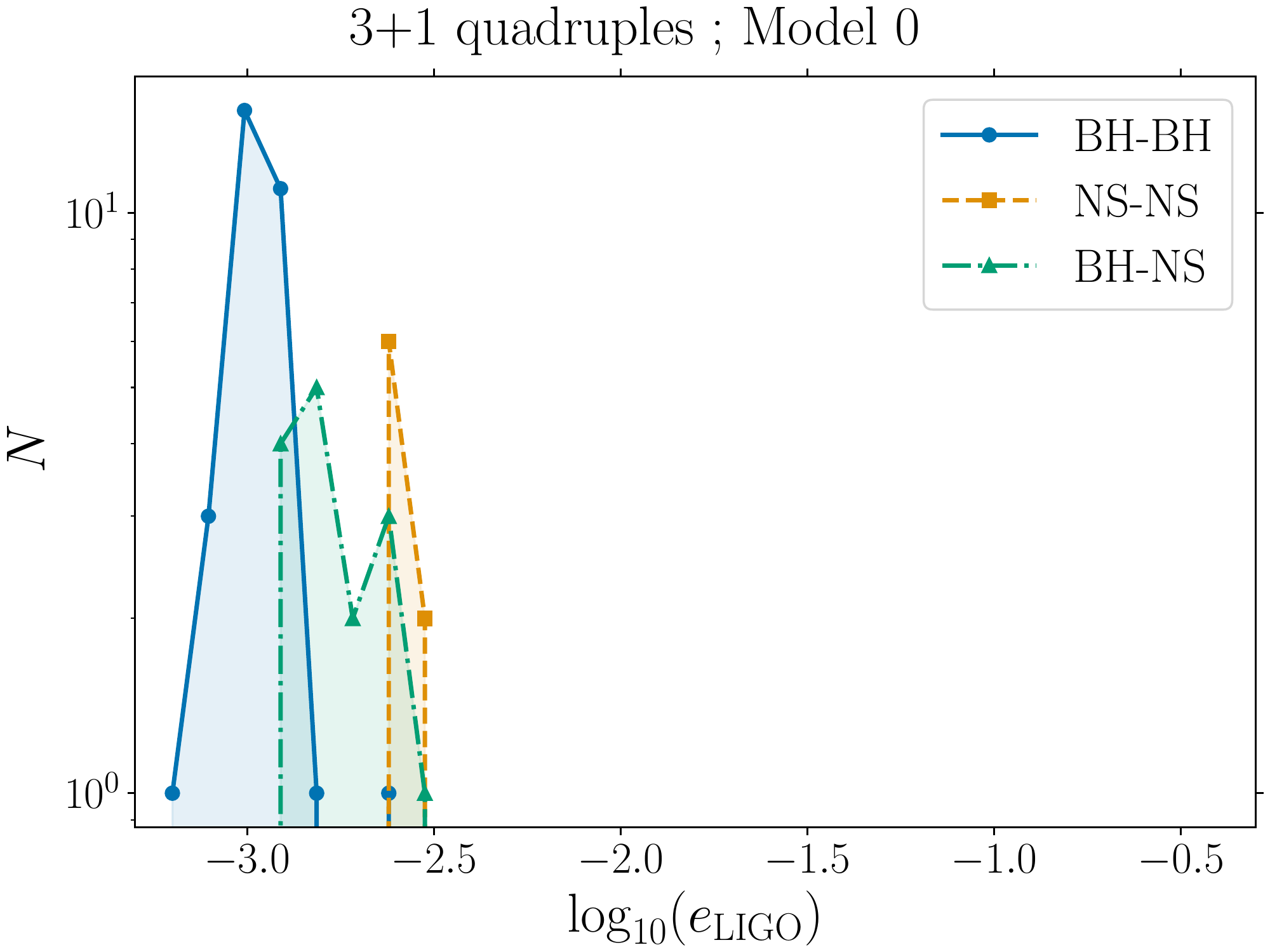}{0.45\textwidth}{(e)}
          \fig{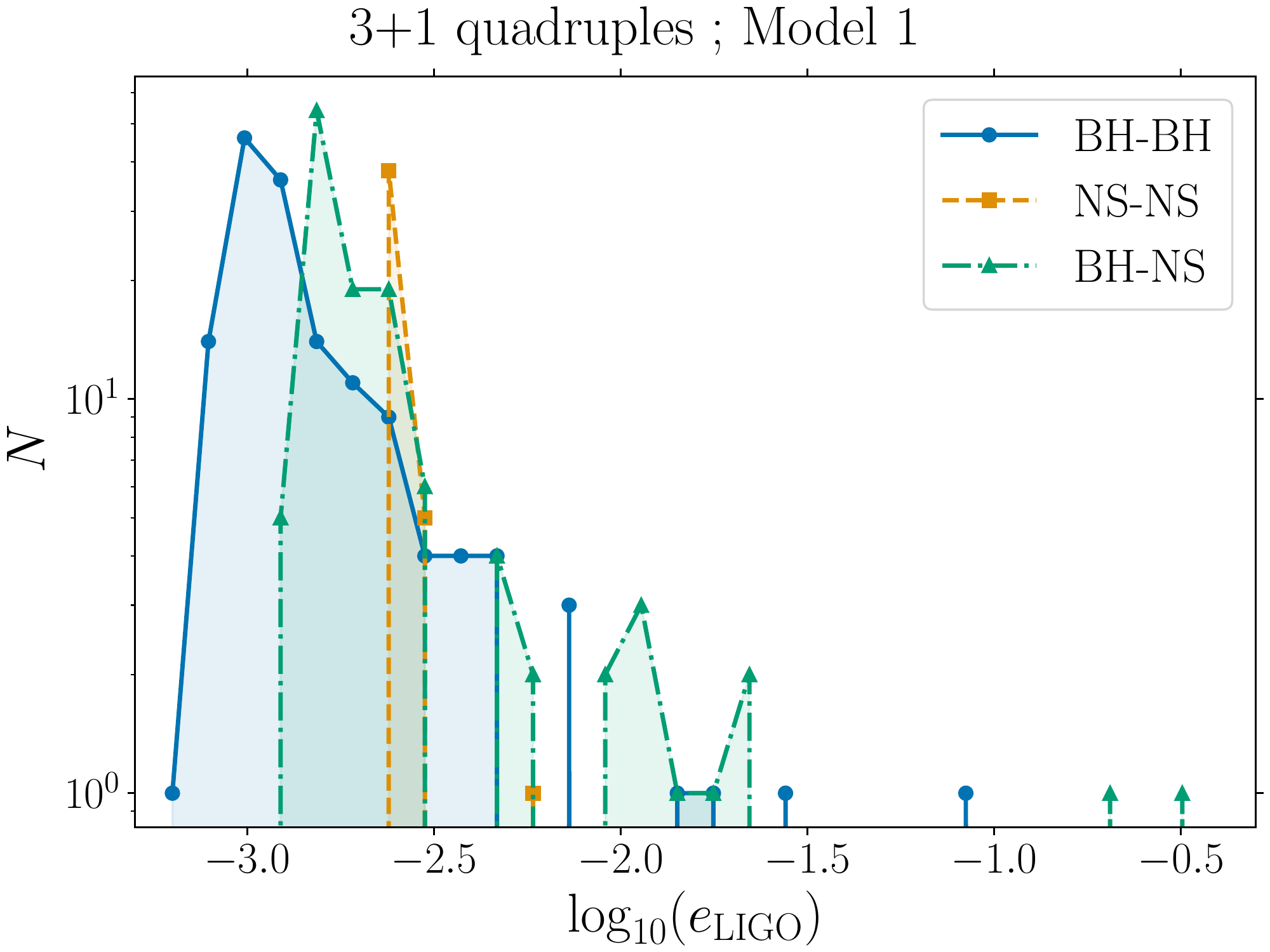}{0.45\textwidth}{(f)}
          }
\caption{Frequency polygon of LIGO-band eccentricities $e_{\mathrm{LIGO}}$. [Rows correspond to 2+2 quadruples, binaries and 3+1 quadruples respectively; Columns correspond to Models 0 and 1 respectively.] The quadruples have a significant tail at high $e_{\mathrm{LIGO}}$ values for Model 1, whereas the binaries do not.
\label{fig:eLIGO}}
\end{figure*}

\begin{figure*}
\gridline{\fig{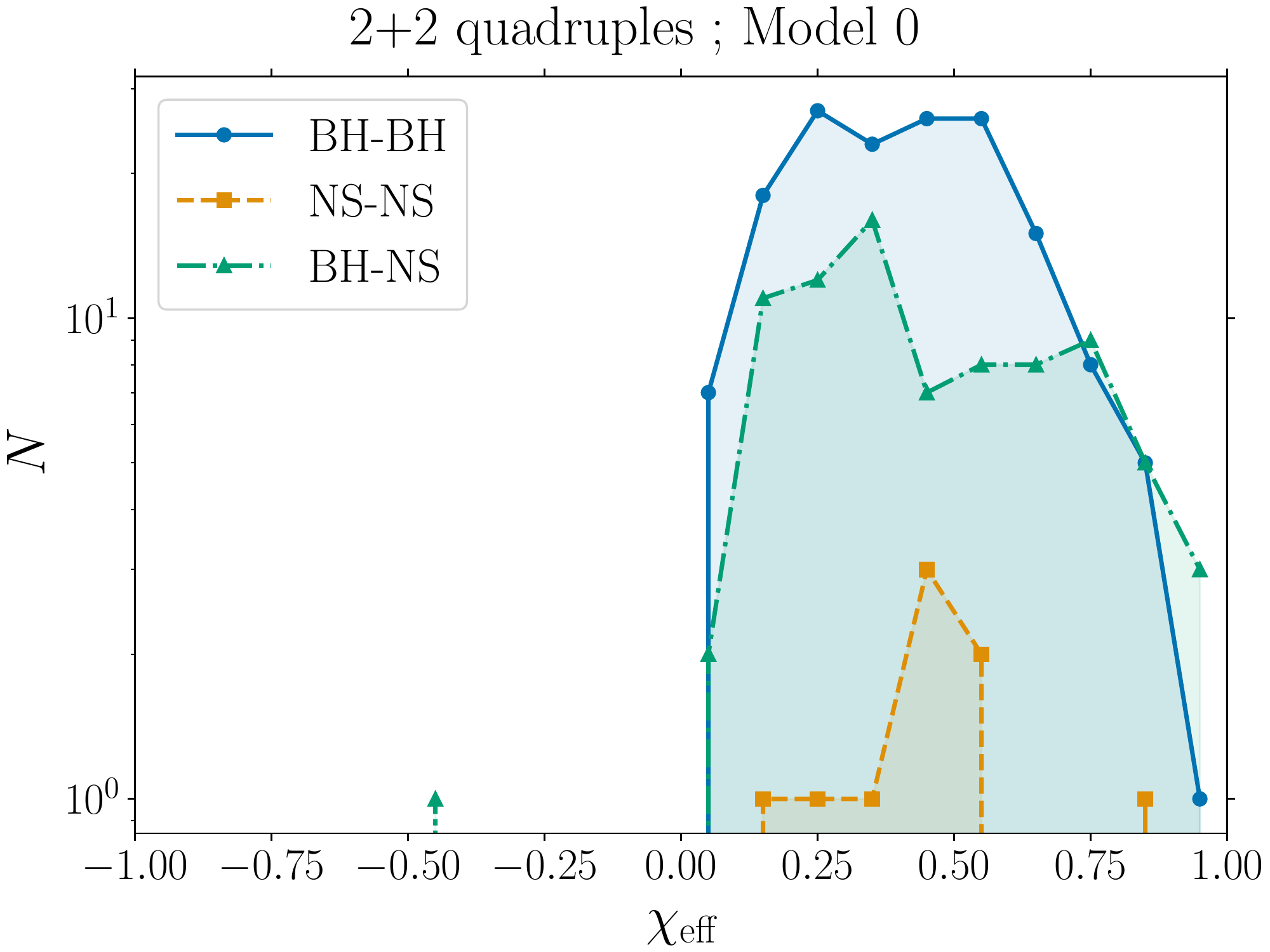}{0.45\textwidth}{(a)}
          \fig{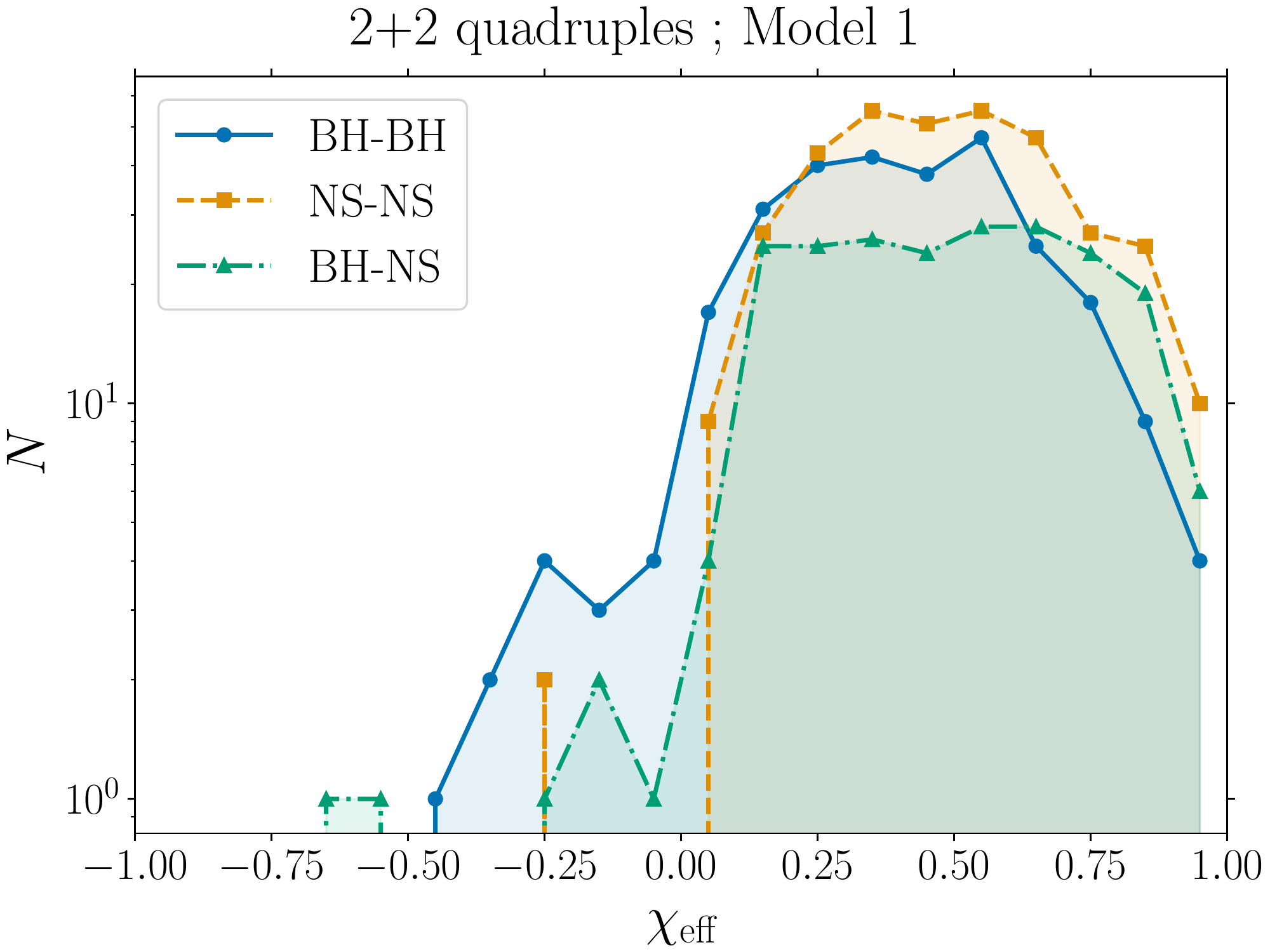}{0.45\textwidth}{(b)}
          }
\gridline{\fig{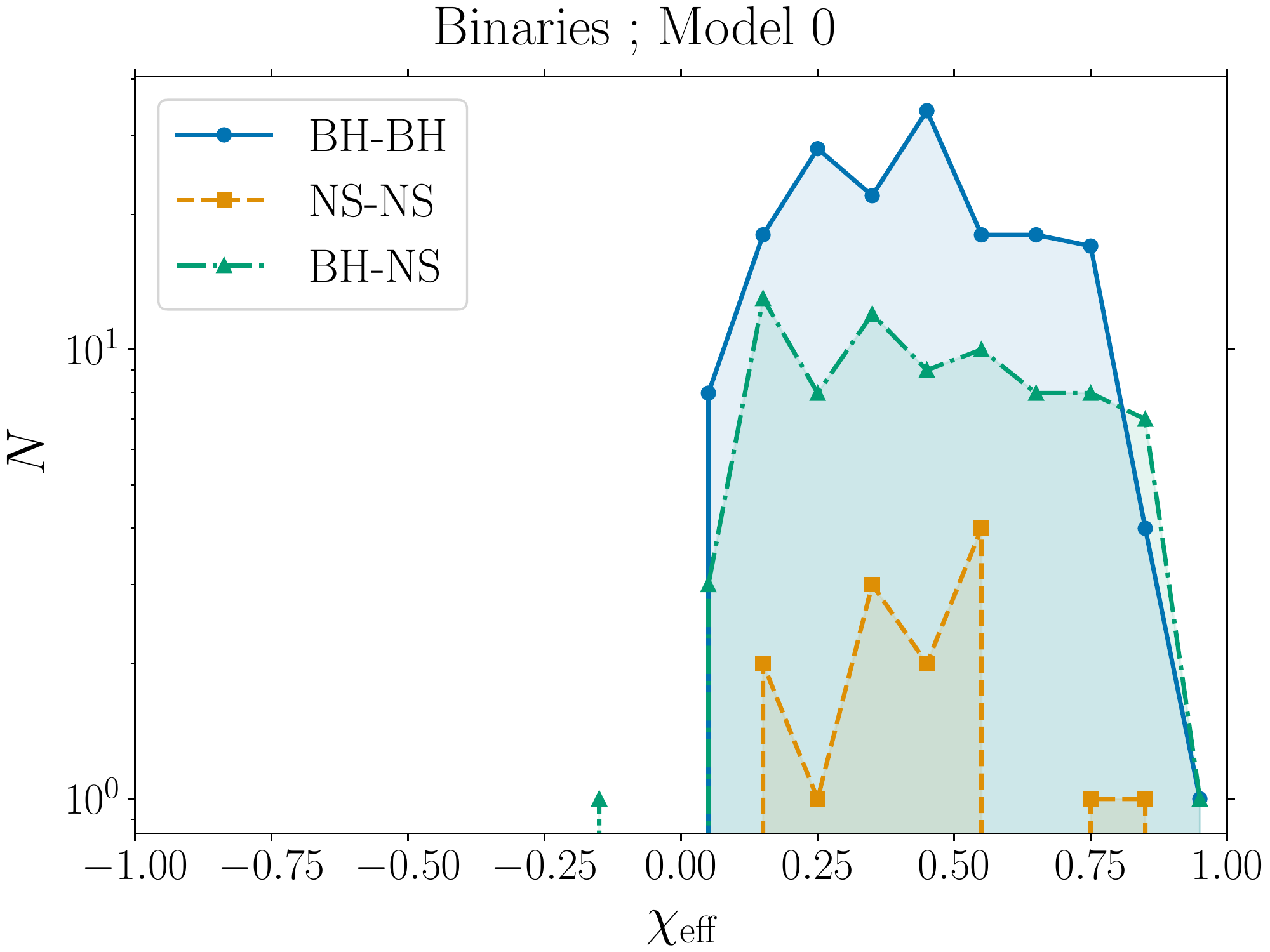}{0.45\textwidth}{(c)}
          \fig{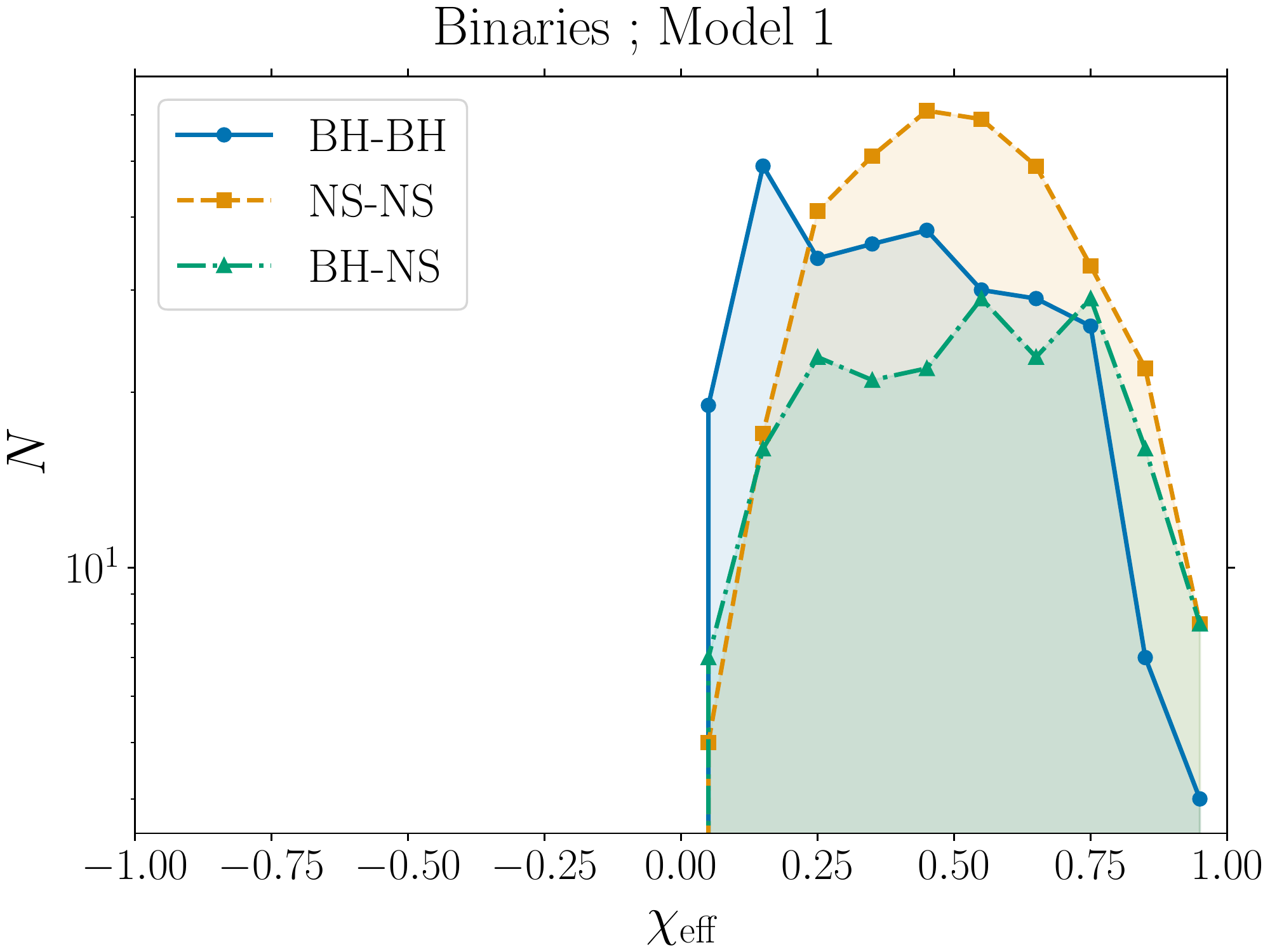}{0.45\textwidth}{(d)}
          }
\gridline{\fig{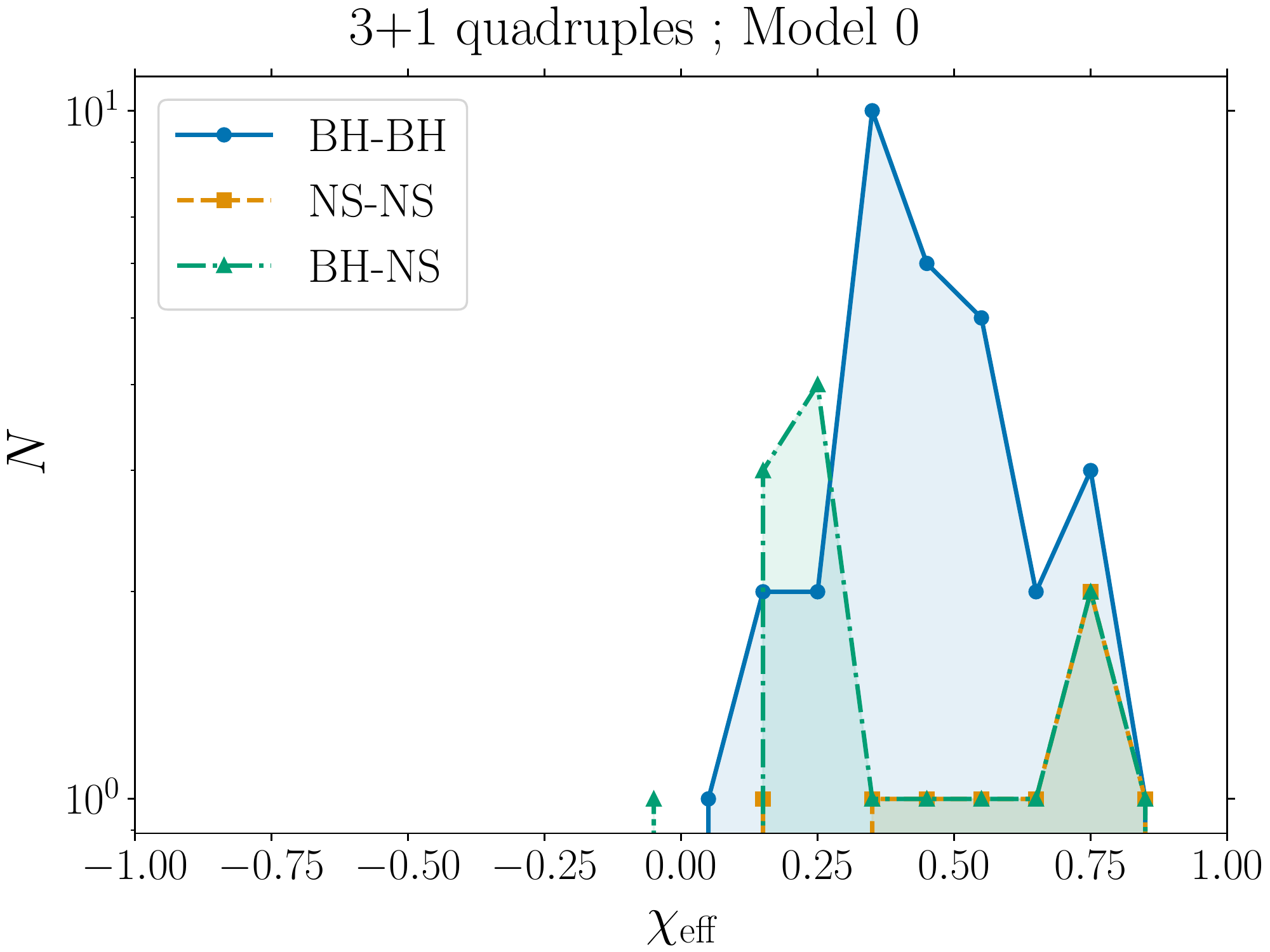}{0.45\textwidth}{(e)}
          \fig{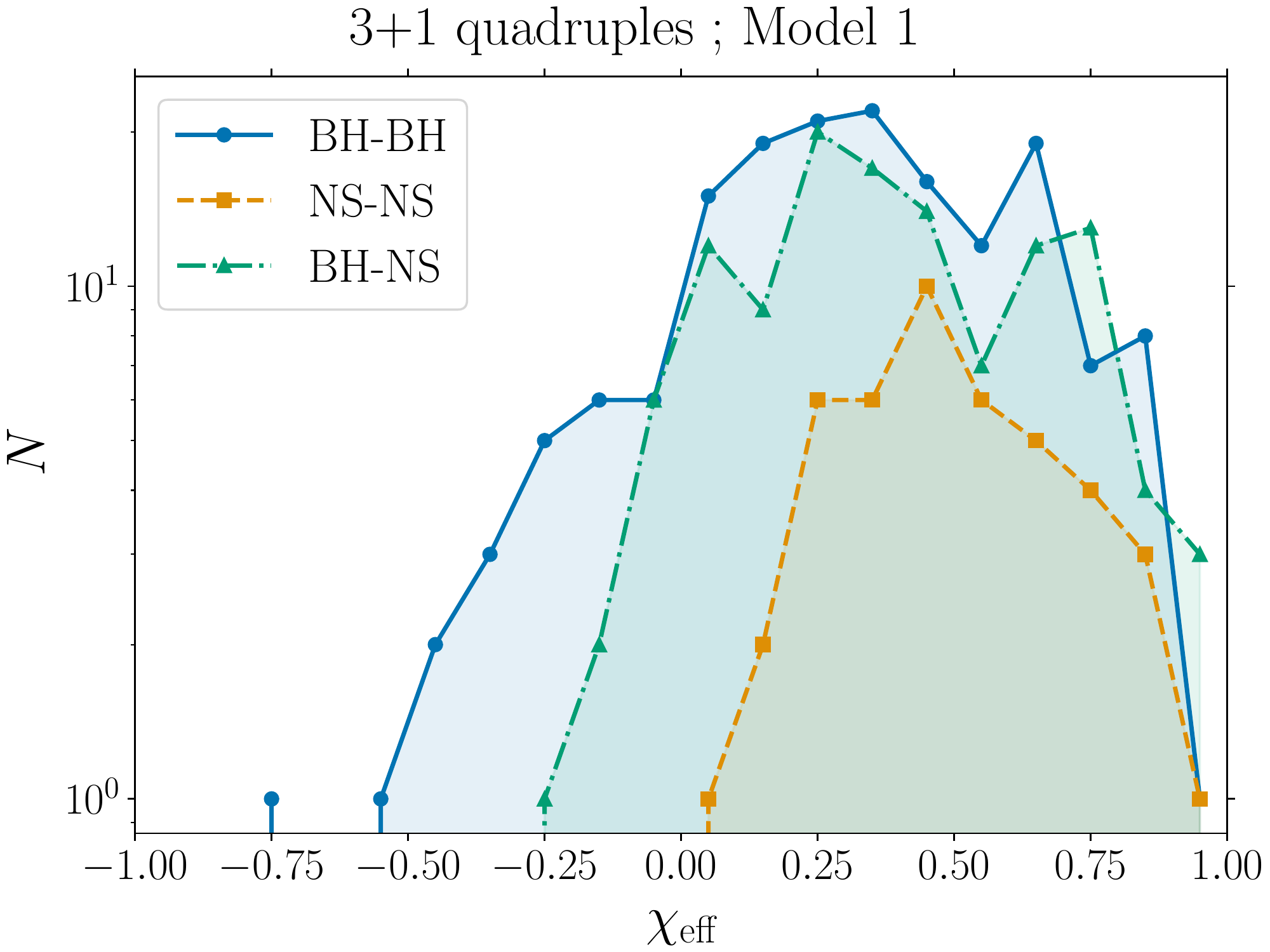}{0.45\textwidth}{(f)}
          }
\caption{Frequency polygon of effective spins $\chi_{\mathrm{eff}}$. [Rows correspond to 2+2 quadruples, binaries and 3+1 quadruples respectively; Columns correspond to Models 0 and 1 respectively.] The quadruples have a significant tail at negative $\chi_{\mathrm{eff}}$ values for Model 1, whereas the binaries do not.
\label{fig:chieff}}
\end{figure*}

\begin{figure}
\gridline{\fig{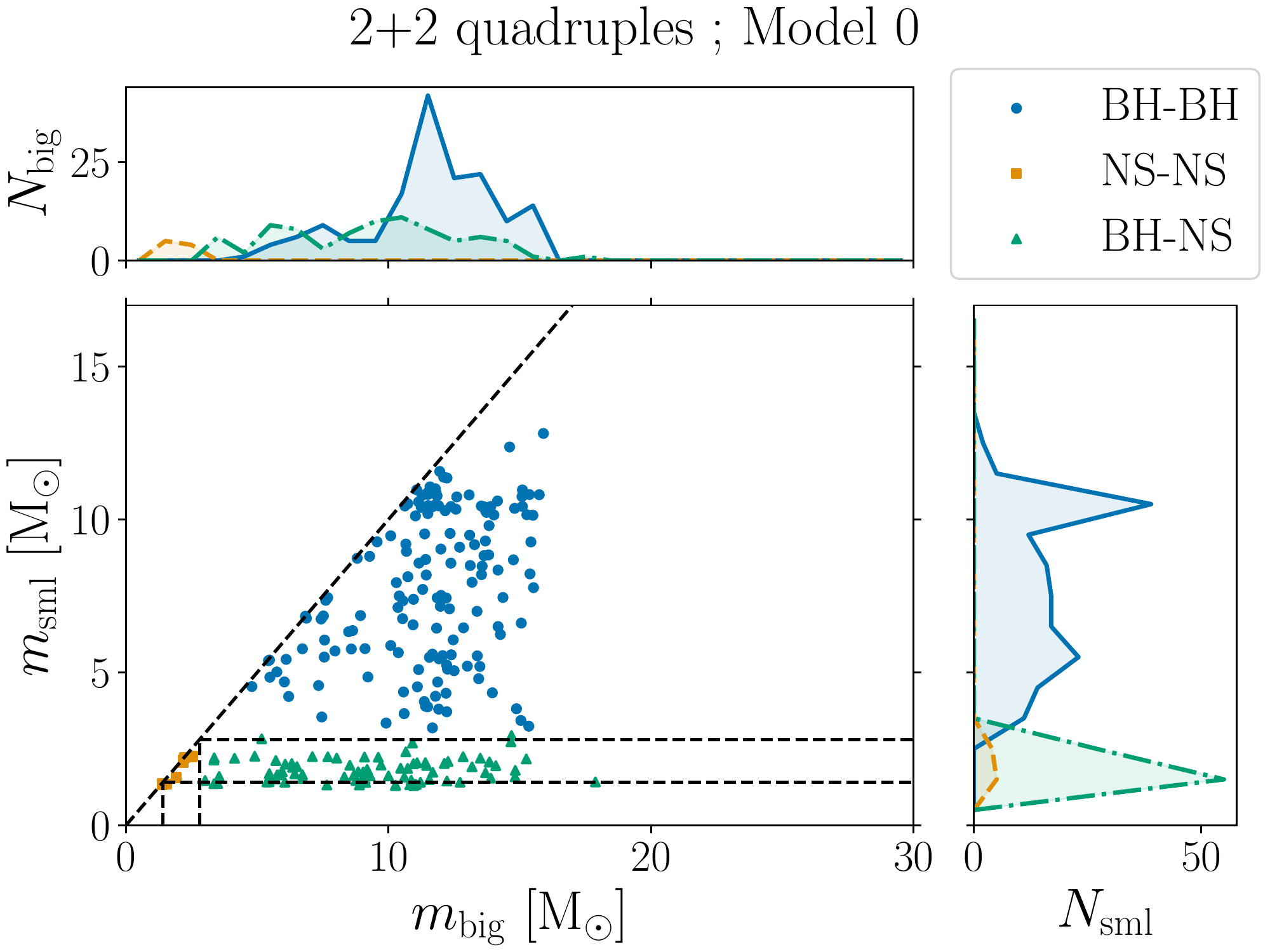}{0.45\textwidth}{(a)}
          }
\gridline{\fig{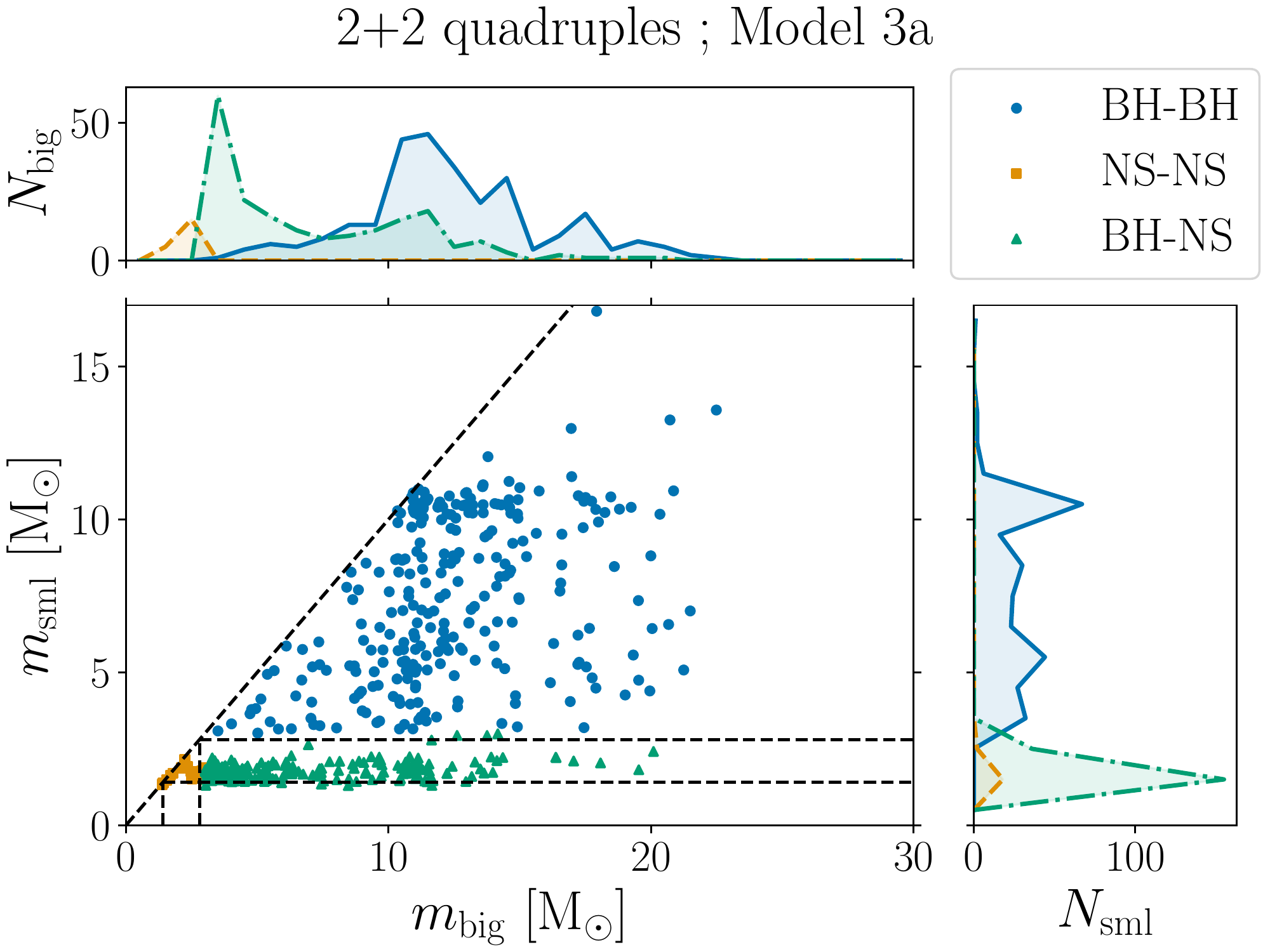}{0.45\textwidth}{(b)}
          }
\gridline{\fig{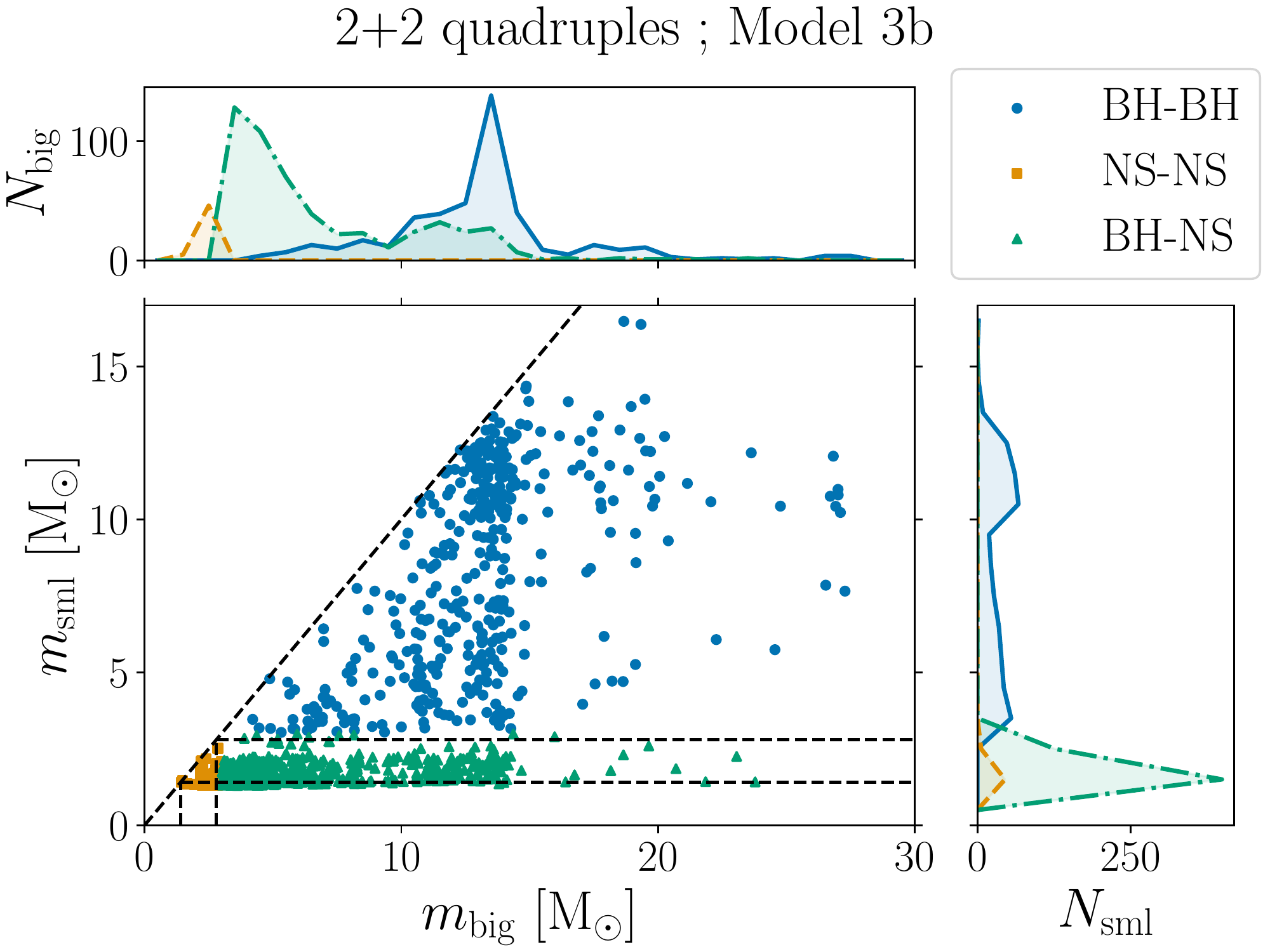}{0.45\textwidth}{(c)}
          }
\caption{Scatter plot of heavier ($m_{\mathrm{big}}$) and lighter ($m_{\mathrm{sml}}$) compact object masses pre-merger. [Rows correspond to Models 1, 3a and 3b respectively.] 
\label{fig:mbigmsml}}
\end{figure}

Let us look at the numbers for each model in detail.
\begin{itemize}
    \item \textit{Model 0:} The number of BH-BH mergers is higher than that of BH-NS mergers, which in turn is significantly higher than that of NS-NS mergers. The very low merger rates for NS-NS binaries can be attributed to their high SNe kicks, which tend to unbind orbits.
    \item \textit{Model 1:} Excluding SNe kicks has a drastic effect on the number of compact object mergers, especially the NS-NS mergers, for both 2+2 quadruples and isolated binaries. This is expected since SNe kicks almost always result in the unbinding of orbits, more so for NSs since their kick distribution has a higher $\sigma$ than that of BHs. Moreover, in 2+2 quadruples, having no SNe kicks means that there can be secular evolution even in the compact object phase. This eccentricity excitation can lead to much higher $e_{\mathrm{LIGO}}$ values and negative $\chi_{\mathrm{eff}}$ values. These outlier systems can be seen in Figures \ref{fig:eLIGO} and \ref{fig:chieff} (b) \& (f) [Model 1 quadruples], but not in (a) \& (e) [Model 0 quadruples] or (d) [Model 1 binaries]. 
    \item \textit{Model 2:} Excluding fly-bys systematically reduces the total number of mergers. This is most prominent in the BH-BH merger numbers for both 2+2 quadruples and isolated binaries. The similar numbers for BH-NS and NS-NS mergers (in the quadruples case) may be attributed to the higher SNe kicks for NSs, which can diminish the effect of fly-bys. It is important to note that, while an external perturbation can destabilize a wide orbit, it can also decrease the outer periapsis distance, triggering stronger secular evolution with its inner orbits. In short, the general consequence of fly-bys is not immediately clear. Parameter distributions for Model 2 are roughly similar to those from Model 0.
    \item \textit{Models 3a and 3b:} Reducing $Z$ from the default $\Zsun = 2\times10^{-2}$ significantly increases the number of mergers. This is expected since lower $Z$ stars are more compact than higher $Z$ stars, thereby reducing the chances of pre-compact object phase merger events. Moreover, the maximum BH masses are also significantly higher than those in Model 0 since lower $Z$ stars lose less mass due to stellar winds. This effect is seen by comparing Figures \ref{fig:mbigmsml} (b) [Model 3a] and (c) [Model 3b] with (a) [Model 0].
    \item \textit{Models 4a and 4b:} Changing $t_{\mathrm{CE}}$ from the default $10^3 \yr$ does not seem to change the overall statistics of the mergers. This may be somewhat unexpected since $t_{\mathrm{CE}}$ affects the number of bound stars after CE evolution. However, since secular evolution does not play a dominant role (see Figure \ref{fig:merger_scn}), whether or not the outer orbits in the multiple system remain bound is not important. Hence, the impact of $t_{\mathrm{CE}}$ tends to be small. Merger numbers for Models 4a and 4b are roughly similar to those from Model 0.
\end{itemize}

\begin{figure*}
\gridline{\fig{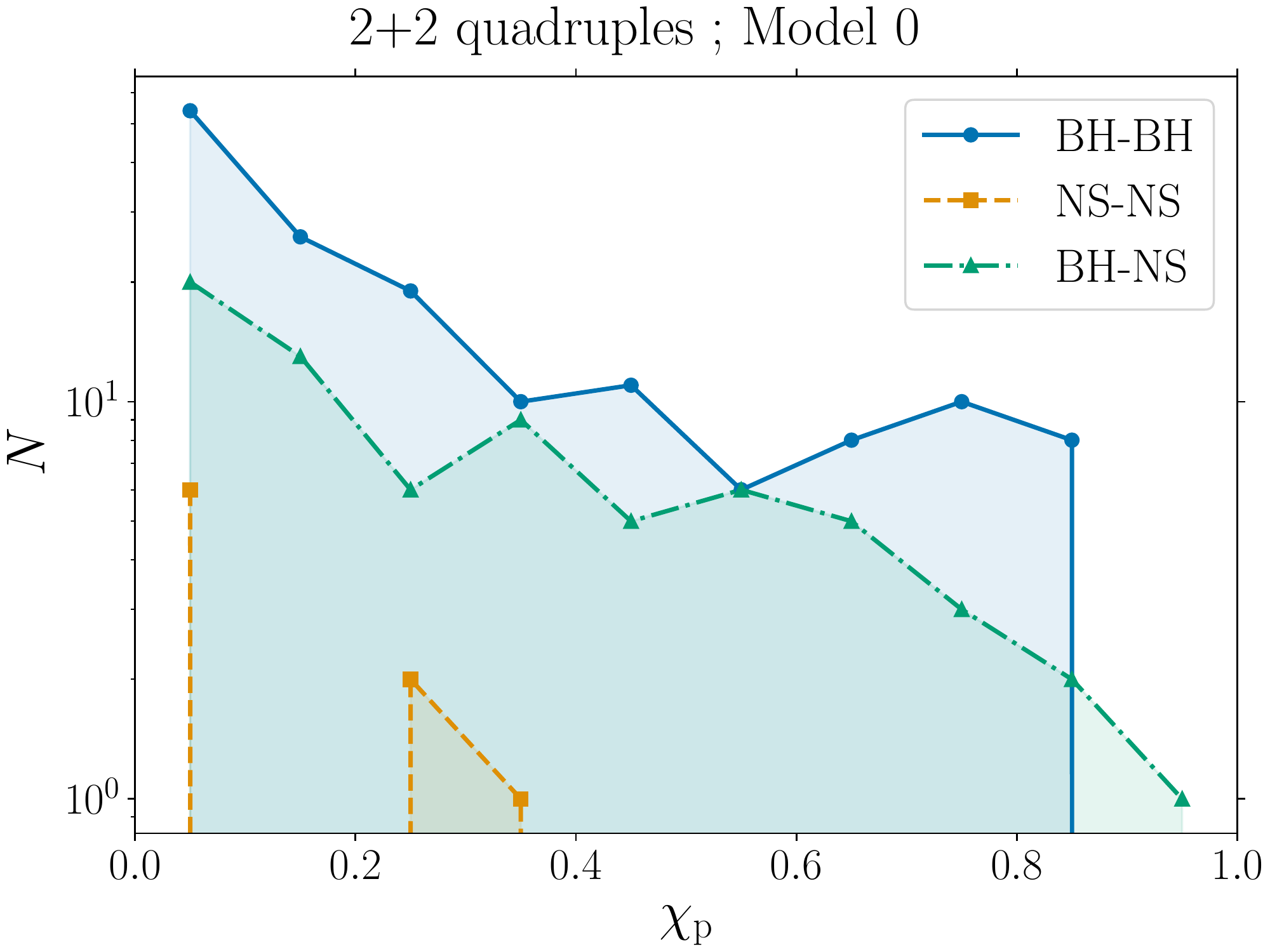}{0.45\textwidth}{(a)}
          \fig{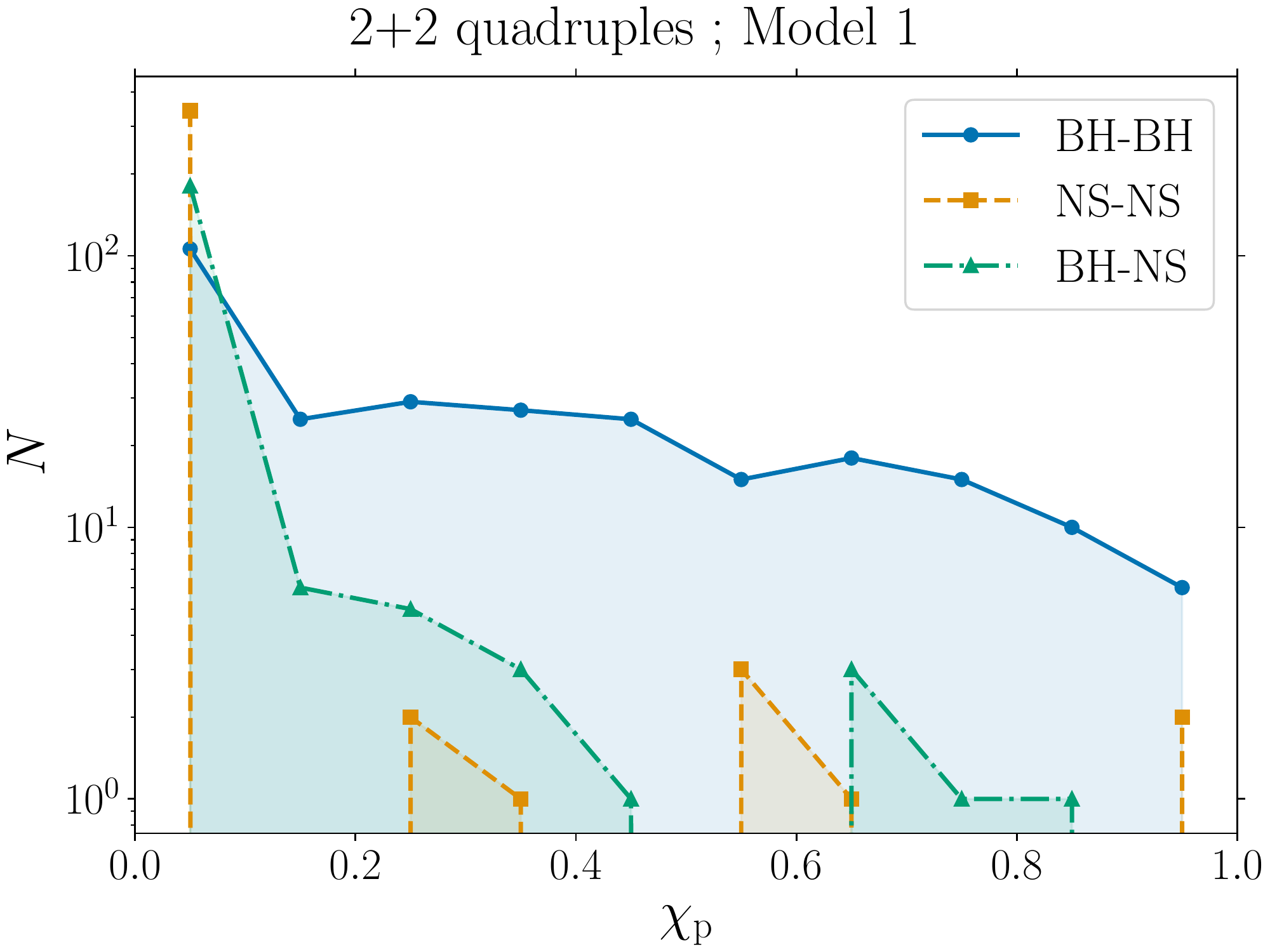}{0.45\textwidth}{(b)}
          }
\gridline{\fig{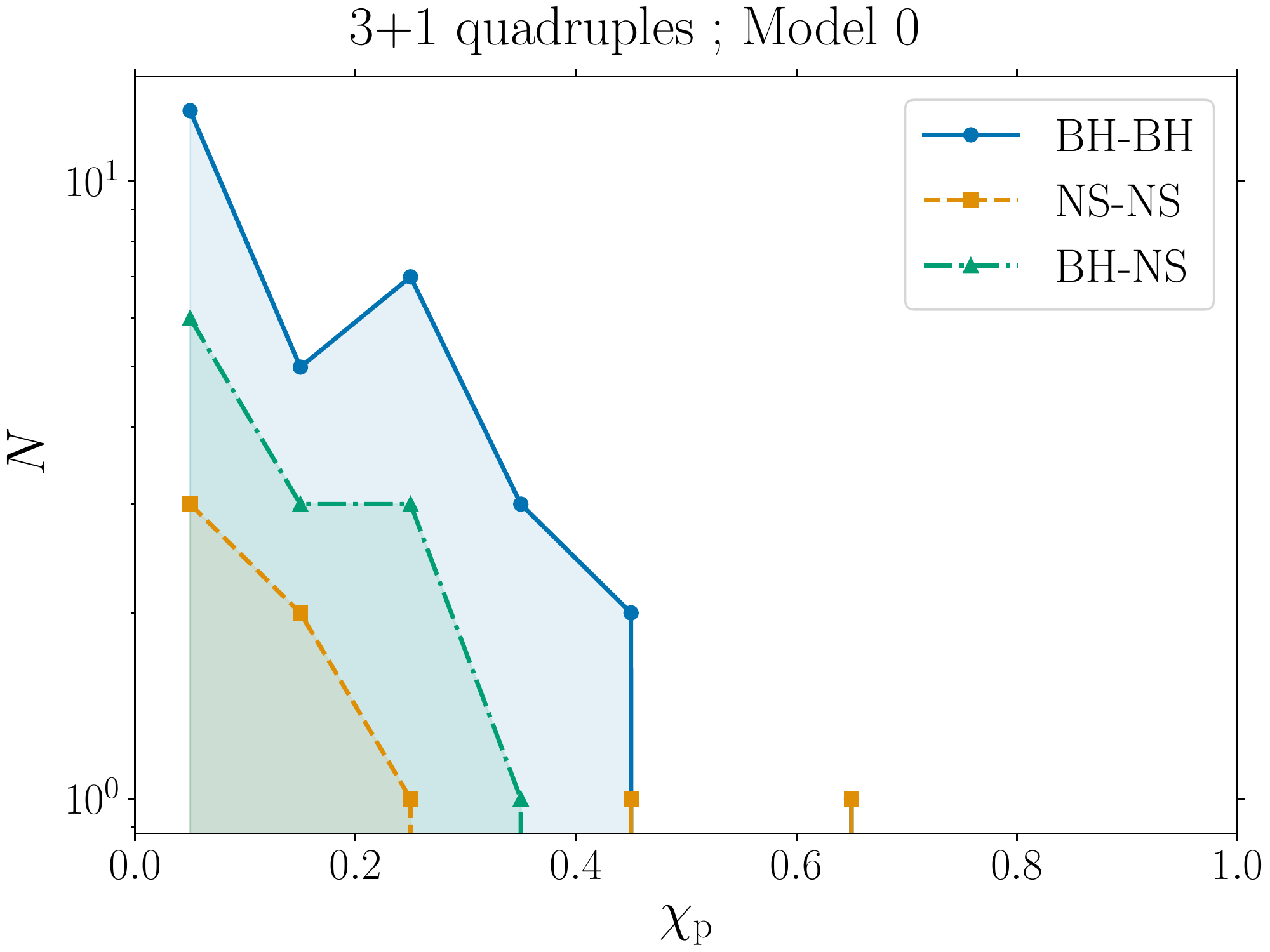}{0.45\textwidth}{(c)}
          \fig{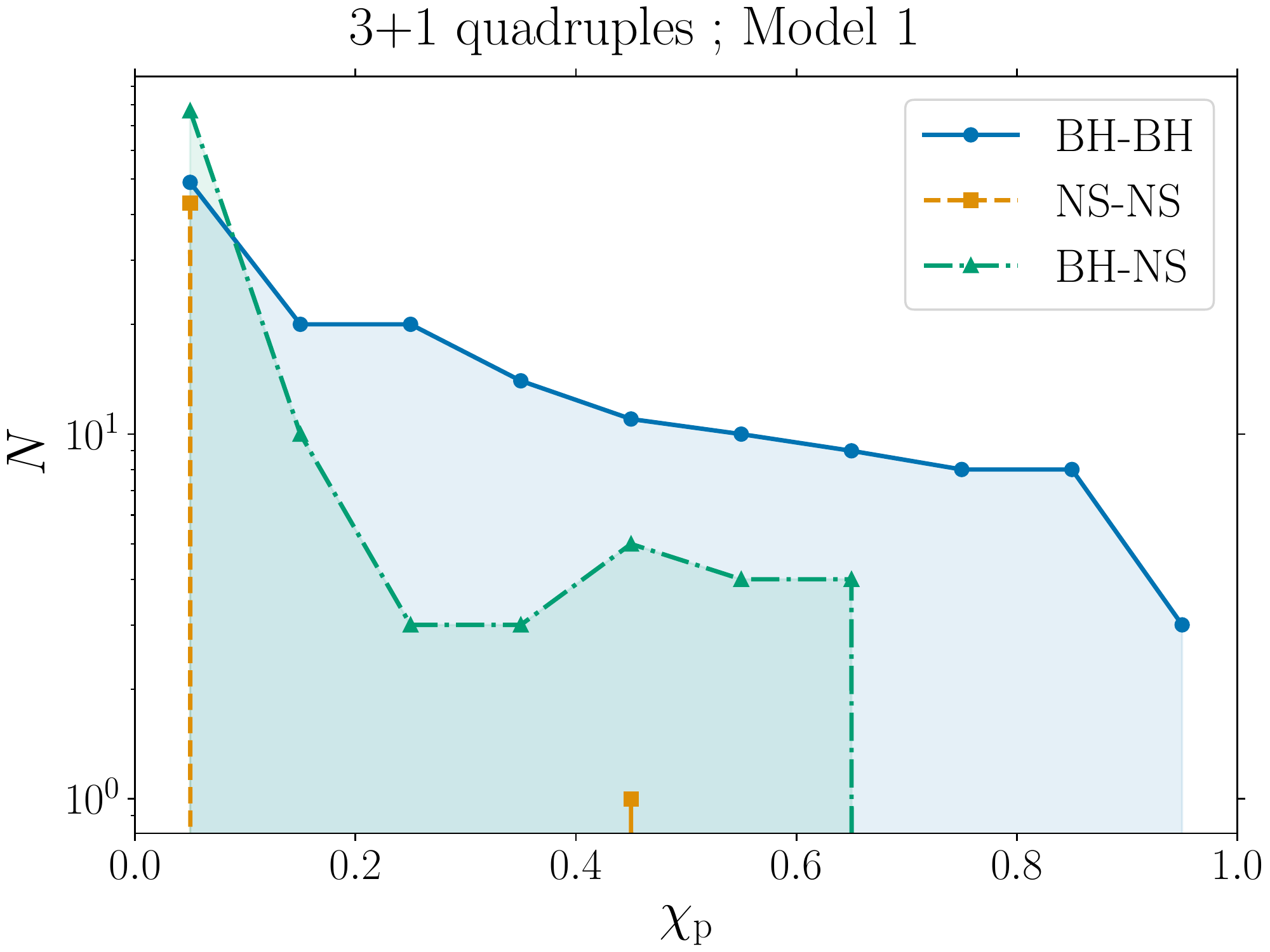}{0.45\textwidth}{(d)}
          }
\caption{Frequency polygon of spin the precession parameters $\chi_{\mathrm{p}}$. [Rows correspond to 2+2 and 3+1 quadruples respectively; Columns correspond to Models 0 and 1 respectively. Isolated binary profiles are similar to 2+2 quadruples.] For comparison, $\chi_{\mathrm{p}}$ for LIGO events GW190814 and GW190425 are $\lesssim 0.07$ and $0.30^{+0.19}_{-0.15}$ respectively.
\label{fig:chip}}
\end{figure*}

\begin{figure*}
\gridline{\fig{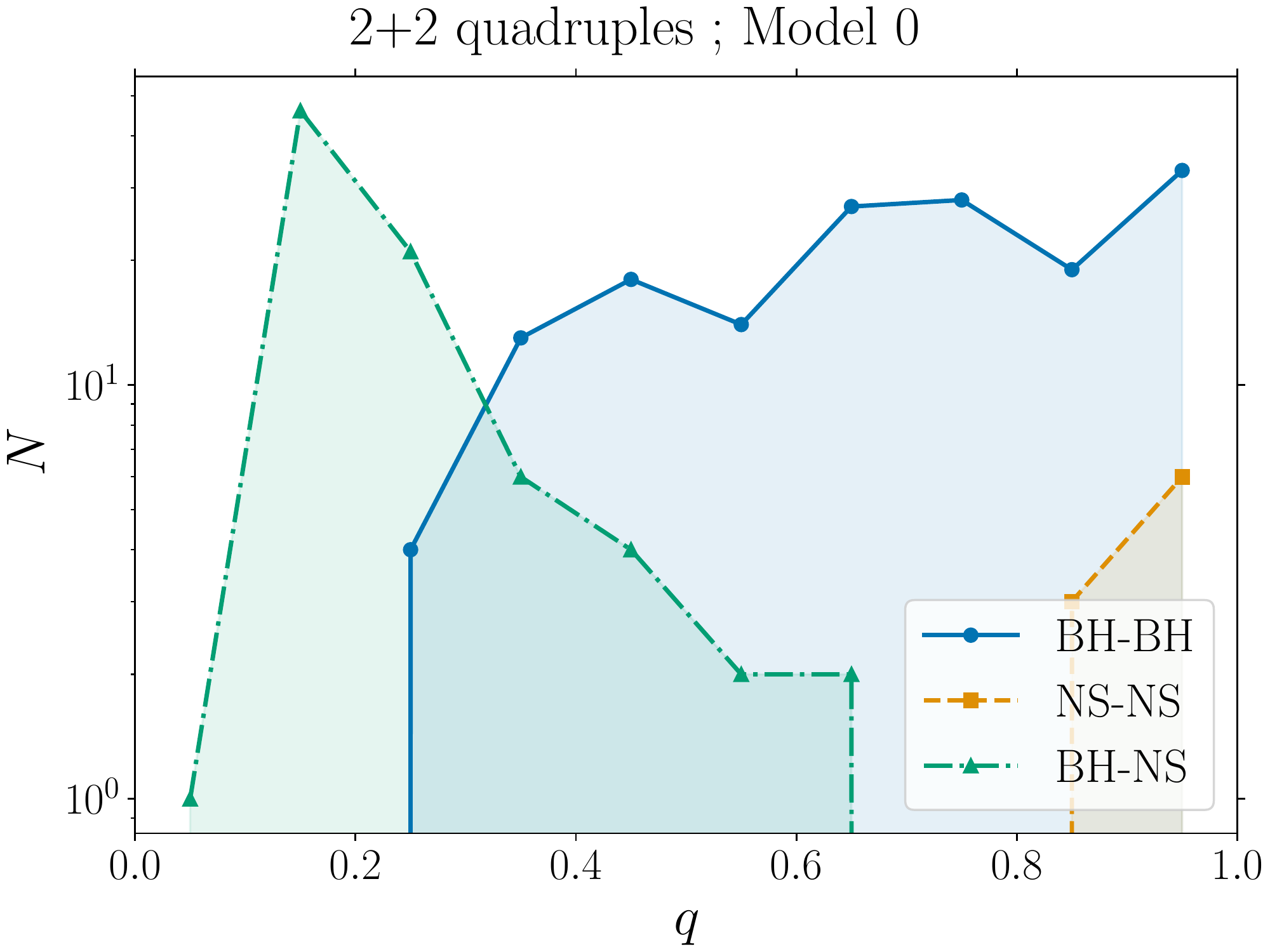}{0.45\textwidth}{(a)}
          \fig{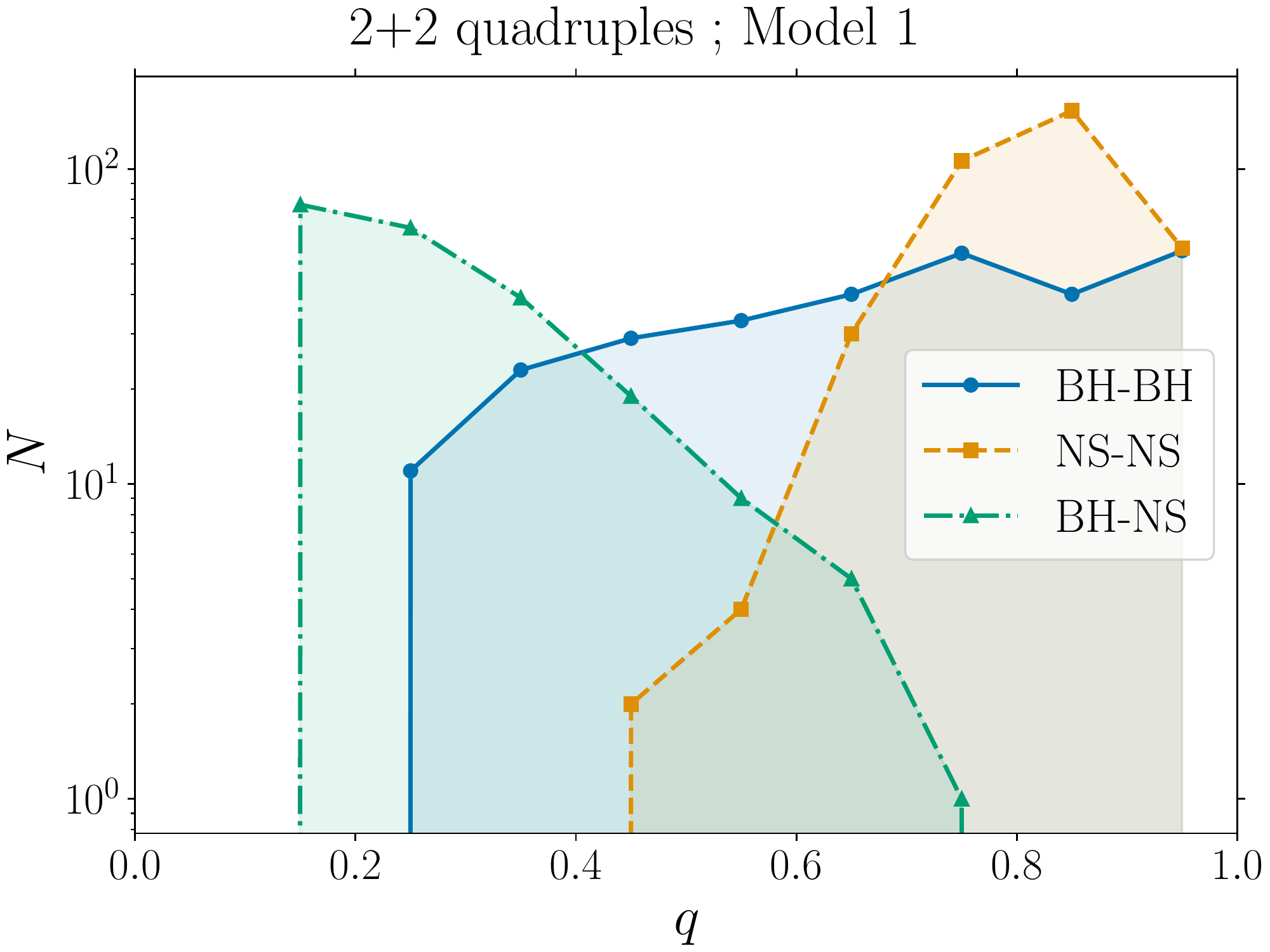}{0.45\textwidth}{(b)}
          }
\gridline{\fig{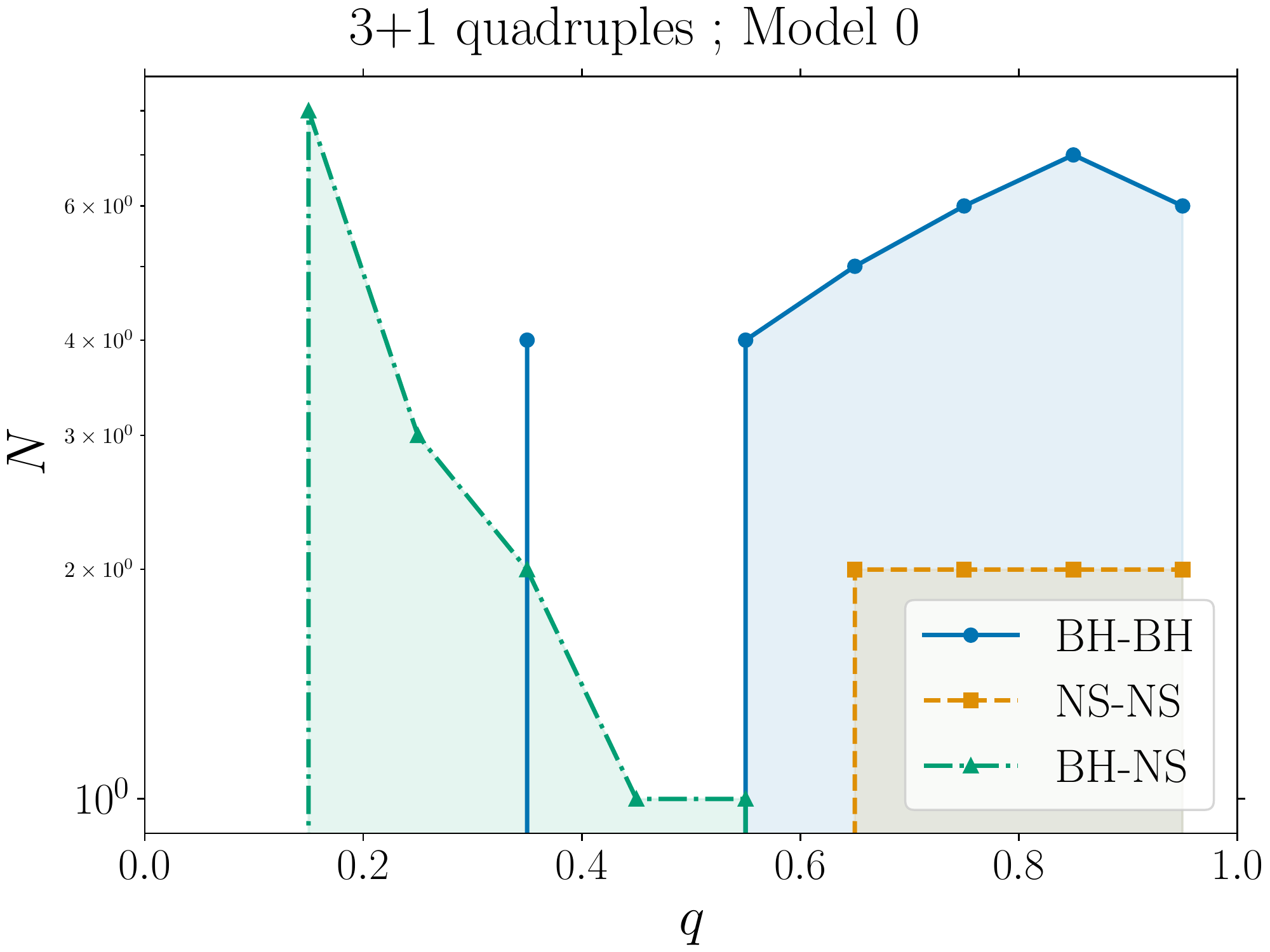}{0.45\textwidth}{(c)}
          \fig{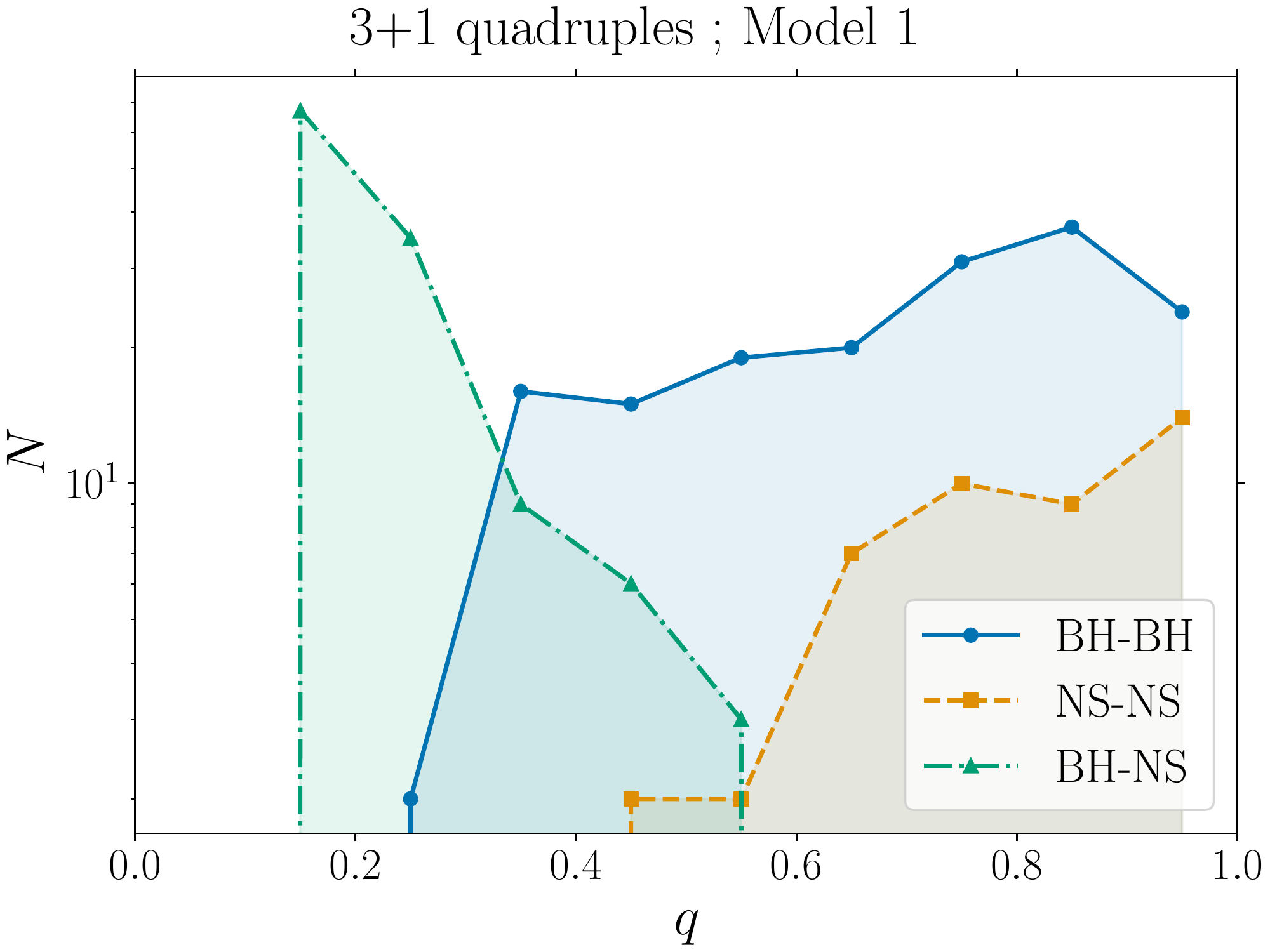}{0.45\textwidth}{(d)}
          }
\caption{Frequency polygon of mass ratios $q$ pre-merger. [Rows correspond to 2+2 and 3+1 quadruples respectively; Columns correspond to Models 0 and 1 respectively. Isolated binary profiles are similar to 2+2 quadruples.] 
\label{fig:mratio}}
\end{figure*}

\begin{figure*}
\gridline{\fig{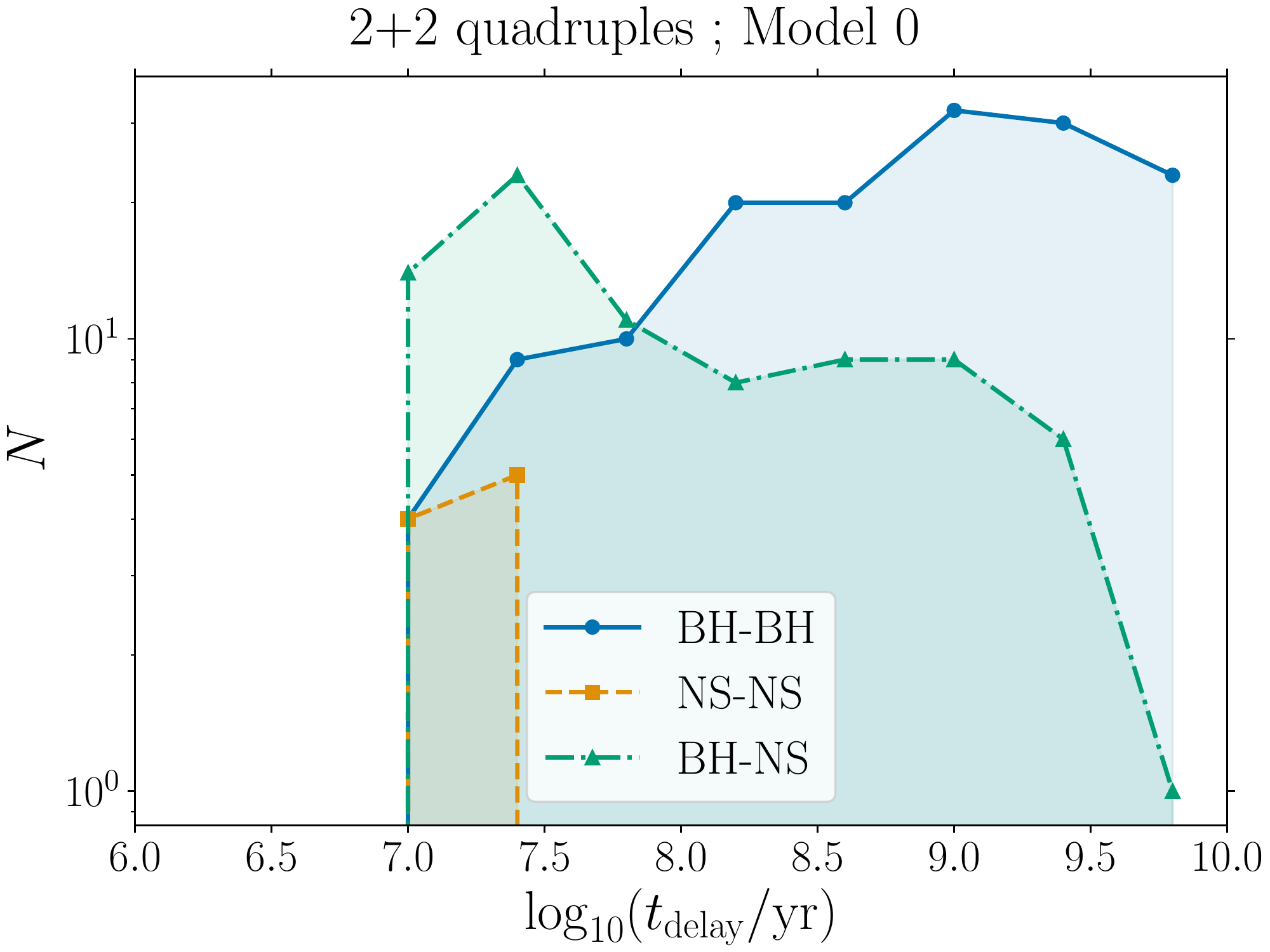}{0.45\textwidth}{(a)}
          \fig{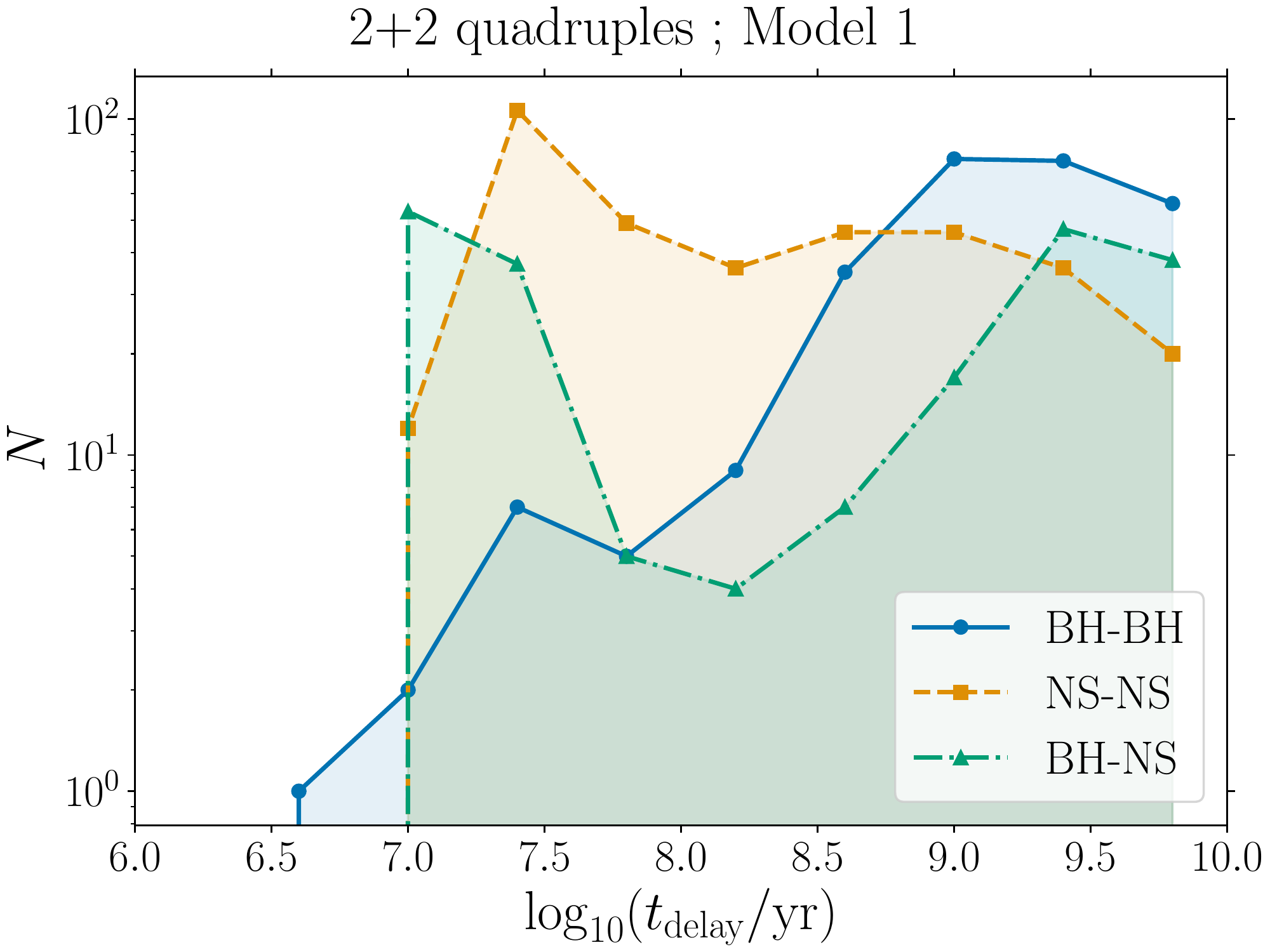}{0.45\textwidth}{(b)}
          }
\gridline{\fig{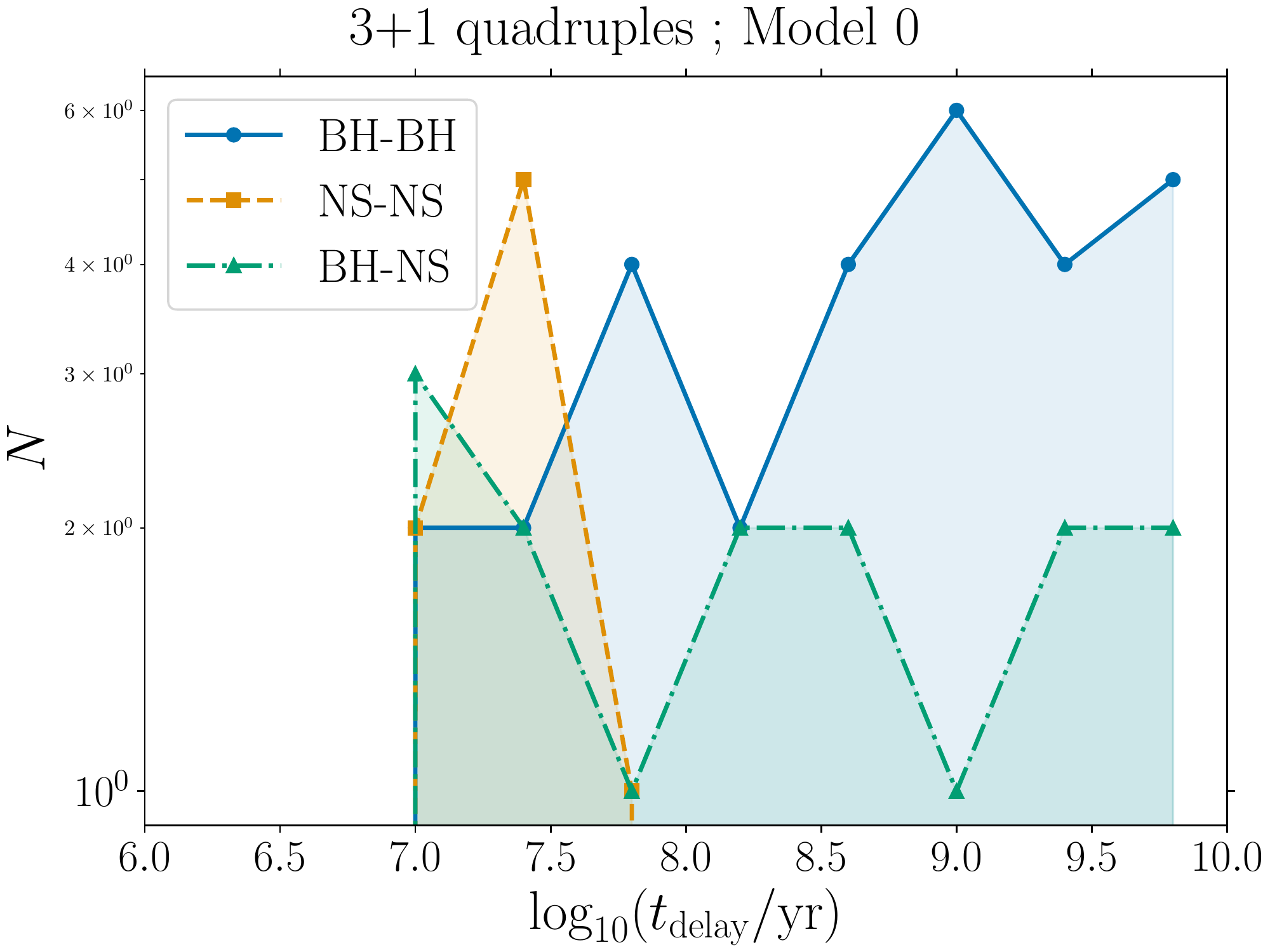}{0.45\textwidth}{(c)}
          \fig{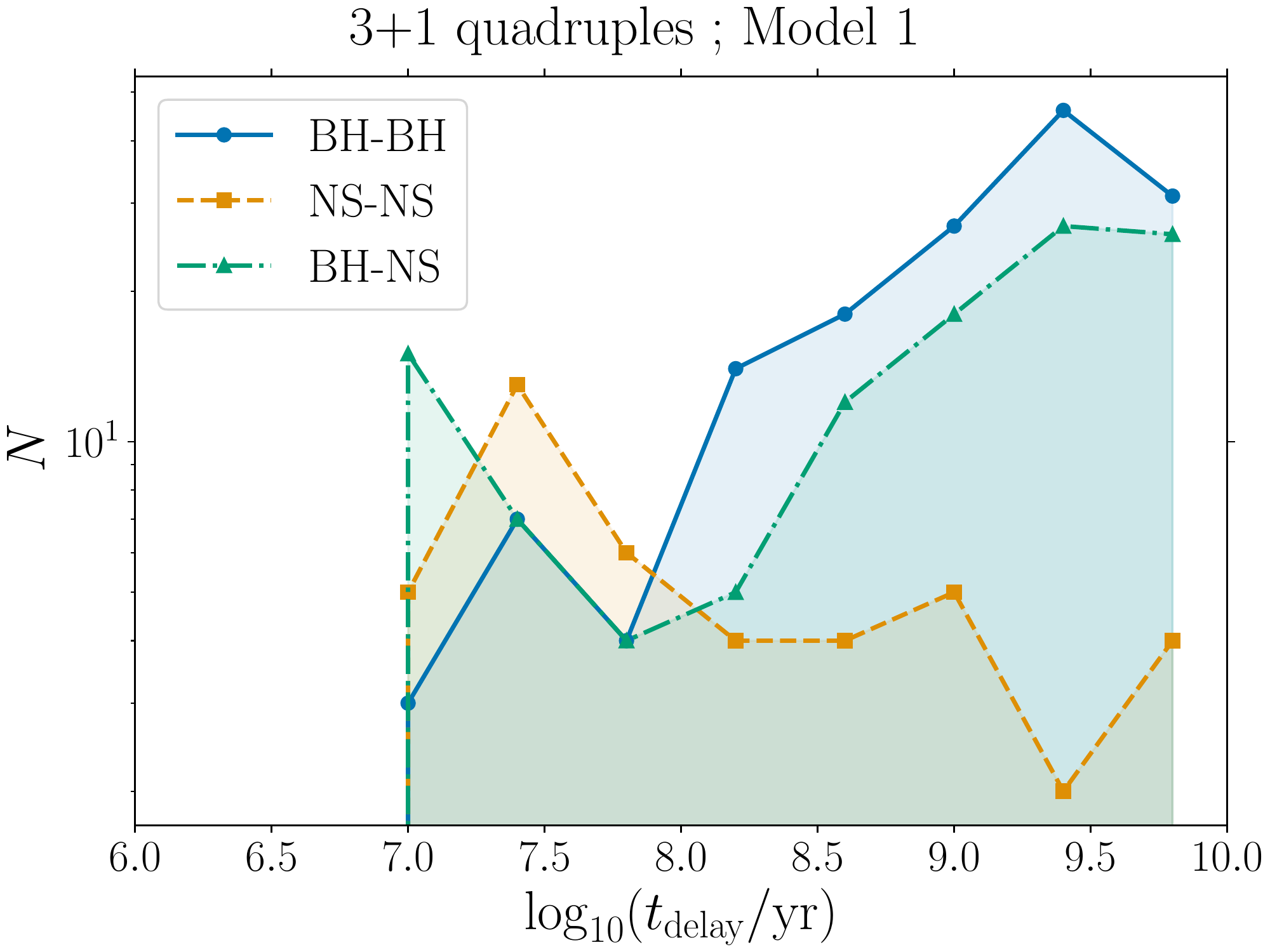}{0.45\textwidth}{(d)}
          }
\caption{Frequency polygon of delay time $t_{\mathrm{delay}}$ of merger. [Rows correspond to 2+2 and 3+1 quadruples respectively; Columns correspond to Models 0 and 1 respectively. Isolated binary profiles are similar to 2+2 quadruples.]
\label{fig:tdelay}}
\end{figure*}

Figures \ref{fig:eLIGO}, \ref{fig:chieff}, \ref{fig:chip} and \ref{fig:mratio} show the $e_{\mathrm{LIGO}}$, $\chi_{\mathrm{eff}}$, $\chi_{\mathrm{p}}$ and $q = m_2/m_1$ ($m_1 \geq m_2$) distributions respectively, for Models 0 and 1 of different system configurations. Figure \ref{fig:mbigmsml} shows the scatter plots of the heavier vs. the lighter compact object masses for Models 0, 3a and 3b of the 2+2 quadruples. Finally, Figure \ref{fig:tdelay} shows the $t_{\mathrm{delay}}$ (time duration between the ZAMS phase and the final merger) distributions for Models 0 and 1 of the quadruple configurations.

We see a clear difference between 2+2 quadruples and isolated binaries in the distributions of $e_{\mathrm{LIGO}}$ for Model 1. Typically, $e_{\mathrm{LIGO}}$ lies in the range $10^{-3.5}$--$10^{-2.5}$, due to the orbits circularising from CE evolution and GW emission. Yet, we see that excluding SNe kicks has the effect of obtaining higher $e_{\mathrm{LIGO}}$ values ($\gtrsim 10^{-2}$). These high values can be solely attributed to secular evolution in the compact object phase (see Example 2 in Section \ref{sec:example}) since a similar effect is not seen in the binaries. 

Likewise, for systems in which CE evolution dominates, the $\chi_{\mathrm{eff}}$ of a merger product is expected to be distributed in the range 0--1. This is because the natal spins $\chi$ of the progenitors are assumed to lie uniformly in the range 0--1, and the spins and orbits were initially assumed to be aligned. This is different when SNe kicks are excluded, in which case negative $\chi_{\mathrm{eff}}$ can be explained by secular evolution -- an orbit's eccentricity and inclination can fluctuate, and hence, the spin-orbit orientation can vary. These secular evolution effects are not seen in Model 0, where SNe kicks are included. The distributions of $e_{\mathrm{LIGO}}$ and $\chi_{\mathrm{eff}}$ are qualitatively similar for BH-BH, BH-NS, and NS-NS mergers.

The $\chi_{\mathrm{p}}$ (Figure \ref{fig:chip}) distribution is dependent on the $q$ values. $\chi_{\mathrm{p}}$ is related to the spin component in the orbital plane (unlike $\chi_{\mathrm{eff}}$, which is related to the perpendicular component), and hence is relevant in quantifying the precession of the orbit. Our distribution shows that $\chi_{\mathrm{p}}$ is spread throughout the parameter space, but with a preference for small $\chi_{\mathrm{p}}$ ($\sim 0$). For Model 0 of 3+1 quadruples, high $\chi_{\mathrm{p}}$ values are not seen, which may be due to the few mergers in this case. The preference for $\chi_{\mathrm{p}} \sim 0$ can be attributed to the dominance of isolated binary evolution: in our simulations, the stellar spins were assumed to be aligned with the orbit, implying initially zero component of the spin to the orbit. Dynamical evolution can change the orbital orientations and hence increase $\chi_{\mathrm{p}}$, and there are indeed more systems with larger $\chi_{\mathrm{p}}$ in Model 1 (no kicks; secular evolution is more important) compared to Model 0.

The $q$ distribution (Figure \ref{fig:mratio}) itself is dependent on the type of merger. NS-Ns mergers have $q \gtrsim 0.5$, peaking at $\sim 1$, since all NSs have masses greater than the Chandrasekhar mass $\sim 1.4 \Msun$ and lower than $\sim 3 \Msun$ (the approximate lower limit for the BH mass). BH-NS mergers typically have low $q$ values because a BH is much heavier than an NS. The BH-BH distribution seems to be roughly flat, with not much of a skew ($q \gtrsim$ 0.3). Metallicity is important in determining the final compact object masses. This is seen in Figure \ref{fig:mbigmsml}, where the models with lower metallicities (Models 3a and 3b) consistently produce very high mass BHs ($\gtrsim 17 \Msun$, up to $\sim 27 \Msun$).

The final distribution we look at is of the delay time $t_{\mathrm{delay}}$ (Figure \ref{fig:tdelay}). All compact object mergers take $\gtrsim 10 \Myr$ to merge. As mentioned before, we limit our simulations up to $14 \Gyr$ (Hubble time). We do not notice any significant preferences relating to $t_{\mathrm{delay}}$.

\subsection{3+1 quadruples} \label{subsec:res3p1}

Table \ref{tab:merger_num} shows the number of compact object mergers, along with the Poisson error, in $10^5$ 3+1 quadruple systems. The table shows that the number of mergers in 3+1 quadruples is significantly lower than in 2+2 quadruples (almost by a factor of 5 for Model 0). This is not surprising since stable 3+1 configurations need to be more hierarchical and can get destabilized more easily. Additionally, we can see that models 1 (zero SNe kicks) and 3a/3b (lower metallicities) host a much higher proportion of mergers than in the case of the 2+2 quadruples. This, again, is due to the susceptibility of 3+1 quadruples to get unbound, for example, due to even small SNe kicks. The details of individual models have similar explanations as in the case of 2+2 quadruples while keeping the above points in mind.

We refer to the Figures \ref{fig:eLIGO}, \ref{fig:chieff}, \ref{fig:chip}, \ref{fig:mratio} and \ref{fig:tdelay} for the distributions of $e_{\mathrm{LIGO}}$, $\chi_{\mathrm{eff}}$, $\chi_{\mathrm{p}}$, $q$ and $t_{\mathrm{delay}}$. The $e_{\mathrm{LIGO}}$ and $\chi_{\mathrm{eff}}$ distributions for Model 1 show the outlier effect (seen in 2+2 quadruples; see section \ref{subsec:res2p2}) characteristic to secular evolution. The other distributions also qualitatively follow the 2+2 quadruples, albeit with fewer data points. Thus, it is not possible to distinguish between 2+2 and 3+1 quadruples based on the parameter distributions alone.

\subsection{Scenarios of mergers} \label{subsec:scenario}
In this section, we look back on the three merger scenarios referred to in Section \ref{sec:example} - \textit{Only CE evolution} (Scenario 1), \textit{Only secular evolution} (Scenario 2) and \textit{Mixture of both} (Scenario 3). An attempt is made to estimate the fraction of merger systems that belong in each of these scenarios. It should, however, be noted that the distinction made between Scenarios 1 and 3 may not be unique, since it is generally difficult to quantify whether or not secular evolution played a decisive role.

The classification is as follows. Any merger system which does not undergo any CE or triple CE event throughout its evolution is considered in Scenario 2. The rest of the systems have a CE event at some point in their evolution. The distinction between Scenarios 1 and 3 is more arbitrary. To see if a system has been affected by secular evolution, we check for changes in the periapsis distance $r_{\mathrm{p}}$ in the early stages of evolution. More specifically, the following conditions need to be satisfied by the inner or intermediate (in 3+1 quadruples) binaries to be classified under Scenario 3: $r_{\mathrm{p}} < 0.8 \,r_{\mathrm{p,0}}$ and $r_{\mathrm{p,0}} > 5 \au$. Here, $r_{\mathrm{p,0}}$ is the inner or intermediate periapsis at $t=0$, and this condition is checked for the first three log entries of MSE (a time step in MSE is `logged' when there are important events -- stellar type changes, supernovae, RLOF or CE events, collisions, dynamical instabilities, etc.). Any system which does not satisfy these conditions is considered in Scenario 1.

\begin{figure}
\plotone{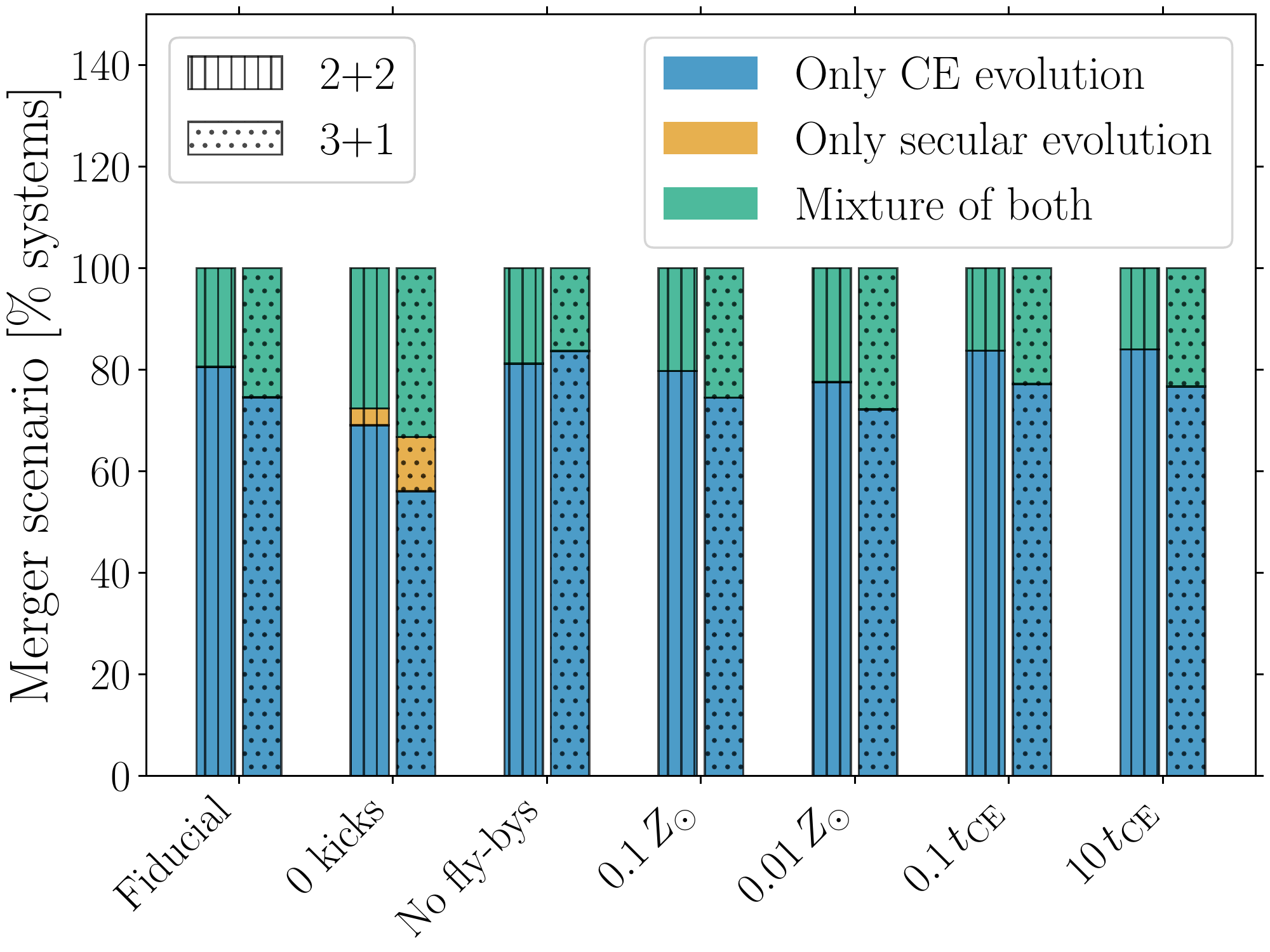}
\caption{Bar graph of percentages of the three scenarios of compact object mergers in 2+2 and 3+1 quadruples. \label{fig:merger_scn}}
\end{figure}

The resulting scenario fraction are presented in Figure \ref{fig:merger_scn}. In every model other than Model 1, Scenario 3 contributes to (15--30) \% of all compact object mergers. The rest are contributed by Scenario 1. 3+1 quadruples have a systematically higher contribution from Scenario 2 as well, most likely because there is more room for secular interactions. Unlike in 2+2 quadruples, the 3+1 quadruples have an intermediate orbit whose eccentricity and inclination can change, which can, in turn, affect the eccentricity and inclination of the inner orbit. Model 1 is the only one where Scenario 2 is observed -- 3 \% and 10 \%, respectively, in 2+2 and 3+1 quadruples. This exemplifies how much SNe kicks affect the evolution of quadruple-star systems.

We stress that our estimated percentages may not represent the true fraction of systems affected by secular evolution. Firstly, our periapsis condition is checked only for the first few log entries. Any future secular evolution effects due to changed inclinations are not taken into account. Thus, the Scenario 3 fraction might be underestimated. Secondly, secular evolution does not aid all the merger systems in Scenario 3. For example, in 2+2 quadruples, one of the inner binary eccentricities might be enhanced, while the actual merger takes place in the other inner binary. In these cases, the Scenario 3 fraction might be overestimated. Nonetheless, Scenario 2 is unambiguously defined and, in particular, shows that the overall fraction of systems in which `clean' secular evolution leads to compact object mergers is very small.

\subsection{Merger rate calculation} \label{subsec:rate}

Our population synthesis systems do not represent the whole parameter space of quadruple-star systems, let alone all types of stellar systems. Thus, we need to take into consideration the assumptions made to get a realistic estimate for compact object merger rates. We employ a rate calculation method similar to the one used by \cite{2013MNRAS.430.2262H}. 

The star formation rate (SFR) at redshift $z=0$ is assumed to be $R_{\mathrm{SF}} = 1.5 \times 10^7 \MsunGpcyr$ (given by \citealp{2014ARA&A..52..415M}). The overall merger rate $R_{\mathrm{merge;conf}}$ for a given model of a given quadruple configuration can be expressed by the following equation:
\begin{align}
\begin{split}
    R_{\mathrm{merge;conf}} &\sim R_{\mathrm{SF}} \frac{N_{\mathrm{merge;conf}}}{M_{\mathrm{sim}}} \\
    &\sim \frac{R_{\mathrm{SF}}}{N_{\mathrm{sample}}} \frac{N_{\mathrm{merge;conf}} F_{\mathrm{quad;conf}}}{M_{\mathrm{avg}}}
\label{eq:rate}
\end{split}
\end{align}
where $N_{\mathrm{sample}} = 10^5$ (total number of sampled systems in each model), $N_{\mathrm{merge;conf}}$ values are given in Table \ref{tab:merger_num} for given model and configuration (2+2 or 3+1) of quadruples, $M_{\mathrm{sim}}$ is the total mass represented by our simulation, $M_{\mathrm{avg}}$ is the average system mass of \textit{all types} (singles, binaries, triples and quadruples) and $F_{\mathrm{quad;conf}}$ is the fraction of parameter space represented by a configuration of quadruples. In our rate normalisation calculations, we neglect the contribution of quintuples and higher-order systems.

\begin{figure}
\plotone{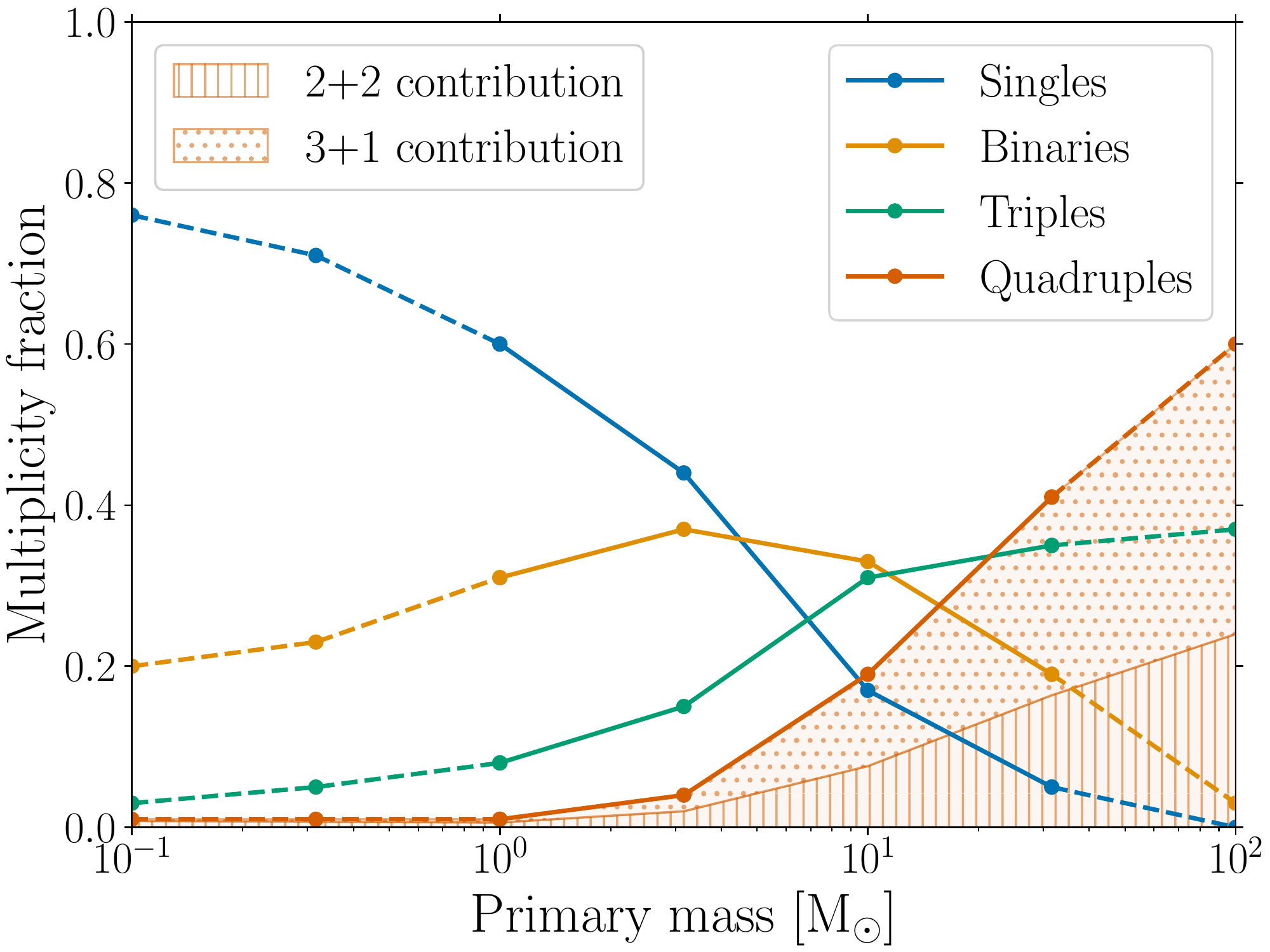}
\caption{Distribution of multiplicity fraction as a function of primary mass. The solid lines are approximate interpolated values from \cite{2017ApJS..230...15M}, while the dotted lines are extrapolations. \label{fig:fmultiple}}
\end{figure}

\begin{figure}
\plotone{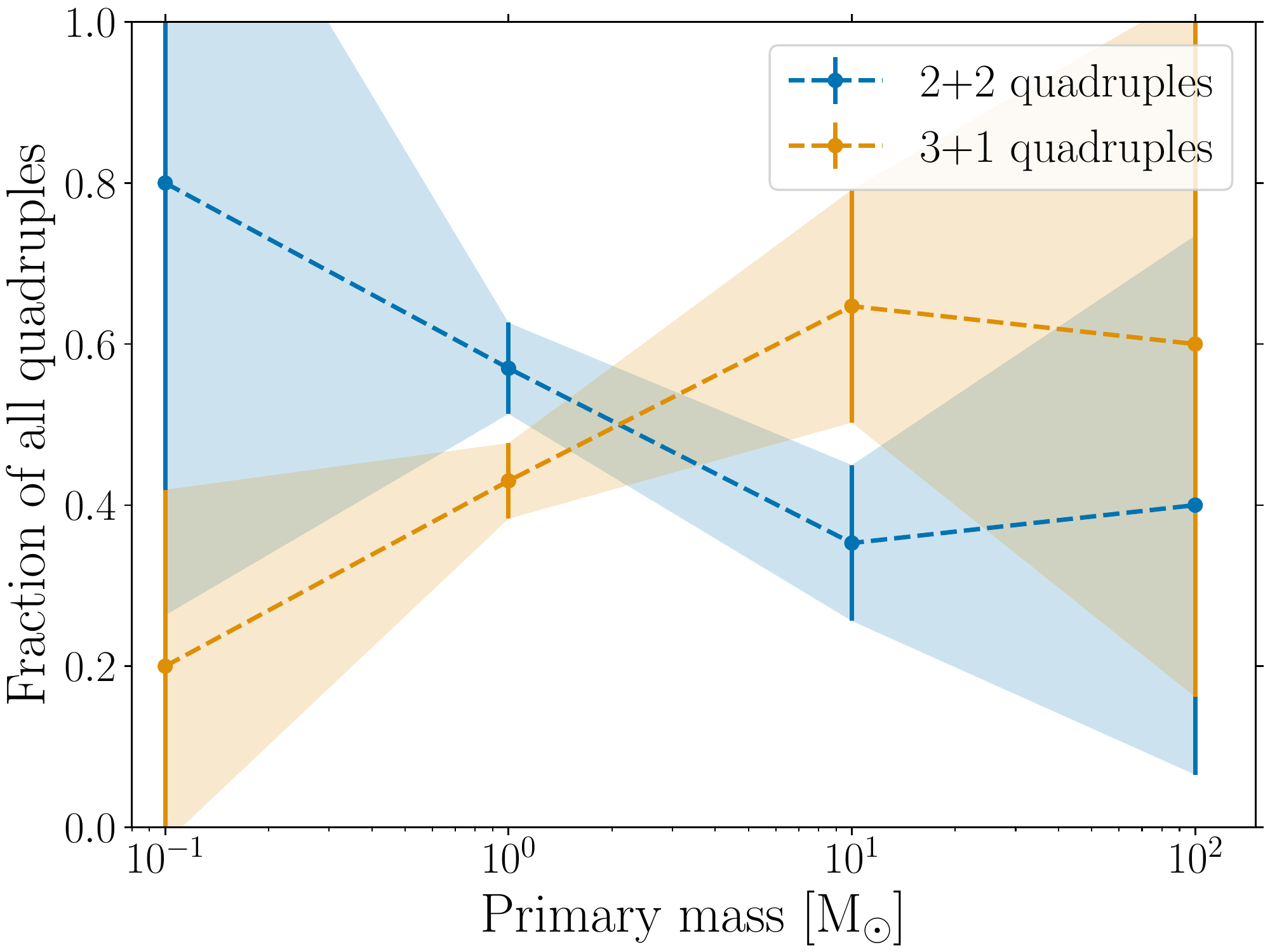}
\caption{Distribution of 2+2 and 3+1 quadruple fraction as a function of primary mass. Systems from the Multiple Star catalog are binned, and the errors are represented by the shaded regions. \label{fig:f_quad}}
\end{figure}

Let us first start with the masses. We assume that the universal initial mass function (IMF) is the Kroupa distribution ($\mathrm{d}N/\mathrm{d}m \propto m^{-1.3}$, $m<0.5\Msun$ and $\mathrm{d}N/\mathrm{d}m \propto m^{-2.3}$, $m>0.5\Msun$; \citealp{2001MNRAS.322..231K}). We then sample single, binary, triple and quadruple (both types) systems from this IMF as done in Section \ref{subsec:ICs} and calculate their average masses $\mu_{\mathrm{sin}}$, $\mu_{\mathrm{bin}}$, $\mu_{\mathrm{trip}}$ and $\mu_{\mathrm{quad}}$ in different mass bins. Each mass bin also occupies a fraction $f$ of the IMF. Next, we use interpolated and extrapolated values of multiplicity fractions from \cite{2017ApJS..230...15M} to calculate the contributions of singles $\alpha_{\mathrm{sin}}$, binaries $\alpha_{\mathrm{bin}}$, triples $\alpha_{\mathrm{trip}}$ and quadruples $\alpha_{\mathrm{quad}}$ to the average system mass $M_{\mathrm{avg}}$ in these mass bins. These fractions are shown in Figure \ref{fig:fmultiple}. In the case of quadruples, we further separate the contributions $\lambda_{2+2}$ and $\lambda_{3+1}$ of 2+2 and 3+1 quadruples respectively. To do this, we analyze the quadruple systems in the comprehensive Multiple Star catalog (MSC) \citep{1997A&AS..124...75T,2018ApJS..235....6T,2018yCat..22350006T} and calculate fractions of 2+2 and 3+1 quadruples in four mass bins. This is shown in Figure \ref{fig:f_quad}. It should be mentioned that the MSC suffers from observational biases, but it is currently the best source to infer fractions of the two types of quadruples. Finally, we put these contributions together to give:
\begin{equation}
\begin{split}
    M_{\mathrm{avg}} = \sum_{m_{\mathrm{bins}}} \left( f_{\mathrm{sin}}\alpha_{\mathrm{sin}}\mu_{\mathrm{sin}} + f_{\mathrm{bin}}\alpha_{\mathrm{bin}}\mu_{\mathrm{bin}} + f_{\mathrm{trip}}\alpha_{\mathrm{trip}}\mu_{\mathrm{trip}} \right. \\
    \left. +  f_{\mathrm{quad;2+2}}\alpha_{\mathrm{quad}}\lambda_{\mathrm{2+2}}\mu_{\mathrm{quad}} + f_{\mathrm{quad;3+1}}\alpha_{\mathrm{quad}}\lambda_{\mathrm{3+1}}\mu_{\mathrm{quad}} \right)
\label{eq:mavg}
\end{split}
\end{equation}

The other quantity in Equation \ref{eq:rate} is $F_{\mathrm{quad;conf}}$:
\begin{equation}
    F_{\mathrm{quad;conf}} = \sum_{m_{\mathrm{bins}}} f_{\mathrm{quad;conf}}\alpha_{\mathrm{quad}}\lambda_{\mathrm{conf}}
\label{eq:Fconf}
\end{equation}

The final calculated rates are shown in Table \ref{tab:merger_rate}. The errors indicate the Poisson errors.

\begin{deluxetable*}{cc|ccc|ccc}
\tablecaption{Merger rates of compact object mergers in 2+2 and 3+1 quadruples. \label{tab:merger_rate}}
\tablewidth{0pt}
\tablehead{
\colhead{Model} & \colhead{Description} & \multicolumn3c{2+2 quadruples [$\Gpcyr$]} & \multicolumn3c{3+1 quadruples [$\Gpcyr$]} \\
\colhead{} & \colhead{} & \colhead{BH-BH} & \colhead{BH-NS} & \colhead{NS-NS} &
\colhead{BH-BH} & \colhead{BH-NS} & \colhead{NS-NS}
}
\startdata
0 & Fiducial & 10.8 \tiny $\scriptscriptstyle \pm$ 0.9 & 5.7 \tiny $\scriptscriptstyle \pm$ 0.6 & 0.6 \tiny $\scriptscriptstyle \pm$ 0.2 & 2.9 \tiny $\scriptscriptstyle \pm$ 0.5 & 1.4 \tiny $\scriptscriptstyle \pm$ 0.4 & 0.7 \tiny $\scriptscriptstyle \pm$ 0.3 \\
1 & 0 kicks & 19.7 \tiny $\scriptscriptstyle \pm$ 1.2 & 14.9 \tiny $\scriptscriptstyle \pm$ 1.0 & 24.3 \tiny $\scriptscriptstyle \pm$ 1.3 & 14.8 \tiny $\scriptscriptstyle \pm$ 1.2 & 10.8 \tiny $\scriptscriptstyle \pm$ 1.0 & 4.0 \tiny $\scriptscriptstyle \pm$ 0.6 \\
2 & No fly-bys & 7.5 \tiny $\scriptscriptstyle \pm$ 0.7 & 5.5 \tiny $\scriptscriptstyle \pm$ 0.6 & 0.7 \tiny $\scriptscriptstyle \pm$ 0.2 & 2.2 \tiny $\scriptscriptstyle \pm$ 0.5 & 1.5 \tiny $\scriptscriptstyle \pm$ 0.4 & 0.7 \tiny $\scriptscriptstyle \pm$ 0.3 \\
3a & 0.1 $\Zsun$ & 19.0 \tiny $\scriptscriptstyle \pm$ 1.2 & 13.2 \tiny $\scriptscriptstyle \pm$ 1.0 & 1.4 \tiny $\scriptscriptstyle \pm$ 0.3 & 6.5 \tiny $\scriptscriptstyle \pm$ 0.7 & 5.8 \tiny $\scriptscriptstyle \pm$ 0.7 & 1.5 \tiny $\scriptscriptstyle \pm$ 0.4 \\
3b & 0.01 $\Zsun$ & 29.7 \tiny $\scriptscriptstyle \pm$ 1.5 & 36.3 \tiny $\scriptscriptstyle \pm$ 1.6 & 3.5 \tiny $\scriptscriptstyle \pm$ 0.5 & 7.8 \tiny $\scriptscriptstyle \pm$ 0.8 & 17.9 \tiny $\scriptscriptstyle \pm$ 1.3 & 2.8 \tiny $\scriptscriptstyle \pm$ 0.5 \\
4a & 0.1 $t_{\mathrm{CE,0}}$ & 10.2 \tiny $\scriptscriptstyle \pm$ 0.8 & 5.0 \tiny $\scriptscriptstyle \pm$ 0.6 & 1.0 \tiny $\scriptscriptstyle \pm$ 0.3 & 2.7 \tiny $\scriptscriptstyle \pm$ 0.5 & 1.7 \tiny $\scriptscriptstyle \pm$ 0.4 & 0.7 \tiny $\scriptscriptstyle \pm$ 0.3 \\
4b & 10 $t_{\mathrm{CE,0}}$ & 10.3 \tiny $\scriptscriptstyle \pm$ 0.8 & 4.7 \tiny $\scriptscriptstyle \pm$ 0.6 & 0.7 \tiny $\scriptscriptstyle \pm$ 0.2 & 3.2 \tiny $\scriptscriptstyle \pm$ 0.5 & 1.5 \tiny $\scriptscriptstyle \pm$ 0.4 & 0.7 \tiny $\scriptscriptstyle \pm$ 0.3 \\
\enddata
\tablecomments{Refer to Table \ref{tab:modelparam} for detailed model specifications.}
\end{deluxetable*}

\subsection{Systems not considered} \label{subsec:toolong}

\begin{figure}
\plotone{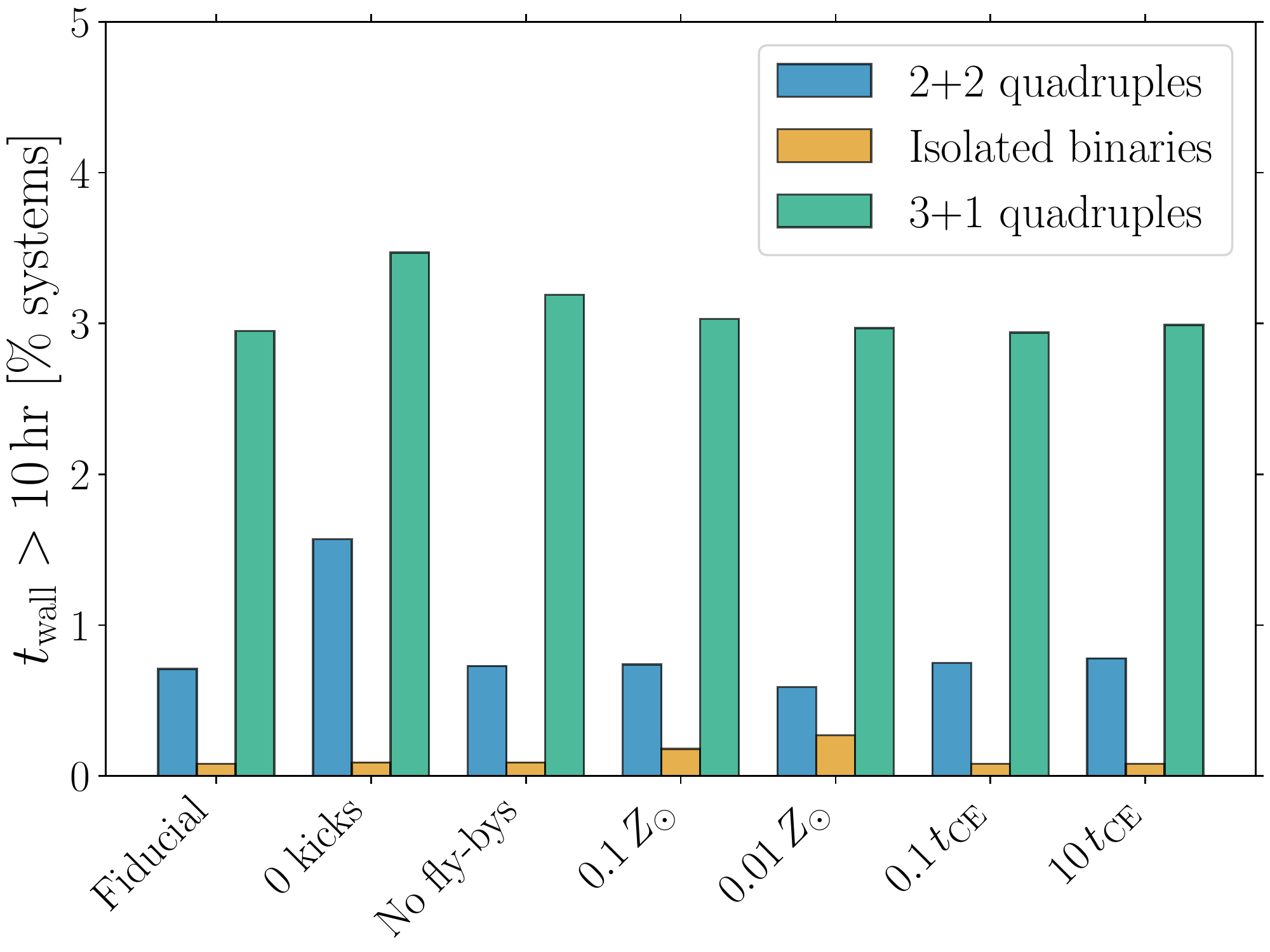}
\caption{Bar graph of percentage of systems for which $t_{\mathrm{wall}} > 10 \hr$ in 2+2 quadruples, isolated binaries and 3+1 quadruples. \label{fig:toolong}}
\end{figure}

Not all of the systems ran completely. For such systems, the code gets stuck either in the direct $N$-body or secular integration modes. Thus, we terminate any system which takes longer than a CPU wall time of $t_{\mathrm{wall}} = 10 \hr$ to run. We note that this is a reasonable time limit given that many systems take a few seconds to a few minutes to run. For completeness, in Figure \ref{fig:toolong}, we present the percentage of systems which take longer than $t_{\mathrm{wall}}$ to run.

The isolated binary models have the least proportion of such systems ($0.1 \%$--$0.3 \%$) and the 3+1 quadruples have the most ($2.9 \%$--$3.5 \%$). The proportion for 2+2 quadruples is midway ($0.6 \%$--$1.6 \%$). This is expected since dynamical integration is straightforward in binaries while it is most complicated in 3+1 quadruples. In the case of the quadruples, it can be seen that Model 1 has the highest proportion of systems with $t_{\mathrm{wall}} > 10 \hr$. This is because zero SNe kicks aid in keeping more systems bound, and hence, there is a higher chance for the gravitational dynamical integration to be slow.

However, more than $96\%$ of the systems in all our models do run completely. Moreover, the offending
systems are not clustered but spread throughout our
initial parameter space. Hence, our overall statistics are not affected.

\section{Discussion} \label{sec:discuss}

\subsection{Merger rate comparisons} \label{subsec:rate_compare}

To put our estimated compact object merger rates in quadruples in context, we mention the compact object merger rates derived from LIGO detections. \cite{2021ApJ...913L...7A} determined the GWTC-2 BH-BH and NS-NS merger rates to be $23.9^{+14.3}_{-8.6} \Gpcyr$ and $320^{+490}_{-240} \Gpcyr$ respectively. \cite{2021ApJ...915L...5A} carried out a similar analysis for BH-NS mergers and found the rate to be $45^{+75}_{-33} \Gpcyr$ (assuming the two BH-NS detections are representative of the whole population) or $130^{+112}_{-69} \Gpcyr$ (assuming a broader distribution of masses). The rates we observe are either in agreement with the LIGO rates (for BH-BH mergers) or much lower (for NS-NS mergers, and BH-Ns mergers to a lesser extent). Merger rates also depend on star formation rates, which depend on redshift \citep{2014ARA&A..52..415M}.

Let us now compare with the other theoretical channels of GW emission. Most of these studies only consider BH-BH mergers, of which quadruples can constitute a significant fraction. The high NS-NS merger rates inferred from LIGO detections are not reproduced in many population synthesis studies due to the high SNe kicks attributed to NSs. This is true for our study as well. Generally, dynamical merger channel can have high LIGO band eccentricities $e_{\mathrm{LIGO}}$, and possibly anti-aligned (negative) effective spins $\chi_{\mathrm{eff}}$. Many of the below-mentioned rates are uncertain and should be taken as order of magnitude estimates.
\begin{itemize}
    \item \textit{Isolated binary channel:} \cite{2016Natur.534..512B} estimated a high BH-BH merger rate of $\sim 200 \Gpcyr$ within redshift $z \sim 0.1$, for their standard model of isolated binary evolution of very massive stars ($\gtrsim 40 \Msun$). In contrast, \cite{2016MNRAS.460.3545D} predicted a local BH-BH merger rate of $\sim 10 \Gpcyr$, and a rate of $\sim 20 \Gpcyr$ at redshift $z \sim 0.4$, for chemically homogeneous evolution in tidally distorted massive binaries. Since there is no secular evolution in isolated binaries, LIGO band eccentricities $e_{\mathrm{LIGO}}$ is very small and  $\chi_{\mathrm{eff}}$ is aligned with their orbits.
    \item \textit{Dynamical channel in star clusters:} \cite{2016PhRvD..93h4029R} predicted a local BH-BH merger rate of $\sim 5 \Gpcyr$ in globular clusters using a Monte Carlo approach, with 80 \% of them being in the mass range $\sim$ (30--60)$\Msun$. They also found that nearly all of the BH-BH systems circularize and have $e_{\mathrm{LIGO}} \lesssim 10^{-3}$, similar to isolated binaries. In comparison, \cite{2017MNRAS.467..524B} used direct $N$-body evolution to estimate a LIGO merger BH-BH rate of $\sim 13 \,\mathrm{yr}^{-1}$ within a radius of $1.5 \Gpc$, which is equivalent to $\sim 3 \Gpcyr$. BH-NS and NS-NS mergers are unlikely since NSs do not efficiently segregate to the center (as BHs do).
    \item \textit{Dynamical channel in galactic nuclei:} \cite{2017ApJ...846..146P} estimated BH-BH, BH-NS and NS-NS merger rates of $\lesssim 15 \Gpcyr$, $\lesssim 0.4 \Gpcyr$ and $\lesssim 0.02 \Gpcyr$ respectively for binaries in the sphere of influence of the central massive BH (MBHs) in galactic nuclei. They also noted that the fraction of systems that reach extremely high eccentricities ($1-e \sim$ $10^4$--$10^6$) is $\sim$ (10--100) higher than in spherical clusters. The predictions of \cite{2018ApJ...865....2H} agree with the above, with their most optimistic BH-BH merger rate being $\sim 12 \Gpcyr$. \cite{2018ApJ...856..140H} predicted lower BH-BH merger rates of $\sim$ (1--3)$\Gpcyr$ using Monte Carlo simulations.
    \item \textit{Dynamical channel in triple systems:} \cite{2017ApJ...836...39S} used a simple triple BH assumption to estimate a merger rate of $\sim 6 \Gpcyr$. On the other hand, \cite{2017ApJ...841...77A} combined stellar evolution and dynamics to predict BH-BH merger rates of $\sim$ (0.3--1.3)$\Gpcyr$, with many of the mergers having $e_{\mathrm{LIGO}} > 10^{-2}$. For BH-NS mergers, \cite{2019MNRAS.486.4443F} estimated rates of $\sim$ ($1.0\times10^{-3}$--$3.5\times10^{-2}$)$\Gpcyr$ when natal kicks are included.
    \item \textit{Dynamical channel in quadruple systems:} \cite{2019MNRAS.483.4060L} showed that interactions between the two inner binaries in 2+2 quadruples can result in resonances which can increase the number of mergers by almost an order of magnitude compared to similar triples. Another study by \cite{2019MNRAS.486.4781F} (assuming all components are BHs) showed that this factor can be $\sim$ (3--4). \cite{2021MNRAS.506.5345H}, who carried a simplified population synthesis of 2+2 quadruples, inferred optimistic rates of $\sim$ (10--100)$\Gpcyr$, of which the lower limits are consistent with our study. They also predicted second-generation mergers (see later) at a rate $\sim 10^{-5} \Gpcyr$, which is not seen in our population synthesis (possibly due to low resolution). As in the case of triples, all these studies see a non-negligible fraction of mergers with high $e_{\mathrm{LIGO}}$. Moreover, \cite{2021MNRAS.506.5345H} noted some merger products with negative $\chi_{\mathrm{eff}}$.
    \item \textit{AGN disk channel:} \cite{2017ApJ...835..165B} and \cite{2017MNRAS.464..946S} calculated BH-BH merger rates of $\sim 1.2 \Gpcyr$ and $\sim 3 \Gpcyr$ in AGN disks. This channel is interesting due to potential electromagnetic counterparts in the form of X-rays or $\gamma$-rays, emitted due to super-Eddington accretion.
\end{itemize}

\subsection{Caveats} \label{subsec:caveats}

As emphasized several times, the main takeaway from our estimated merger rate is that SNe kicks are a major deciding factor. The kick distribution we have chosen is a Maxwellian one. Even though NS kicks have been constrained, to some extent, from the motion of observed isolated pulsars (though it is not clear how this generalizes to natal kicks in multiple-star systems), BH kicks are still poorly constrained. Assuming different distributions can hence significantly affect merger rates, and the values of $e_{\mathrm{LIGO}}$ and $\chi_{\mathrm{eff}}$.

Another caveat is that we looked at quadruple-star systems in the field, where stellar encounters are typically weak and infrequent. The evolution of quadruples in high-density environments, such as star clusters or galactic nuclei, will ostensibly be more dependent on fly-bys. For example, in the cores of globular clusters, binary-binary scatterings \citep[e.g.,][]{1983MNRAS.203.1107M,1993Natur.364..423S} can dynamically form triple systems, with one of the stars escaping. Meanwhile, stable triples in clusters are much more likely to be destroyed by strong interactions. Similarly, quadruples can be dynamically formed by scattering processes\citep[e.g.,][]{2007MNRAS.379..111V} and can be destroyed by strong encounters. \cite{2016MNRAS.456.4219A} performed scattering experiments of binary-binary, triple-single and triple-binary interactions to quantify and characterize the formation of triples in clusters. \cite{2016ApJ...816...65A} studied dynamically formed triple systems in globular clusters using Monte Carlo models for the cluster coupled with an $n$-body integrator. They found that the timescales for angular momentum change can become comparable to the orbital periods, thereby making the secular approximation inaccurate. They also found eccentric LIGO band mergers. \cite{2020ApJ...903...67M} performed a similar study and derived conservative BH-BH merger rates of $\sim 0.35 \Gpcyr$. However, a study on quadruples in clusters is yet to be carried out and is beyond the scope of this paper.

We also looked at the effect of the CE mass-loss timescale $t_{\mathrm{CE}}$ and concluded that it does not affect merger rates meaningfully. Other studies \citep{2012ApJ...759...52D,2021MNRAS.tmp.2473B,2021ApJ...918L..38F} have shown that the CE efficiency parameter $\alpha_{\mathrm{CE}}$ has a more significant effect. Moreover, CE  evolution still faces significant uncertainties.

The main question this work has tried to address is the contribution of secular evolution to compact object mergers in quadruple-star systems. In Section \ref{subsec:scenario}, we described the different scenarios of mergers and concluded that secular evolution indeed has an effect, although perhaps not as significant as could be expected based on secular dynamics alone. However, our classification does not represent systems that undergo pre-compact object phase stellar collisions or have dynamical instabilities. Here, we briefly mention a few other examples from our population synthesis, leading to compact object mergers, not shown in Section \ref{sec:example}:
\begin{itemize}
    \item In 2+2 quadruples, the two inner binaries can be close enough that they merge in their main-sequence (MS) or giant phases. Alternatively, eccentricity enhancements cause them to merge prematurely. Now, we have a binary system of merger products, typically very massive stars, which can evolve into BHs and merge. There can also be systems in which there is only one inner binary merger in the pre-compact object phase. The resulting system evolves as a triple-star system.
    \item In 3+1 quadruples, the intermediate orbit's eccentricity can be enhanced to an extent that the resulting configuration becomes dynamically unstable. One of two things can happen -- either there is a collision, or one of the stars gets ejected from the system, resulting in a triple hierarchy. This triple system can now evolve and produce a merger.
\end{itemize}

Such examples show the wide complexity in the evolution of quadruples, and also explain why second-generation compact object mergers (see \citealp{2020ApJ...895L..15F,2021MNRAS.506.5345H} for examples) are rare. In fact, in our simulations, none of such mergers have been observed, not even in the zero SNe kicks case. Nevertheless, we can expect to see a few if the number of systems run was significantly higher. For a second-generation merger to occur, we need to have at least three compact objects in a bound stable configuration, which in itself never happens in our non-zero SNe kick models. Then, the first merger should occur, either due to interaction with the tertiary or due to a prior CE evolution and GW emission-aided orbit shrinkage. Now, the two resulting compact objects are far enough that they will not merge within a Hubble time, and there is no companion to enhance the merger. 

The possibility of having four bound compact objects is even more scarce than having three of them. Even then, the orbit alignments and separations need to be just right to have a second-generation merger within a Hubble time. Thus, quadruple systems probably cannot explain recent GW detections of higher mass BH ($\gtrsim 40\Msun$) mergers.

\section{Conclusion} \label{sec:conclude}
We used the population synthesis code MSE \citep{2021MNRAS.502.4479H} to look for black hole (BH) and neutron star (NS) mergers in quadruple-star systems, from their birth as main-sequence (MS) stars to their death as compact objects, taking into account a wide range of physical processes. We looked at the two configurations of quadruples -- 2+2 and 3+1 -- and compared the 2+2 quadruples with `isolated' binaries. We also compared seven different models (with altered parameters) for each of the configurations. Our main conclusions are listed below:
\begin{enumerate}
    \item Quadruple-star systems contribute to a significant fraction of BH-BH mergers, with their merger rates being on the order of the LIGO rate. BH-NS mergers also do occur in quadruples, but not as many as LIGO rates predict. On the other hand, NS-NS merger rates are extremely low due to SNe kicks in these systems. The measured rates for our fiducial model (includes both types of quadruples) are $13.7 \Gpcyr$, $7.1 \Gpcyr$, and $1.3 \Gpcyr$ respectively for BH-BH, BH-NS, and NS-NS mergers.
    \item 2+2 quadruples have similar merger numbers to the isolated binaries, although the latter have higher merger numbers in all models except our model with zero SNe kicks. This indicates that, although the effects of secular evolution are seen in (15 --30) \% of systems, they can either aid or hinder compact object mergers.
    \item Only in $\sim$ (3--10) \% of cases (when SNe kicks are excluded) is a compact object merger not associated with CE evolution, and a compact object multiple system forms successfully. In such cases, the mergers are due to dynamically induced high eccentricities.
    \item A comparison of the two types of quadruples 2+2 and 3+1 shows that the former has many more BH-BH and BH-NS than the latter (by a factor of 3--4). The NS-NS merger rates are comparable (with SNe kicks).
    \item SNe kicks are the most important factor for determining merger rates. Excluding kicks increases the number of mergers by factors of $\sim$ 2--40 for 2+2 quadruples, and $\sim$ 5--8 for 3+1 quadruples. The large outlier (factor $\sim$ 40 increase) is the NS-NS merger rate in 2+2 quadruples. The increase in BH-BH and BH-NS mergers is more significant in 3+1 quadruples, however.
    \item Metallicity $Z$ is another parameter that significantly affects the merger numbers, similar to isolated binaries. For $Z = 0.1 \Zsun$, merger numbers are scaled up by factors of $\sim$ 2--4. For $Z = 0.01 \Zsun$, they are scaled up by factors of $\sim$ 3--10. The greatest increase is seen in BH-NS mergers ($\sim$ 3--4 and $\sim$ 6--10 for the two metallicities respectively) for both types of quadruples, but more drastically in the 3+1 quadruples.
    \item For quadruples, the LIGO band ($f_{\mathrm{GW}} \sim 10 \Hz$) eccentricities $e_{\mathrm{LIGO}}$ lie in the range $10^{-3.5}$ to $10^{-2.5}$ in all the models where supernova (SNe) kicks are included. In the latter case, $e_{\mathrm{LIGO}}$ can be high, even up to 0.3 (for both 2+2 and 3+1 systems). This can be attributed to secular evolution in the compact object phase. 
    \item Effective spin parameters $\chi_{\mathrm{eff}}$ of the compact object mergers, for quadruples, mostly lie in the range 0--1 in all the models where supernova (SNe) kicks are included. In the latter case, a significant fraction of systems have negative $\chi_{\mathrm{eff}}$. Similarly to $e_{\mathrm{LIGO}}$, the outliers, with negative $\chi_{\mathrm{eff}}$, can be explained by secular evolution.
    \item BH masses pre-merger can go up to $\sim 17 \Msun$ for Solar metallicity systems. Lower metallicities can produce even higher masses, up to $\sim 27 \Msun$ for $Z = 0.01 \Zsun$. We find no second-generation mergers, which could be due to their low formation rate and the limited number of systems considered here.
    \item Excluding fly-bys decreases the number of BH-BH mergers by a factor of $\sim 0.7$ for both types of quadruples. However, this decrease is not seen for BH-NS and NS-NS mergers. In the case of binaries, there is a noticeable decrease in all three types of mergers.
    \item The common envelope mass-loss timescale $t_{\mathrm{CE}}$ does not alter merger rates much. Scaling or descaling $t_{\mathrm{CE}}$ by a factor of 10 does not change merger numbers beyond the Poisson error deviation. This can be understood by noting that isolated binary evolution (in particular, CE evolution) is the dominant factor in driving compact object mergers in quadruples. Therefore, whether or not outer orbits in the multiple system remain bound after a CE event is not very important, since secular evolution on its own only drives a small fraction of compact object mergers. 
\end{enumerate}

\acknowledgments

We thank Maxwell Moe for his comments regarding the relation of multiplicity fraction with primary mass. We also thank Patrick Neunteufel for providing comments on the manuscript. A.S.H. thanks the Max Planck Society for support through a Max Planck Research Group.

\vspace{5mm}
\facilities{\textit{ada} cluster at the Max Planck Institut für Astrophysik}

\software{ 
          MSE \citep{2021MNRAS.502.4479H} 
          }

\bibliography{quad_popsyn}
\bibliographystyle{aasjournal}

\end{document}